\documentclass{thesis}
\usepackage{psfig,latexsym,mathrsfs,axodraw}
\usepackage[dvips]{epsfig}
\begin{document}
\def\Tr{\mathop{\rm Tr}\nolimits}
\def\bea{\begin{eqnarray}}
\def\eea{\end{eqnarray}}
\submissionmonth{April}
\submissionyear{2002}
\author{\bf {INDRAJIT MITRA} \\ THEORY GROUP \\ SAHA INSTITUTE OF NUCLEAR PHYSICS \\ KOLKATA}
\title{SOME ASPECTS OF \\PARTICLE PHYSICS \\IN THERMAL BACKGROUND}
\maketitle

%
%
%
\begin{acknowledgements}
{\it I am indebted to Prof.\ Palash Baran Pal for guidance, encouragement, collaboration
and numerous stimulating discussions. I am also grateful to Prof.\ Jos\'e F.\ Nieves for
collaboration and many valuable discussions via e-mail.

Interactions with many other members, both past and present, of the Theory Group of SINP,
including Dr.\ Ananda DasGupta, Prof.\ Asit K.\ De, Prof.\ Binayak Dutta-Roy, Dr.\ Abhee K.\
Dutt-Mazumder, Prof.\ A.\ Harindranath and Prof.\ Samir Mallik, have helped me learn
quantum field theory and thermal field theory. I thank them all.

I express my gratitude to all members of the Theory Group for their help and
cooperation. The Research Fellows, in particular, have helped me in every possible way
during my stay here.

Last, but not the least, I thank my mother for her support and understanding, without
which this work could not have been done.}

\end{acknowledgements}

%

\tableofcontents
\listoffigures

%
%

\chapter{Introduction}\label{intro}
\section{Importance of particle physics in thermal background}
The study of elementary particles in vacuum is based on quantum
 field theory. However, there are situations in which
it is necessary to consider these particles in a non-trivial background
 consisting of a large number of particles at finite temperature and/or density.
Such temperatures and densities of significance occur in stellar interiors and were also
present in the
 early universe. Another relevant situation occurs in the laboratory
during heavy-ion collisions. The formalism needed for the study of these
 situations is provided by thermal field theory, also known 
as finite temperature field theory \cite{kapusta, lebellac, adas}.

From a theoretical point of view (as we will see in the course of the present
work), two major features are induced by the background medium. These are the
occurrence of an additional four-vector in the form of the four-velocity
of the medium, and the breaking of discrete symmetries like $C$, $P$ and $T$
even when the underlying Lagrangian respects these symmetries. As a result,
self-energies, vertex functions and other Green's functions involve
new tensor structures and associated form factors, together with a relaxation
of some of the constraints restricting the form factors. A third important
feature induced by the medium (also to be encountered in the present work) is the
breaking of flavour symmetry. All these features give rise to rich physics which
was absent in the vacuum. A modified self-energy, for example, gives rise to
modification of the dispersion relation (energy-momentum relation). This, in
turn, among other things, can open up the phase space, and allow processes which
were kinematically forbidden in the vacuum to take place in the medium.

Let us now turn to some specific illustrations of changes in the
properties of elementary particles due to a thermal background.
Gauge symmetry spontaneously broken in vacuum
can be restored by finite temperature corrections. The masses of the fermions
and gauge bosons of the Standard Model vanish when this
 happens \cite{linde}.
In the phase of restored chiral invariance, one finds the
 generation of thermal effective masses of the particles, different effective
 masses of left-handed and right-handed leptons, and different dispersion
relations of transverse and longitudinal modes of gauge bosons
\cite{qed:weldon, weldonboson}.  Modification of photon dispersion relations
leads to Cerenkov radiation (emission of a photon) by even a single neutrino
in a medium \cite{sawyer}.
The dispersion
relation of neutrinos become flavour dependent in a medium containing electrons
but not the other charged leptons \cite{notzold, josenieves}. 
In such a flavor asymmetric background, the
electromagnetic \cite{magnu} and gravitational \cite{gravnu} interactions of the neutrinos in response to
external fields are also flavor dependent, while the radiative decay of a
 massive neutrino is strongly enhanced compared to the decay in vacuum
(as the leptonic GIM mechanism becomes inoperative) \cite{prl}. 
When neutrinos travel in a background of electrons
with a static external magnetic field also present, the latter causes an
anisotropy of the neutrino dispersion relation \cite{magnu}. 
An entirely new property
of the Standard Model neutrino acquired in a background of electrons is an
effective electric charge \cite{charge}. 
A $CPT$-asymmetric background medium causes rotation of the plane of polarization
of light passing through it \cite{optac}, a phenomenon known as optical activity.

There are a number of extremely important applications of background-induced
 particle physics, some of which we now mention. 
The temperature dependence of the effective potential
for the Higgs field gives rise to the inflationary scenario in the early
universe \cite{infl}. The flavour dependence of the neutrino dispersion relation in matter
leads to the MSW effect, with consequences for the solar neutrino
 problem \cite{msw}. The
decay of a photon propagating through a medium, known as plasmon, into
neutrino-antineutrino pair is the major cooling mechanism in white dwarfs
and red giants \cite{cool, braaten}. The anisotropy of neutrino dispersion relation in matter
with background magnetic field has been suggested as a possible explanation for
the large birth velocities of pulsars \cite{segre}.
\section{The problems considered in this thesis and their motivations}
The major problem considered in this thesis is the gravitational interaction
of charged leptons in a medium.
Gravitational interactions (in the vacuum) are universal in the sense that the
ratio of the inertial and the gravitational masses of any particle is
a constant.  This fact, expressed in the form of the equivalence principle,
is one of the basic axioms of the general theory of relativity.
Although this is a feature of the theory at the classical level, it
has been shown by Donoghue, Holstein and Robinett
(DHR) \cite{DHR84} that the corresponding linearized quantum
theory of gravity also respects this ratio, at least to $O(\alpha)$.

However, it was also shown by DHR \cite{DHR84, DHR86} that this property
is lost when the particles are in the presence of a thermal background
rather than the vacuum.  To arrive at this idea, the inertial and the
gravitational masses must be defined in the context of quantum field
theory. We consider in Chapters \ref{qed} and \ref{gr} their precise
definitions in terms of the particle propagator and the gravitational
vertex, which we will need in the subsequent work.  For the moment,
let us denote the mass in vacuum by $m$, the inertial mass by $M$ and the gravitational
mass by $M'$, and
summarize the results of DHR. These authors considered the case of the electron, and
calculated $M_e$ and $M'_e$ to $O(\alpha)$, 
in a background with a
temperature $T\ll m_e$ 
and zero chemical potential. Thus, the
background contained only photons, but not electrons or any other
matter particles.
Although in those calculations only the case of the
electron was considered explicitly, the results are equally
applicable to other charged leptons, namely, the muon and the tau lepton.
Thus for any charged lepton $\ell$, 
the DHR results imply
\begin{eqnarray}
M_{\ell} &=& m_{\ell} + \frac{\alpha\pi T^2}{3m_{\ell}}\,,
\label{DHR:M}\\*
M'_{\ell} &=& m_{\ell}-\frac{\alpha\pi T^2}{3m_{\ell}}\,,
\label{DHR:Mg}\\*
{M'_{\ell} \over M_{\ell}} &=& 1 - \frac{2\alpha\pi T^2}{3m{_{\ell}^2}}
\end{eqnarray}
to $O(\alpha)$. Therefore, not only the inertial
and gravitational masses of a given charged lepton cease to be equal when the
background effects are taken into account, but, in addition, the ratio
of these two quantities is no longer the same for all the particles;
i.e., universality is lost as well. 

Now, the DHR results, as mentioned before, were obtained by considering only
photons in the thermal background. 
But in a matter background with a non-zero chemical potential (such as
the Sun), one can no longer neglect the matter particles in medium,
even when $T \ll m_e$.
It is then possible that the matter contribution to the charged lepton masses
will dominate over the photon background contribution (the latter being,
as the above results indicate,
quite small for $T \ll m_e$). Moreover, since a medium with non-zero chemical
potential is asymmetric with respect to charge conjugation,
the matter-induced corrections will not be the same for the 
corresponding antileptons.

Motivated by these considerations, in this work we calculate the
leading matter-induced QED corrections to the inertial and gravitational masses of
charged fermions in a medium that consists of a photon background and
a matter background of electrons and nucleons \cite{gravlep}. These represent the
dominant corrections for charged leptons and antileptons (weak interaction
corrections being negligible in comparison). For
strongly interacting particles such as the quarks, gluon exchange
corrections are expected to be even stronger and our results will not
apply. After finding the general expressions for the
matter-induced corrections in a generic
matter background, we give explicit formulas for the corrections in
terms of the macroscopic parameters of the background medium for a few
special cases of the background gases.

While these calculations are performed for low temperatures and densities,
in the last part of the thesis, we turn to a background at high temperature.
We consider high-temperature QED with exact chiral invariance and at zero
chemical potential, and investigate gauge independence of th one-loop electron
dispersion relation in this situation.

It is well-known that the one-loop electron dispersion relation is gauge independent
to leading order in the temperature $T$. This was demonstrated by Weldon
\cite{qed:weldon}, by showing that the leading part of the one-loop electron
self-energy (the self-energy being the field-theoretic input to the dispersion
relation) goes like $T^2$ and is gauge independent, while the gauge dependent
part goes like $T$. Now,
a somewhat similar situation occurs for the one-loop self-energy of
 neutrinos in material medium, which is gauge independent at $O(1/M{_W^2})$ but gauge
dependent when the $O(1/M{_W^4})$ terms are included. However, it has been shown
 \cite{josenieves} that the one-loop dispersion relation following from the latter is still
independent of the gauge parameter.
This raises the possibility that the gauge dependence of the part of the
 one-loop electron self-energy subleading in $T$ may also not show up in the one-loop
 dispersion relation. This possibility is explored in the present work in two
different limiting cases \cite{qed:indra}. One of theses cases is the limit of momenta much larger
than $eT$. The other is the zero momentum limit, when we investigate the gauge
independence of the effective mass.

It may be mentioned in this context that
the demonstration of gauge independence of the properties of elementary particles
in material medium has turned out to be of extreme importance in 
thermal field theory. In particular,
the gauge dependence of the gluon damping rate at one loop was a long-standing
problem. In Ref.\ \cite{kkr}, the Ward identities determining the gauge dependence
of the gluon dispersion relations were deduced and used to prove gauge independence
in a self-consistent perturbative expansion. In Refs.\
\cite{bp1} and \cite{ft}, it was shown that such a consistent expansion requires
resummation of the hard thermal loops. The use of this idea finally
led to a gauge-independent gluon damping rate to leading order in the coupling
constant \cite{bp2}. It will be seen that the
gauge dependence identities of Ref.\ \cite{kkr}, although dealing with gluons,
are still relevant for the case studied by us.
\section{Outline of the thesis}
In Chapter \ref{th}, we introduce the real-time formulation of thermal field
theory. After working out the thermal averages of bilinears of ladder operators,
we deduce the time-ordered thermal propagators for various fields. Additional 
vertex and propagators in the real-time formulation are then briefly mentioned.

In Chapter \ref{lg}, we discuss the linearized theory of gravity. The
linearization of the metric and how it gives rise to the graviton are explained.
Photons and spinors in curved space are discussed, leading to the QED Lagrangian
of curved space. This Lagrangian is then linearized, and the Feynman rules for
the lowest order interaction of the graviton with photons and fermions are
deduced.

In Chapter \ref{qed}, we consider charged leptons in thermal QED. General 
expressions for the dispersion relation, the inertial mass, the spinor and the
wavefunction renormalization factor are obtained at $O(e^2)$ for a fermion
as well as an antifermion. Relations between the fermion and antifermion properties
are then deduced for both $C$-symmetric and $C$-asymmetric backgrounds. Explicit
$O(e^2)$ calculation of the inertial masses and wavefunction renormalization
factors of charged leptons and antileptons are performed.

In Chapter \ref{gr}, we turn to gravitational couplings in a medium. We show
in detail how the gravitational mass is determined from the gravitational vertex
function, and obtain an operational formula for the gravitational mass at $O(e^2)$.
We then calculate the gravitational vertex of the leptons, and use it to find
the gravitational masses of charged leptons and antileptons in a medium.
Since the terms involving the fermion distribution function cannot be evaluated
exactly, we evaluate the corrections in two different limits, viz., the classical
and the degenerate limits for the electron gas.

In Chapter \ref{high}, we consider some aspects of high-temperature QED. The 
expression for the one-loop self-energy of the electron in a general linear 
covariant gauge is deduced. The gauge dependence relation for the one-loop electron
dispersion relation is obtained, and gauge independence is proved for momenta
much larger than $eT$. After obtaining a formula for the effective mass of the
electron, gauge independence of the one-loop effective mass is investigated.

In Chapter \ref{conclu}, we present our conclusions.
\section{Some notations and conventions}
We now summarize some notations and conventions which will be frequently used
in this thesis. Although these will be mentioned again in the appropriate places,
it was felt that a separate summary of them may be useful as a ready reference 
in going through the present work.

Throughout this work, we will denote four-momenta by lower case letters, and the
corresponding three-momenta by upper case letters. For example, we write
$p^\mu=(p^0,\vec P)$. The magnitude of $\vec P$ is denoted by $P$. Also,
we use the notation $E_P\equiv\sqrt{P^2+m^2}$, where $m$ is the mass of the particle
concerned.

The mass in the vacuum, the inertial mass in a medium and the gravitational mass
in a medium are denoted by $m$, $M$ and $M'$ respectively. They come with various
subscripts: $f$ denotes a generic fermion, 
$p$ a proton, $\ell$ a charged lepton and $\bar\ell$ a charged antilepton.
Further, $\ell$ can be $e$ or $\mu$ or $\tau$. 

The notations for the thermal corrections to $m$, as defined in
Eqs.\ (\ref{Melldef}) and (\ref{Mell'def}), contain the additional subscripts 1 and 2.
Here 1 stands for the contribution from the photon background, and 2 stands for the
contribution from the electron background. The symbols 1 and 2 are used in this
way for various other medium effects also. Thus, it is used to denote
the contributions from a diagram: the B1 contribution, for example, denotes the photon
background contribution to the diagram B, and the A2 contribution denotes the electron
background contribution to the diagram A.

Finally, in our work on high-temperature QED, the effective mass of the electron is 
denoted by $M$, while $M_0$ denotes the effective mass to leading order in temperature.

\chapter{Thermal field theory}\label{th}
\section{Introduction}\label{th:intro}
Thermal field theory is the theoretical framework for studying a
large number of quantum mechanical and relativistic particles in
thermodynamic equilibrium. The two well-known versions of this subject
are the imaginary-time formalism \cite{kapusta, lebellac, adas, landsman, 
altherr} and the 
real-time formalism \cite{lebellac, adas, landsman, 
altherr,  nieves}, of which
we are going to discuss only the latter. The real-time formalism has
the following advantages. The calculations are manifestly covariant,
the effects of the medium are separated out at the very beginning from
the part already present  in the vacuum, and it can be easily understood
that the medium does not introduce any new ultraviolet divergence.
The particular approach to the real-time formalism to be discussed by us
is the canonical approach \cite{nieves, lebellac2}. This approach 
involves a
generalization of the canonical quantization formalism of ordinary
quantum field theory, by defining the Green's functions as thermal 
averages instead of vacuum expectation values.
\section{Thermodynamic equilibrium}\label{th:TE}
Thermal field theory makes use of quantum field theory as well as
statistical mechanics. As mentioned in Sec.\ \ref{th:intro}, we are 
going to use the canonical
quantization formalism of quantum field theory. For doing statistical
mechanics, we have to choose a particular ensemble according to 
our convenience. 
In the present case, each of the systems making up the ensemble is in
some state of the Fock space, and the calculations are the simplest 
(see below) if the ensemble contains all the states of the Fock space, 
without any restriction
on the total number of particles in a system.
Consequently, we consider the grand canonical ensemble, which
is characterized by the temperature
\bea
T\equiv \frac{1}{\beta}\,
\eea
and the chemical potential
$\mu$. (We have chosen our units such that the Boltzmann constant is equal to 
unity.) The thermal expectation value (ensemble average) of any operator
$\cal O$ is
\bea
\langle{\cal O}\rangle=\rm{Tr}(\rho_G {\cal O})\,,    \label{TE1}
\eea
with the density operator $\rho_G$ in the grand canonical ensemble given by
\bea
\rho_G=\frac{e^{-\beta J}}{\Tr\,e^{-\beta J}}\,,       \label{TE2}
\eea
where
\bea
J=H-\mu Q.                                                    \label{TE3}
\eea
Here $H$ is the normal-ordered hamiltonian, and $Q$ the normal-ordered charge
operator, i.e., the total number operator for the
particle minus that for the antiparticle.

We shall now consider the case of free fields. Let us begin with a real
scalar field, for which (and, in general, for any self-conjugate field) $\mu=0$. 
For the sake of convenience,
we shall consider quantization with discrete momenta 
in this section. (From Sec.\ \ref{th:TP} onward, we shall consider
quantization with continuous momenta; Appendix \ref{appdisc} relates the two
approaches.) We therefore have
\bea
H=\sum\limits_{\vec K} E_K a^\dagger_{\vec K} a_{\vec K}\,.              \label{TE4}
\eea
Here
\bea
E_K\equiv\sqrt{K^2+m^2}\,,                                               \label{EK}
\eea
$m$ being the mass of the quantum of the scalar field, and 
$a^\dagger_{\vec K} a_{\vec K}$ is the number operator for quanta of momentum
$\vec K$.

We are now going to find out the thermal average of the bilinears of the ladder
operators. First consider
\bea
\left<a^\dagger_{\vec P} a_{\vec P}\right>=
         \frac{\Tr[a^\dagger_{\vec P}  a_{\vec P}\,
         {\rm exp}(-\beta\sum\limits_{\vec K} E_K  a^\dagger_{\vec K} a_{\vec K})]}
         {\Tr[{\rm exp}(-\beta\sum\limits_{\vec K} E_K 
          a^\dagger_{\vec K}  a_{\vec K})]}\,,                       \label{TE5}
\eea
where Eqs.\ (\ref{TE1})-(\ref{TE4}) (with $\mu=0$) were used. The trace is in the 
Fock space, which is spanned by the direct product of the eigenstates of the number
operators $a^\dagger_{\vec K}  a_{\vec K}$ for the various modes $\vec K$. 
Therefore for any operator $\cal O\,'$,
\bea
\Tr\,{\cal O\,'}=\sum\limits_{n_{\vec K_1},n_{\vec K_2},\cdots}
   \langle n_{\vec K_1}|\langle n_{\vec K_2}|\cdots {\cal O\,'}\cdots
   | n_{\vec K_2}\rangle |n_{\vec K_1}\rangle\,.                     \label{TE6}
\eea
Here $\langle n_{\vec K_1}|\langle n_{\vec K_2}|\cdots $ denotes the direct product
of the states $\langle n_{\vec K_1}|$, $\langle n_{\vec K_2}|$, $\cdots\,.$ Also, 
since we are considering a bosonic field, the sum over each eigenvalue
$n_{\vec K}$ runs from zero to $\infty$. Use of Eq.\ (\ref{TE6}) in both the
numerator and denominator of Eq.\ (\ref{TE5}) leads to
\bea
\left<a^\dagger_{\vec P} a_{\vec P}\right>&=&
      \frac{\sum\limits_{n_{\vec P}}n_{\vec P}e^{-\beta E_P n_{\vec P}}
      \prod\limits_{\vec K\neq\vec P}\sum\limits_{n_{\vec K}}e^{-\beta E_K n_{\vec K}}}
      {\prod\limits_{\vec K}\sum\limits_{n_{\vec K}}e^{-\beta E_K n_{\vec K}}}
                                                        \nonumber\\
& =&\frac{\sum\limits_{n_{\vec P}}n_{\vec P}e^{-\beta E_P n_{\vec P}}}
    {\sum\limits_{n_{\vec P}}e^{-\beta E_P n_{\vec P}}}\\
&=&-\frac{\partial}{\partial(\beta E_P)}\ln
   \sum\limits_{n_{\vec P}}e^{-\beta E_P n_{\vec P}}                         \label{TE6'}
\eea
The sum over $n_{\vec P}$ is an infinite geometric series, and equals 
$1/(1-e^{-\beta E_P})$. So
\bea
\left<a^\dagger_{\vec P}  a_{\vec P}\right>
                           =\frac{1}{e^{\beta E_P}-1}\,,              \label{TE7}
\eea
which is the Bose distribution function. Using $\left[a_{\vec P},
a^\dagger_{\vec P}\right]_-=1$, this leads to
\bea
\left<a_{\vec P}a^\dagger_{\vec P} \right>=1+\frac{1}{e^{\beta E_P}-1}\,.
                                                                      \label{TE8}
\eea
Also using Eqs.\ (\ref{TE1})-(\ref{TE4}) and Eq.\ (\ref{TE6}), it follows at once that
\bea
\left<a^\dagger_{\vec P}   a_{\vec P'}\right>
&=& \left<a_{\vec P}a^\dagger_{\vec P'}  \right>
=0 ~~{\rm for}~~\vec P\neq\vec P'                           \label{TE9}\\
\left<a_{\vec P}a_{\vec P'}\right>
&=&\left<a^\dagger_{\vec P}a^\dagger_{\vec P'}\right>
=0\,.                                                                 \label{TE9a}
\eea
 
For a complex scalar field with the number operators $a^\dagger_{\vec K} 
a_{\vec K}$ for the particle and $b^\dagger_{\vec K} 
b_{\vec K}$ for the antiparticle of momentum $\vec K$, Eq.\ (\ref{TE3}) gives
\bea
J=\sum\limits_{\vec K}[(E_K-\mu)a^\dagger_{\vec K}a_{\vec K}
      +(E_K+\mu)b^\dagger_{\vec K} b_{\vec K}]\,,                   \label{TE10}
\eea
The Fock space states are now the direct product of the 
eigenstates of $a^\dagger_{\vec K}  a_{\vec K}$ and $b^\dagger_{\vec K}  
b_{\vec K}$ for various $\vec K$. The only change in Eqs.\ (\ref{TE7})~--~(\ref{TE9a})
is then found to be $E_P\rightarrow E_P-\mu$. For the bilinears of the 
antiparticle ladder operators, relations similar to (\ref{TE7})~--~(\ref{TE9a})
hold with $E_P\rightarrow E_P+\mu$. Also, averages of mixed operators, like 
$\left<a_{\vec P} b_{\vec P'}\right>$, are all zero.

We next turn to the case of the Majorana fermion field, which is analogous to
the real scalar field, and has $\mu=0$ necessarily. The important difference from 
the case of the real scalar field relevant for our purpose is that the number
operator $c^\dagger_{\vec K,s} c_{\vec K,s}$ for any given mode can only
have the eigenvalues 0 and 1. Consequently, the geometric series represented by 
the sum in Eq.\ (\ref{TE6'}) now contains only two terms and
equals $1+e^{-\beta E_P}$. So
\bea
\left<c^\dagger_{\vec P,s} c_{\vec P,s}\right>=\frac{1}{e^{\beta E_P}+1}\,,
                                                                   \label{Fermi1}
\eea
which is the Fermi distribution function. Using $\left[c_{\vec P,s}, 
{c_{\vec P,s}}^\dagger\right]_+=1$, this leads to
\bea
\left<c_{\vec P,s}c^\dagger_{\vec P,s}\right>=1-\frac{1}{e^{\beta E_P}+1}\,.
                                                                        \label{Fermi2}
\eea
The other thermal averages are zero, as for the real scalar field.

Finally, for a fermion field which is not self-conjugate,
\bea
J=\sum\limits_{\vec K,s}[(E_K-\mu)c^\dagger_{\vec K,s} c_{\vec K,s}
                 +(E_K+\mu)d^\dagger_{\vec K,s} d_{\vec K,s}]
\eea
(analogous to Eq.\ (\ref{TE10})), so that the replacements $E_P\rightarrow
E_P\mp\mu$ result, just as in the case of the complex scalar field.
\section{Time-ordered thermal propagators}\label{th:TP}
\subsection{Scalar field}
Time-ordered thermal propagators are defined by replacing the vacuum 
expectation value in the definitions of the vacuum propagators by a
thermal average. 
Beginning with scalar fields, we shall first work out the more complicated
case of the complex scalar field in detail, and then state the result for
the case of the real scalar field. We shall use the Fourier expansion
\bea
\phi(x)=\int\frac{d^3P}{\sqrt{(2\pi)^3 2E_P}}
     \Big(a(\vec P)e^{-ip\cdot x}
     +b^\dagger(\vec P) e^{ip\cdot x}\Big),                        \label{TP1}
\eea
for the complex scalar field, where $p^\mu=(p^0,\vec P)$ and $p^0=E_P$. 
Since we are now considering 
continuous momenta, the thermal averages read
\bea
\left<a^\dagger(\vec P) a(\vec P\,')\right>&=&\delta^{(3)}(\vec P-\vec P')
               f_B(E_P,\mu)\,,                                       \label{TP2}\\
\left<a(\vec P)a^\dagger(\vec P\,')\right>&=&\delta^{(3)}(\vec P-\vec P')
              [1+ f_B(E_P,\mu)]\,,                                       \label{TP3}\\
\left<b^\dagger(\vec P) b(\vec P\,')\right>&=&\delta^{(3)}(\vec P-\vec P')
               f_B(E_P,-\mu)\,,                                       \label{TP4}\\
\left<b(\vec P)b^\dagger(\vec P\,')\right>&=&\delta^{(3)}(\vec P-\vec P')
              [1+ f_B(E_P,-\mu)]\,,                                       \label{TP5}
\eea
with the other thermal averages equalling zero. Here the Bose distribution function
for the particle is given by 
\bea
f_B(E_P,\mu)=\frac{1}{e^{\beta(E_P-\mu)}-1}\,.                           \label{TP6}
\eea
Eqs.\ (\ref{TP2}) and (\ref{TP3}), consistent with Eq.\ (\ref{ETCR}), correspond to Eqs.\
(\ref{TE7})--(\ref{TE9}). 

The time-ordered thermal propagator for the complex scalar field is defined by
\bea
iD(x-y)&\equiv&\langle T[\phi(x)\phi^\dagger(y)] \rangle\\
  &=&\theta(x_0-y_0)\langle\phi(x)\phi^\dagger(y) \rangle
    +\theta(y_0-x_0)\langle\phi^\dagger(y)\phi(x) \rangle\,.                  \label{TP7}
\eea
Using Eqs.\ (\ref{TP1})~--~(\ref{TP5}), we can obtain $\langle\phi(x)\phi^\dagger(y)
\rangle$
and $\langle\phi^\dagger(y)\phi(x)\rangle$. After substituting these in Eq.\ (\ref{TP7}),
let us split up the resultant expression as
\bea
iD(x-y)=iD_F(x-y)+D_T(x-y)\,,                                             \label{TP8}
\eea
where $iD_F(x-y)$ is the part independent of $f_B$, and $D_T(x-y)$ is the part arising 
out of the medium. Now, if $f_B$ were set equal to zero in Eqs.\ (\ref{TP2})~--~(\ref{TP5}),
the thermal averages would reduce to the vacuum expectation values, and the thermal 
propagator to the propagator in the vacuum. So, we can immediately write down
\bea
D_F(x-y)&=&\int\frac{d^4 p}{(2\pi)^4}e^{-ip\cdot(x-y)}D_F(p)\,,\nonumber\\*
D_F(p)&=&\frac{1}{p^2-m^2+i\epsilon}\,.                                      \label{TP9}
\eea
On the other hand,
\bea
D_T(x-y)=\int\frac{d^3P}{(2\pi)^32E_P}
[f_B(E_P,\mu)e^{-ip\cdot(x-y)}+ f_B(E_P,-\mu)e^{ip\cdot(x-y)}]\,.     \label{TP9'}
\eea
(One can also get Eq.\ (\ref{TP9'}) by using Eq.\ (\ref{TP1'}) and the discretized 
versions of Eqs.\ (\ref{TP2})~--~(\ref{TP5}), and finally using Eq.\ (\ref{is}) in 
reverse.) Now, Eq.\ (\ref{TP9'}) can be rewritten as
\bea
D_T(x-y)=\int\frac{d^4p}{(2\pi)^32E_P}
\!\!\!\!&[\delta(p_0-E_P)f_B(p_0,\mu)e^{-ip_0(x_0-y_0)+i\vec P\cdot(\vec x-\vec y)}
\nonumber\\
&+\delta(p_0+E_P)f_B(-p_0,-\mu)e^{-ip_0(x_0-y_0)-i\vec P\cdot(\vec x-\vec y)}]\,.
\eea
Changing $\vec P$ to $-\vec P$ in the second term, and using
\bea
\theta(\pm p_0)\delta(p^2-m^2)=\frac{1}{2E_P}\delta(p_0\mp E_P)\,,
\eea
we obtain
\bea
D_T(x-y)&=&\int\frac{d^4 p}{(2\pi)^4}e^{-ip\cdot(x-y)}D_T(p)\,,\nonumber\\*
D_T(p)&=&2\pi\delta(p^2-m^2)\eta_B(p)\,,\nonumber\\*
\eta_B(p)&=&\theta(p_0)f_B(p_0,\mu)+\theta(-p_0)f_B(-p_0,-\mu)\,.        \label{TP10}       
\eea

Thus, the thermal propagator is given by Eqs.\ (\ref{TP8}), (\ref{TP9}) and (\ref{TP10}).
However, this form is not manifestly covariant since the distribution function (\ref{TP6})
was written in the rest frame of the medium. Therefore we introduce the four-velocity
$v^\mu$ of the centre-of-mass of the medium. In the rest frame of the medium,
\begin{eqnarray}
v^\mu = (1, \vec 0 \,)\,. 
\label{v}
\end{eqnarray}
Then carrying out the replacement
\bea
p_0\rightarrow p\cdot v                                        \label{pdotv}
\eea
in Eq.\ (\ref{TP10}), we arrive at the manifestly covariant form of the thermal 
propagator.

For the real scalar field, we have to use the Fourier expansion
\bea
\phi(x)=\int\frac{d^3P}{\sqrt{(2\pi)^3 2E_P}}
     \Big(a(\vec P)e^{-ip\cdot x}
     +a^\dagger(\vec P) e^{ip\cdot x}\Big),        
\eea
and the thermal averages (\ref{TP2}) and (\ref{TP3}) with $\mu=0$. This
leads to the thermal propagator
\bea
\langle T[\phi(x)\phi(y)]\rangle=\int\frac{d^4p}{(2\pi)^4} e^{-ip\cdot(x-y)}
\left[\frac{i}{p^2-m^2+i\epsilon}
+2\pi\delta(p^2-m^2)\frac{1}{e^{\beta\, |p\cdot v|} - 1}\right]\,,
                                                                    \label{TP11}
\eea
which is seen to be the thermal propagator of the complex scalar field at $\mu=0$. 
\subsection{Fermion field}
For the fermion field, we shall use the Fourier expansion
\bea
\psi(x)=\int\frac{d^3P}{(2\pi)^{3/2}}\sum\limits_s
     \Big(c_s(\vec P)u_s(\vec P)e^{-ip\cdot x}
     +d^\dagger_s(\vec P)v_s(\vec P)e^{ip\cdot x}\Big)             \label{psiFour}
\eea
and normalize the spinors as
\bea
u^\dagger_s(\vec P)u_{s'}(\vec P)=\delta_{ss'},~   
v^\dagger_s(\vec P)v_{s'}(\vec P)=\delta_{ss'}\,.                     \label{TP12}
\eea
(This normalization will be later extended to the spinors in a medium; see Eqs.\
(\ref{Unorm}) and (\ref{Vnorm}).) The thermal averages then read
\bea
\left<c{^\dagger_s}(\vec P) c_{s'}(\vec P\,')\right>
     &=&\delta_{ss'}\delta^{(3)}(\vec P-\vec P') f_f(E_P)\,,          \label{TP13}\\ 
\left<c_{s}(\vec P)c{^\dagger_{s'}}(\vec P\,')\right>
     &=&\delta_{ss'}\delta^{(3)}(\vec P-\vec P') [1-f_f(E_P)]\,,          \label{TP14}\\ 
\left<d{^\dagger_s}(\vec P) d_{s'}(\vec P\,')\right>
     &=&\delta_{ss'}\delta^{(3)}(\vec P-\vec P') f_{\bar f}(E_P)\,,          \label{TP15}\\ 
\left<d_{s}(\vec P)d{^\dagger_{s'}}(\vec P\,')\right>
     &=&\delta_{ss'}\delta^{(3)}(\vec P-\vec P') [1-f_{\bar f}(E_P)]\,,     \label{TP16} 
\eea
with the other thermal averages equalling zero.
Here the distribution functions for a fermion and an antifermion
are given by 
\begin{eqnarray}
\label{distfunctions}
f_{f,\bar f}(E_P) = \frac{1}{e^{\beta(E_P \mp \mu_f)} + 1} \,,
\end{eqnarray}
respectively. (We have denoted the fermion chemical potential by $\mu_f$, since we will
later deal with several species of the fermion $f$.) 
Eqs.\ (\ref{TP13}) and (\ref{TP14}), consistent with Eq.\ (\ref{ETACR}),
correspond to Eqs.\ (\ref{Fermi1}) and (\ref{Fermi2}).

The thermal propagator for the fermion is defined by
\bea
iS_{fab}(x)&\equiv&\langle T[\psi_a(x)\bar\psi_b(y)] \rangle\\
           &=&\theta(x_0-y_0)\langle\psi_a(x)\bar\psi_b(y)\rangle
           -\theta(y_0-x_0)\langle \bar\psi_b(y)\psi_a(x)\rangle\,.
                                                                        \label{TP17}
\eea
Using Eq.\ (\ref{psiFour}), Eqs.\ (\ref{TP13})~--~(\ref{TP16}), and the spin-sum 
relations
\bea
\sum\limits_s u_s(\vec P)\bar u_s(\vec P)
&=&\frac{E_P\gamma_0-\vec P\cdot\vec \gamma+m_f}{2E_P}
\,,\nonumber\\*
\sum\limits_s v_s(\vec P)\bar v_s(\vec P)
&=&\frac{E_P\gamma_0-\vec P\cdot\vec \gamma-m_f}{2E_P}                \label{TP18} 
\eea
following from Eq.\ (\ref{TP12}), we can find the R.H.S. of Eq.\ (\ref{TP17}). The expression
is then split up into the vacuum part and the thermal part:
\bea
S_f(x-y)&=&\int\frac{d^4p}{(2\pi)^4}e^{-ip\cdot (x-y)}S_f(p)\,,\\
iS_f(p)&=& iS_{Ff}(p) + S_{Tf}(p)\,.                           \label{S}
\eea
The vacuum part is
\begin{eqnarray}\label{S0}
S_{Ff} &=& \frac{\rlap/{p} + m_f}{p^2 - m_f^2 + i\epsilon} \,,
\eea
while, following the derivation in the case of the complex scalar field, the thermal part
is found to be 
\bea
S_{Tf}(p) &=& - 2\pi (\rlap/{p} + m_f) \delta(p^2 - m_f^2) \eta_f(p) \,,
                                       \label{S'}       \\
\eta_f(p) &=& \frac{\theta(p\cdot v)}{e^{\beta(p\cdot v - \mu_f)} + 1}
+ \frac{\theta(-p\cdot v)}{e^{-\beta(p\cdot v - \mu_f)} + 1} \,.
                                        \label{etaf}
\eea
\subsection{Photon field}
In this case, the thermal propagator in the Feynman gauge is the same as that
for the real scalar field, except for the facts that the photon is massless,
and that there is an extra factor of $-\eta_{\mu\nu}$ (as in the vacuum). Thus, we have
the propagator
\begin{eqnarray}\label{D}
iD_{\mu\nu}(k) = -\eta_{\mu\nu} \left[ i\Delta_F(k) +
\Delta_T(k) \right] \,,
\end{eqnarray}
where
\begin{eqnarray}
\Delta_F(k) &=& {1 \over k^2+i\epsilon} \,, \label{D0}\\
\Delta_T(k) &=& 2\pi \delta(k^2) \eta_\gamma(k) \,,\label{D'}\\
\eta_\gamma(k) &=& \frac{1}{e^{\beta\, |k\cdot v|} - 1} \,.\label{etab}
\end{eqnarray}
The factor $\eta_\gamma(k)$ arises out of the photon distribution function
\begin{eqnarray}
\label{fgamma}
f_\gamma(K) = \frac{1}{e^{\beta K} - 1} \,,
\end{eqnarray}
where $K$ is the magnitude of the three-momentum of the photon.

We end this section with the following observation. At any finite temperature,
the distribution function contained in the medium-dependent part of the thermal
propagator becomes an exponential damping factor for momenta much larger than $T$.
Consequently, the corresponding loop integration will have an effective cut-off 
of $O(T)$. If $T=0$, we can still have a Fermi gas; in that case, the Fermi momentum
provides a cut-off. Thus, the medium does not introduce any new ultraviolet divergence,
and renormalization in the vacuum suffices to make the theory finite in the presence
of thermal background.
\section{Additional vertex and propagators}
The full structure of the real-time formulation of thermal field
theory is more complicated than the replacement of vacuum propagators
with time-ordered thermal propagators. However, as will be seen later,
the additional features of the full structure do not play a role in
the present work. So, in this section, we shall just outline the main
results involving those additional features \cite{derivations}.

Ultimately,~one has to calculate an $n$-point Green's function like
$\langle T[\phi(x_1)\phi^\dagger(x_2)\break\cdots\phi(x_n)]\rangle$, where
$\phi(x_i)$ are the Heisenberg picture fields. (We shall consider only
complex scalar fields in this section. Similar results hold for other
fields.) In a medium, as in the vacuum, one can prove Wick's theorem
and reduce the $n$-point function to a sum of fully contracted
terms. But the fact that the Green's function in the present case
involves a thermal average over all the Fock space states, rather than
a vacuum expectation value, leads to four types of contractions or
``propagators." They are defined as
\bea
iD_{11}(x-y)&=&\langle T[\phi(x)\phi^\dagger(y)]\rangle\,,\label{D11}\\
iD_{22}(x-y)&=&\langle {\overline T}[\phi(x)\phi^\dagger(y)]\rangle\,,\\
iD_{12}(x-y)&=&\langle \phi^\dagger(y)\phi(x)\rangle\,,\\
iD_{21}(x-y)&=&\langle \phi(x)\phi^\dagger(y)\rangle\label{D21}\,.
\eea
Here $\overline T$ stands for anti-time ordering. In Eqs.\ 
(\ref{D11})~--~(\ref{D21}), the fields are the interaction picture fields, which can
be expressed in terms of the ladder operators. Thus, in addition to the time-ordered
propagator of Sec.\ \ref{th:TP}, we also have anti-time ordered propagator, as well as
propagators without any ordering in time. As given by Eqs.\ (\ref{TP8}), (\ref{TP9})
and (\ref{TP10})
\bea
iD_{11}(p)=\frac{i}{p^2-m^2+i\epsilon} +2\pi\delta(p^2-m^2)\eta_B(p)\,.
\eea
The others are found to be
\bea
iD_{22}(p)&=&-\frac{i}{p^2-m^2-i\epsilon} +2\pi\delta(p^2-m^2)\eta_B(p)\,,\\
iD_{12}(p)&=&2\pi\delta(p^2-m^2)[\eta_B(p)+\theta(-p\cdot v)]\,,\\
iD_{21}(p)&=&2\pi\delta(p^2-m^2)[\eta_B(p)+\theta(p\cdot v)]\,.
\eea
Here it is to be understood that the replacement (\ref{pdotv}) has been carried out
in $\eta_B(p)$.

The Feynman diagrams now contain two types of vertices: the 1-type vertex is the usual one,
while the 2-type vertex has a vertex factor which differs from that for the 1-type vertex
by a sign. Green's functions can now be evaluated with these two types of vertices and
four types of propagators. However, the external legs representing physical particles
can only be connected to 1-type vertices.

The importance of this doubling of degrees of freedom manifests itself in diagrammatic 
calculations in the following way. If one uses only the 1-type vertex and ``11"
propagator, one sometimes ends up with singular expressions involving products of
delta functions with the same argument. However the use of the additional vertex and
propagators leads to the cancellation of such singularities \cite{lebellac3}.

\chapter{Linearized theory of gravity}\label{lg}
\section{Introduction}\label{lg:intro}
Gravity is governed by the general theory of relativity 
which tells us how the geometry of space-time, as given by the metric 
tensor $g_{\mu\nu}(x)$, is affected by the presence of matter. The Einstein
field equations for $g_{\mu\nu}$ in the absence of external source, derived
from the action for pure gravity, are
\begin{eqnarray}
R_{\mu\nu}-\frac{1}{2}Rg_{\mu\nu}=0.                          \label{lg:eineqn}
\end{eqnarray}
Here $R_{\mu\nu}$ is the Ricci tensor and $R$ is the 
curvature scalar. One important point is that these equations are non-linear
in $g_{\mu\nu}$. This is in contrast with the linear free-field equations for the
scalar, fermion and photon fields, where we use the principle of linear 
superposition to write down the general solution before quantizing the field.
The reason for this non-linearity in the case of gravity is that the gravitational
field possesses energy and hence effective mass, and therefore {\it produces} 
gravitational
field. The gravitational field thus contributes to its own source. 
To ignore this feedback effect, we
have to consider a weak gravitational field. One thus considers only small
deviations from the flat space metric, and writes
\begin{eqnarray}
g_{\mu\nu}(x)=\eta_{\mu\nu}+2\kappa h_{\mu\nu}(x)                  \label{lg:hdef}
\end{eqnarray}
with $|\kappa h_{\mu\nu}|\ll 1$. Here $\kappa$ is related to the Newton constant $G$.
The approximation stated in 
Eq.\ (\ref{lg:hdef}) is the starting point of the linearized theory of gravity
\cite{lg:wein1, lg:abs}.
Now $\eta_{\mu\nu}$, and hence from Eq.\ (\ref{lg:hdef}), $h_{\mu\nu}$,
are symmetric flat space tensors, but not coordinate tensors (in
contrast with $g_{\mu\nu}$, which is a symmetric coordinate tensor,
and so, as a special case, also a tensor in flat space).
Consequently, {\it all} tensors in an expression should be regarded
only as flat space tensors {\it after} we use the approximation of
Eq.\ (\ref{lg:hdef}) to perform an expansion of the expression
in powers of $\kappa$, and we will adopt the convention of raising and lowering
the indices of such tensors with the flat space metric.

The precise value of $\kappa$, to be deduced in the next section by identifying
$h_{\mu\nu}$ with the graviton field, is given by
\begin{eqnarray}
\kappa=\sqrt{8\pi G}.                                          \label{lg:kappa} 
\end{eqnarray}
Since $\sqrt G=1/m_{\rm Planck}$ in natural units, it is clear that the weak field 
approximation given by Eq.\ (\ref{lg:hdef}) is valid when all mass-scales in the
problem are much smaller than the Planck mass.
These mass-scales include not only particle masses, but also temperature and chemical
potential, as we shall later perform calculations in the presence of a thermal background.
Since $m_{\rm Planck}$ is $O(10^{19})$ GeV, the weak field
approximation holds in all physical situations except in the very early
universe.

At this point, it is interesting to consider the case of the
Yang-Mills theory for comparison, because the equation of motion for
the non-Abelian gauge field derived from the pure Yang-Mills
Lagrangian is also non-linear. However, there the equation of motion
{\it is} linear when one considers the free gauge fields, setting the
coupling constant equal to zero, and so the need to linearize does not
arise.  In contrast, the flat space metric tensor
$\eta_{\mu\nu}$, being a constant, does not have any dynamics
(and the LHS of Eq.\ (\ref{lg:eineqn}) reduces to zero in this case).
So, to do quantum field theory with gravity it is essential to impose
the condition (\ref{lg:hdef}), which considers linear deviations from
the flat space metric case.
\section{Graviton }\label{lg:graviton}
The field equations (\ref{lg:eineqn}) follow from the Einstein-Hilbert action
given by
\begin{eqnarray}                                     
{\mathscr A}=\int d^4x~{\mathscr L},                                \label{lg:action} \\
{\mathscr L}=\frac{1}{16\pi G}\sqrt{-{\tt g}}R.
\end{eqnarray}
The curvature scalar, the Ricci tensor and the affine connection are respectively
given by
\begin{eqnarray}
R&=&g^{\mu\nu}R_{\mu\nu},                                 \label{lg:cscalar}\\
R_{\mu\nu} &=&\partial_\nu\Gamma^\lambda_{\mu\lambda}
        -\partial_\lambda\Gamma^\lambda_{\mu\nu}
        +\Gamma^\eta_{\mu\lambda}\Gamma^\lambda_{\nu\eta}
        -\Gamma^\eta_{\mu\nu}\Gamma^\lambda_{\lambda\eta},    \label{lg:Ricci}\\
\Gamma^\lambda_{\mu\nu}&=&\frac{1}{2}g^{\lambda\sigma}
         (\partial_\mu g_{\nu\sigma}+ \partial_\nu g_{\mu\sigma} 
         - \partial_\sigma g_{\nu\mu}),                   \label{lg:affine}
\end{eqnarray}
while
\begin{eqnarray}
{\tt g}&\equiv&{\rm det} g_{\mu\nu}.
\end{eqnarray}
In anticipation of the transition to quantum field theory, here and elsewhere 
we always write the action in the form (\ref{lg:action})
(although ${\mathscr L}$ is not a coordinate scalar).
 
From the weak field condition (\ref{lg:hdef}), we obtain
\begin{eqnarray}
g^{\mu\nu}=\eta^{\mu\nu}-2\kappa h^{\mu\nu}+O(\kappa^2)         \label{lg:^munu}
\end{eqnarray}
(since $g^{\mu\nu}g_{\nu\lambda}=\delta^\mu_\lambda$), and
\begin{eqnarray}
\sqrt{-{\tt g}}=1+\kappa h+O(\kappa^2)                         \label{lg:sqrt-g}
\end{eqnarray}
where
\begin{eqnarray}
h\equiv {h^\mu}_\mu=\eta^{\mu\nu}h_{\mu\nu}.
\end{eqnarray}
Note that we are raising and lowering the indices of 
the graviton field
with the flat space metric \cite{lg:wein2}; this is in accord with the convention
stated in Sec.\ \ref{lg:intro}. 

Use of Eqs.\ (\ref{lg:hdef}) and (\ref{lg:^munu}) in Eq.\ (\ref{lg:affine})
leads to
\begin{eqnarray}
\Gamma^\lambda_{\mu\nu}=\kappa(\eta^{\lambda\sigma}-2\kappa h^{\lambda\sigma})
         (\partial_\mu h_{\nu\sigma}+ \partial_\nu h_{\mu\sigma} 
         - \partial_\sigma h_{\nu\mu}) +O(\kappa^3).          \label{lg:Chris}    
\end{eqnarray}
Putting this in Eq.\ (\ref{lg:Ricci}), one can determine $R_{\mu\nu}$ to $O(\kappa^2)$.
Let us write
\begin{eqnarray}
R_{\mu\nu}= {R_{\mu\nu}}^{(1)}+ {R_{\mu\nu}}^{(2)} + O(\kappa^3)
\end{eqnarray}
with ${R_{\mu\nu}}^{(1)}$ and ${R_{\mu\nu}}^{(2)}$ denoting the parts linear 
and quadratic in $\kappa$ respectively. 

Following again the convention stated in Sec.\ \ref{lg:intro}, we shall now raise
and lower the indices of 
${R_{\mu\nu}}^{(1)}$, ${R_{\mu\nu}}^{(2)}$ and $\partial/\partial x^\mu$ 
with the flat space metric \cite{lg:wein2}. Then
\begin{eqnarray}
16\pi G{\mathscr L}&=&\sqrt{-\tt g}g^{\mu\nu}R_{\mu\nu}  \nonumber\\
               &=& {{R^\mu}_\mu}^{(1)}
               +\kappa h {{R^\mu}_\mu}^{(1)}
               -2\kappa h^{\mu\nu} {R_{\mu\nu}}^{(1)}
               + {{R^\mu}_\mu}^{(2)}+O(\kappa^3).                \label{lg:ggR}
\end{eqnarray}
The first term on the RHS is the only term of $O(\kappa)$. In fact,
\begin{eqnarray}
{{R^\mu}_\mu}^{(1)}=2\kappa\partial_\mu(\partial^\mu h-\partial_\lambda
                    h^{\lambda\mu}),
\end{eqnarray}
and this, being a surface term, does not contribute to the Einstein-Hilbert
action. So we now turn to the three $O(\kappa^2)$ terms on the RHS of 
Eq.\ (\ref{lg:ggR}). First consider ${{R^\mu}_\mu}^{(2)}$. 
After leaving out surface terms, we obtain
\begin{eqnarray}
{{R^\mu}_\mu}^{(2)}&=&\kappa^2
                  [(\partial_\mu h_{\eta\lambda}+ \partial_\lambda h_{\mu\eta}
                  -\partial_\eta\ h_{\lambda\mu})
                  (\partial^\mu h^{\eta\lambda}- \partial^\lambda h^{\mu\eta}
                  +\partial^\eta\ h^{\lambda\mu})\nonumber\\
                  &&+(\partial_\mu h)(\partial^\mu h
                  - 2\partial_\lambda h^{\lambda\mu})].
\end{eqnarray}
Carrying out of several partial integrations with discarding of 
surface terms [for example, 
$(\partial_\lambda h_{\mu\eta}) (\partial^\eta h^{\lambda\mu})\rightarrow 
-h_{\mu\eta}(\partial_\lambda\partial^\eta h^{\lambda\mu})\rightarrow 
(\partial^\eta h_{\mu\eta}) (\partial_\lambda h^{\lambda\mu})$]
leads to 
\begin{eqnarray}
{{R^\mu}_\mu}^{(2)}=-\kappa^2
               [(\partial_\mu h_{\eta\lambda})(\partial^\mu h^{\eta\lambda})
               -2(\partial_\lambda h^{\mu\lambda}) (\partial^\eta h_{\mu\eta}) 
               +2(\partial_\mu h)(\partial_\lambda h^{\mu\lambda})
               -(\partial_\mu h)(\partial^\mu h)].          \label{lg:Rmumu2}
\end{eqnarray}
Similarly one can write $\kappa h {{R^\mu}_\mu}^{(1)}$ and 
$2\kappa h^{\mu\nu} {R_{\mu\nu}}^{(1)}$ in terms of the combinations contained
on the R.H.S. of Eq.\ (\ref{lg:Rmumu2}). Putting everything back into 
Eq.\ (\ref{lg:ggR}), we get
\begin{eqnarray}
16\pi G{\mathscr L}=\kappa^2 
               [(\partial_\mu h_{\eta\lambda})(\partial^\mu h^{\eta\lambda})
               -2(\partial_\lambda h^{\mu\lambda}) (\partial^\eta h_{\mu\eta}) 
               +2(\partial_\mu h)(\partial_\lambda h^{\mu\lambda})
               -(\partial_\mu h)(\partial^\mu h)].      
\end{eqnarray}
Here and henceforth it is to be understood that the expression for
$16\pi G{\mathscr L}$ has been written neglecting the $O(\kappa^3)$
terms.

With the aim of arriving at the propagator, we carry out further partial integrations
 to obtain 
\bea
16\pi G{\mathscr L}=\kappa^2 h_{\alpha\beta}{\mathscr P}^{\alpha\beta
             \mu\nu}h_{\mu\nu},
\eea
\bea
{\mathscr P}^{\alpha\beta\mu\nu}&=&-\frac{1}{2}
                  [(\eta^{\alpha\mu}\eta^{\beta\nu}+\eta^{\alpha\nu}\eta^{\beta\mu}
                 -2 \eta^{\alpha\beta}\eta^{\mu\nu})\Box
                 -(\eta^{\beta\nu}\partial^\alpha\partial^\mu
                 +\eta^{\alpha\nu}\partial^\beta\partial^\mu 
                 +\eta^{\beta\mu}\partial^\alpha\partial^\nu\nonumber\\
                 &&+\eta^{\alpha\mu}\partial^\beta\partial^\nu)
                 +4\eta^{\alpha\beta}\partial^\mu\partial^\nu].
\eea                
We have explicitly symmetrized ${\mathscr P}^{\alpha\beta\mu\nu}$
with respect to $\alpha\leftrightarrow \beta$ and with respect to 
$\mu\leftrightarrow\nu$ (such symmetrization helps one to guess the form of
the inverse; see later). But this operator does not possess an inverse! This
is easily seen from the fact that
\bea
{\mathscr P}^{\alpha\beta\mu\nu}\partial_\alpha\partial_\beta\partial_\mu
\partial_\nu \Lambda(x)=0
\eea
where $\Lambda(x)$ is any function of $x$, i.e., the operator has a zero
eigenvalue.
   
The problem is similar to the one encountered in quantizing the photon field,
and the solution also is the same, viz., gauge-fixing. To this end, one chooses
the harmonic or de Donder gauge
\bea
F^\mu\equiv \partial_\nu h^{\mu\nu}-\frac{1}{2}\partial^\mu h=0
                                                            \label{lg:gauge1}
\eea
and adds to $16\pi G{\mathscr L}$ the gauge-fixing term \cite{lg:berends}
\bea
16\pi G{\mathscr L}_{\rm GF}=2\kappa^2 F^\mu F_\mu.          \label{lg:gauge2}
\eea
Before proceeding further, let us dwell a little on gauge invariance.

While the Einstein-Hilbert action is invariant under general coordinate 
transformations, the allowed coordinate transformations in the present case are 
only those which sustain the weak field approximation. Thus, we consider
\bea
x^\mu\rightarrow x^{\prime\,\mu}=x^\mu+2\kappa \epsilon^\mu(x)  \label{lg:gauge3}
\eea
and demand that 
\bea
g^{\prime\,\mu\nu}=\eta^{\mu\nu}-2\kappa h^{\prime\,\mu\nu}+O(\kappa^2).
\eea
Using these relations in
\bea
g^{\prime\,\mu\nu}=\frac{\partial x^{\prime\,\mu}}{\partial x^\lambda}
                 \frac{\partial x^{\prime\,\nu}}{\partial x^\rho}
                 g^{\lambda\rho},                          
\eea
we arrive at the following gauge transformation of $h_{\mu\nu}$
which keeps the action invariant:
\bea
h^\prime_{\mu\nu}=h_{\mu\nu}-(\partial_\mu\epsilon_\nu
                    +\partial_\nu\epsilon_\mu),                 \label{lg:gauge4}
\eea
where $\epsilon_\mu\equiv \eta_{\mu\nu}\epsilon^\nu$.
It can easily be checked that if $h_{\mu\nu}$ does not satisfy the gauge 
condition stated in Eq.\ (\ref{lg:gauge1}), $h^\prime_{\mu\nu}$ does, provided
\bea
\Box \epsilon^\mu=F^\mu.
\eea
So it is always possible to implement the harmonic gauge condition. Another
point to be confirmed is that the gauge-fixing term given in Eq.\ (\ref{lg:gauge2}) 
does break the gauge
invariance of the action. This readily follows from 
\bea
g^{\mu\nu}\Gamma^\lambda_{\mu\nu}=-4\kappa F^\lambda +O(\kappa^2).
\eea
Since
the affine connection is not a coordinate tensor, the square of the L.H.S. is
not a coordinate scalar. (Note that ${\mathscr A}_{\rm GF}$ involves
$d^4 x$ which also is
not a scalar. But the transformation of $d^4x$ cannot compensate the breaking
of gauge invariance, as $16\pi G{\mathscr L}_{\rm GF}$ is already of $O(\kappa^2)$.)

We now explore the consequences of fixing the gauge. 
Addition of the gauge non-invariant term given in Eq.\ (\ref{lg:gauge2}) 
immediately yields the correctly normalized kinetic terms for the massless
spin-2 field, to be identified as the graviton field:
\begin{eqnarray}
{\mathscr L}+{\mathscr L}_{\rm GF}=
         \frac{1}{2}(\partial_\mu h_{\nu\lambda})(\partial^\mu h^{\nu\lambda})
         -\frac{1}{4}(\partial_\mu h)(\partial^\mu h),
\end{eqnarray}
provided we set $\kappa$ equal to the value stated in Eq.\ (\ref{lg:kappa}).

We conclude this section by obtaining the graviton propagator. Partial integration of the last 
expression leads to
\bea 
{\mathscr A}+{\mathscr A}_{\rm GF}=\int d^4x~\frac{1}{2}h_{\alpha\beta}
            {\mathscr Q}^{\alpha\beta\mu\nu} h_{\mu\nu}, \\
{\mathscr Q}^{\alpha\beta\mu\nu}=-\frac{1}{2}
                  (\eta^{\alpha\mu}\eta^{\beta\nu}+\eta^{\alpha\nu}\eta^{\beta\mu}
                 - \eta^{\alpha\beta}\eta^{\mu\nu})\Box.
\eea
So the graviton propagator $iD_{\alpha\beta,\mu\nu}$ is $i/k^2$ times the inverse of 
\bea
M^{\alpha\beta,\mu\nu}\equiv \frac{1}{2}
                  (\eta^{\alpha\mu}\eta^{\beta\nu}+\eta^{\alpha\nu}\eta^{\beta\mu}
                 - \eta^{\alpha\beta}\eta^{\mu\nu}).
\eea
The inverse must be determined from 
\bea
M^{\alpha\beta,\mu\nu}(M^{-1})_{\alpha\beta.\lambda\rho}
            =\frac{1}{2}(\delta{^\mu_\lambda}\delta{^\nu_\rho}+
            \delta{^\mu_\rho}\delta{^\nu_\lambda}).
\eea
Guided by the fact that $M^{-1}$ must 
be symmetric under $\alpha\leftrightarrow \beta$ and (separately) under
$\lambda\leftrightarrow \rho$, one can write down
\bea
(M^{-1})_{\alpha\beta.\lambda\rho}=
                  A(\eta_{\alpha\lambda}\eta_{\beta\rho}+
                  \eta_{\alpha\rho}\eta_{\beta\lambda})
                  + B\eta_{\alpha\beta}\eta_{\lambda\rho}.
\eea
We then obtain $A=1/2=-B$. Therefore the
graviton propagator is given by
\bea
iD_{\alpha\beta,\lambda\rho}=i\frac{\eta_{\alpha\lambda}\eta_{\beta\rho}
                +\eta_{\alpha\rho}\eta_{\beta\lambda}
                -\eta_{\alpha\beta}\eta_{\lambda\rho}}{2k^2}.
\eea
One should now compare the photon propagator in the Feynman gauge, given by 
$iD_{\mu\nu}=-i\eta_{\mu\nu}/k^2$. Clearly $D_{00}$ is negative, while
$D_{00,00}$ is positive. This has the consequence that the fermion-fermion
interaction in the non-relativistic limit
is repulsive when mediated by the photon and attractive when mediated 
by the graviton \cite{lg:peskin}.
\section{Photons in curved space}\label{lg:photon}
For the electromagnetic field, the Lagrangian in curved space is given by
\bea
\mathscr L{^{(\gamma)}_{\rm G}}=-\frac{1}{4}\sqrt{-\tt g}F_{\mu\nu}F_{\alpha\beta}
              g^{\mu\alpha}g^{\nu\beta}                 \label{lg:photon1}
\eea
where
\bea
F_{\mu\nu}=\partial_\mu A_\nu-\partial_\nu A_\mu.       \label{lg:photon2}     
\eea
This has been written down following the standard prescription of
passing from flat space to curved space, i.e., by
replacing all flat space tensors with coordinate tensors, all ordinary
derivatives with covariant derivatives 
(by which we shall mean gravity-covariant derivatives)
and the flat space metric 
tensor with $g_{\mu\nu}$. The additional $\sqrt{-\tt g}$ in
Eq.\ (\ref{lg:photon1}) comes because $d^4 x$ in the action gets replaced 
with the coordinate scalar $d^4x\sqrt{-\tt g}$. It may be noted that
the use of covariant derivative and ordinary derivative in 
Eq.\ (\ref{lg:photon2}) are equivalent due to the symmetry of the Christoffel 
symbol in the two lower indices.
\section{Spinors in curved space}\label{lg:spinor}
The procedure of replacing flat space tensors with coordinate tensors
cannot be extended to the the case of spinors. The procedure works with
tensors because the tensor representations of GL(4,R), the group
of $4\times 4$ real matrices, behave like tensors
under the subgroup SO(3,1). Thus, considering the
vector representation 
as an example, $V^{\prime\,\mu}(x^\prime)=(\partial x^{\prime\,\mu}
/\partial x^\nu) V^\nu(x)$ becomes 
$V^{\prime\,\mu}(x^\prime)={\Lambda^\mu}_\nu V^\nu(x)$ if we restrict
ourselves to $x^{\prime\,\mu}={\Lambda^\mu}_\nu x^\nu$. But there are
no representations of GL(4,R) which behave like spinors under SO(3,1)
(there is no function of $x$ and $x^\prime$ which reduces to $D(\Lambda)$
for $x^{\prime\,\mu}={\Lambda^\mu}_\nu x^\nu$).

Consequently, spinors have to be considered as scalars under general
coordinate transformations. But the problem of incorporating the flat
space spinorial property into curved space remains. This is tackled by
making use of the tangent space formalism, to be described now
\cite{lg:weinvier, lg:weinsugra, lg:GSW}.

At any given point of the curved space, we can choose a locally inertial
coordinate system. Let the locally inertial coordinates
at the point $x^\mu$ be be $\xi^a(x)$ with $a=0,1,2,3.$ Since the
locally inertial coordinate system
at each point may be redefined by an arbitrary Lorentz transformation
(LT),
the action must be invariant under the local LT
\bea
\xi^a(x)\rightarrow\xi^{\prime\,a}(x)={\Lambda^a}_b(x)\xi^b(x),
                                                         \label{lg:lLT}
\eea
with ${\Lambda^a}_b(x)$ a real matrix satisfying 
\bea
\eta_{ab}{\Lambda^a}_c(x){\Lambda^b}_d(x)=\eta_{cd}.     \label{lg:Lorentz}  
\eea
The action, of course, has to be also invariant under the general
coordinate transformation $x^\mu\rightarrow x^{\prime\,\mu}$, which is
independent of the local LT. The requirement of invariance under local 
LT does not affect the actions discussed earlier, since
coordinate tensors like $g_{\mu\nu}$ and $A_\mu$ are taken to be
local Lorentz scalars (i.e. scalars under
local LT).

Let us now consider the vierbein or tetrad, defined by  
\bea
{e^a}_\mu(X)\equiv\frac{\partial\xi^a(x)}{\partial x^\mu}{\Bigg|}_{x=X}.
                                                          \label{lg:vier}
\eea
These constitute a set of four coordinate vectors, forming a basis for the 
(flat) tangent space to the curved
space at the point $x=X$. Under $x^\mu\rightarrow x^{\prime\,\mu}$, they
transform as
\bea
{e^{\prime\, a}}_\mu(x^\prime)=\frac{\partial x^\nu}
             {\partial x^{\prime\,\mu}}{{e^a}_\nu}(x),
\eea
since $\xi^{\prime\,a}(x^\prime)=\xi^a(x)$. Under local LT, the vierbein
transforms just like $\xi^a(x)$:
\bea
{e^{\prime\, a}}_\mu(x)={\Lambda^a}_b (x){e^b}_\nu(x),
                                                    \label{lg:viertrans}
\eea
since $x^\mu$ does not transform. 
From the definition given by Eq.\ (\ref
{lg:vier}), it follows that raising or lowering of the local Lorentz index $a$
(in general, a latin index) and the general coordinate index $\mu$ (in general, a 
greek index) of the vierbein must be done with $\eta_{ab}$
and $g_{\mu\nu}$ respectively. 

The metric $g_{\mu\nu}$ is given by
\bea
g_{\mu\nu}(x)=\eta_{ab}{e^a}_\mu(x){e^b}_\nu(x).         \label{lg:ortho1}
\eea
Therefore $\delta{^\mu_\nu}={e_b}^\mu{e^b}_\nu$, i.e., ${e_b}^\mu$ is the
inverse of ${e^b}_\mu$. We can rewrite this as ${\bf 1}_{\mu\nu}
=({\bf E}^{-1})_{\mu b}{\bf E}_{b\nu}$, where the indices now denote
matrix indices. It then also follows that ${\bf 1}_{a b}
={\bf E}_{a\mu}({\bf E}^{-1})_{\mu b}$. This means that 
$\delta{^a_b}={e^a}_\mu{e_b}^\mu$, or, 
\bea
\eta^{ab}=g^{\mu\nu}(x){e^a}_\mu(x){e^b}_\nu(x).         \label{lg:ortho2} 
\eea
Eq.\ (\ref{lg:ortho1}), or equivalently, Eq.\ (\ref{lg:ortho2}), expresses the
orthonormality of the vierbeins.

The vierbeins turn out to be necessary for constructing the curved space Lagrangian 
for the Dirac fermion. This will involve modifying the flat space Dirac 
Lagrangian 
\bea
{\mathscr L}{_0^{(f)}}=\Big[\frac{i}{2}\bar\psi\gamma^\mu\partial_\mu\psi
          +{\rm h.c.}\Big]-m_f\bar\psi\psi                \label{lg:flatD}
\eea
such that the action is both a coordinate scalar and a local Lorentz scalar.
With this aim in mind, we turn to the transformation properties of the spinor
field.

Spinor fields are coordinate scalars which transform under local LT as
\bea
\psi_\alpha(x)\rightarrow \psi{^\prime_\alpha}(x)
            =D_{\alpha\beta}(\Lambda(x)) \psi_\beta(x). \label{lg:lLTspin}
\eea
Here $D_{\alpha\beta}(\Lambda)$ is the spinor representation
of the 
Lorentz group. Note that we use $\alpha$ and $\beta$ to denote the components of
$\psi$, $D(\Lambda)$ and other quantities associated with the spinor
representation; they must not be confused with general coordinate indices.
Eq.\ (\ref{lg:lLTspin}) is the relation which incorporates the flat space spinorial 
property into general relativity.
 
Since $\Lambda$ is a function of $x^\mu$, $\partial_\mu
\psi_\alpha$ does not transform like $\psi_\alpha$ under local LT.
Hence one defines the covariant derivative 
\bea
{\mathscr D}_\mu\psi_\alpha\equiv\partial_\mu\psi_\alpha
              +[\Omega_\mu]_{\alpha\beta}\psi_\beta
\eea
with the connection matrix $\Omega_\mu$ transforming as
\bea
\Omega{^\prime_\mu}=D(\Lambda)\Omega_\mu D^{-1}(\Lambda)
             -(\partial_\mu D(\Lambda))D^{-1}(\Lambda),
                                                    \label{lg:connprime}    
\eea
so that
\bea
{\mathscr D}_\mu\psi_\alpha(x)\rightarrow D_{\alpha\beta}(\Lambda(x))
                  {\mathscr D}_\mu\psi_\beta(x).    \label{lg:Dtrans}
\eea
The connection matrix can be written as
\bea
[\Omega_\mu]_{\alpha\beta}(x)=\frac{i}{2}[S_{ab}]_{\alpha\beta}
                \,{\omega_\mu}^{ab}(x)                    \label{lg:conn}
\eea
where $S_{ab}$ are the generators of the Lorentz group in the spinor
representation:
\bea
S_{ab}=\frac{1}{2}{\sigma}_{ab}=\frac{i}{4}[\gamma_a,\gamma_b]
\eea
(with $\gamma_a$ denoting the ordinary gamma matrices), and
${\omega_\mu}^{ab}$ is a coordinate vector field antisymmetric in
$a$ and $b$, called the spin connection.

From Eq.\ (\ref{lg:Dtrans}), it follows that 
$\bar\psi\gamma^a{\mathscr D}_\mu\psi$ is a local Lorentz
vector (just like $\bar\psi\gamma^a\psi$). It is also a coordinate vector
(as $\psi$ is a coordinate scalar).
So when we
contract it with ${e_a}^\mu$, we get a quantity
which is a local Lorentz  
scalar and a coordinate scalar. We can now modify Eq.\ (\ref{lg:flatD})
to write down the curved space Lagrangian
\bea
{\mathscr L}{_{\rm G}^{(f)}}&=&\sqrt{-\tt g}\Big[\Big(\frac{i}{2}\bar\psi\gamma^a
              {e_a}^\mu\mathscr D_\mu\psi
          +{\rm h.c.}\Big)-m_f\bar\psi\psi\Big] \\
          &=&\sqrt{-\tt g}\Big[\Big(\frac{i}{2}\bar\psi\gamma^a
              {e_a}^\mu\Big(\partial_\mu+\frac{i}{4}
          \sigma^{bc}\omega_{\mu bc}\Big)\psi
          +{\rm h.c.}\Big)-m_f\bar\psi\psi\Big]       \label{lg:curveD}
\eea
The latin indices of the spin connection are by definition raised and
lowered with the flat space metric.

We now explain why we started with the explicitly hermitian form 
$\mathscr L{_0^{(f)}}$ of the flat space Dirac Lagrangian. As long
as we stay in flat space, the form  
$\mathscr L{_0^{(f)\,\prime}}=\bar\psi(i\gamma^\mu\partial_\mu -m_f)\psi$
can also be used, since it differs from $\mathscr L{_0^{(f)}}$
by $\frac{i}{2}\partial_\mu(\bar\psi\gamma^\mu\psi)$ which contributes
a surface term to the action. But this difference is modified to
$\frac{i}{2}\sqrt{-\tt g}{e_a}^\mu\partial_\mu(\bar\psi\gamma^a\psi)$
in curved space, which is no longer a surface term. Thus we would have
missed some contribution if we naively modified $\mathscr L{_0^{(f)\,\prime}}$
to make the transition to curved space. 

Our next aim is to express the spin connections in terms of the vierbeins.
The procedure to be adopted by us will involve the covariant derivative
of the vierbein. So we first discuss the covariant derivative of a
local Lorentz vector.

We have already discussed the covariant derivative of a local Lorentz spinor,
and similar considerations apply for other representations of the
Lorentz group. The analogues of Eqs.\ (\ref{lg:lLTspin})-(\ref{lg:conn})
for a local Lorentz vector field $V^a$ are
\bea
V^a(x)\rightarrow V^{\prime\,a}(x)={\Lambda^a}_b(x)V^b(x),
\eea
\bea
{\mathscr D}_\mu V^a\equiv\partial_\mu V^a
       +{[\Omega{^V_\mu}]^a}_b V^b,              \label{lg:covvec}
\eea
\bea
{[\Omega{^{V\,\prime}_\mu}]^a}_b 
       ={\Lambda^a}_c{[\Omega{^V_\mu}]^c}_d{(\Lambda^{-1})^d}_b
        -(\partial_\mu{\Lambda^a}_c){(\Lambda^{-1})^c}_b
                                                   \label{lg:connvprime}  
\eea
\bea
{\mathscr D}_\mu V^a(x)\rightarrow {\Lambda^a}_b(x)
              {\mathscr D}_\mu V^b(x),
\eea
\bea
{[\Omega{^V_\mu}]^a}_b(x)=\frac{i}{2}{[J_{cd}]^a}_b
                \,{\omega_\mu}^{cd}(x).            \label{lg:connvgen}
\eea 
Here $\Omega{^V_\mu}$ is the connection matrix and $J_{cd}$ the Lorentz
group generators in the vector representation:
\bea
[J_{cd}]^{ab}=i(\delta{^a_c}\delta{^b_d}-\delta{^a_d}\delta{^b_c}).
                                                   \label{lg:genvec}
\eea
The spin connection is representation-independent and the same as
that in Eq.\ (\ref{lg:conn}). Eqs.\ (\ref{lg:connvgen}) and (\ref{lg:genvec})
yield
\bea
[\Omega{^V_\mu}]^{ab}=-{\omega_\mu}^{ab}            \label{lg:connconn}
\eea  
and so Eq.\ (\ref{lg:covvec}) reduces to                
\bea
{\mathscr D}_\mu V^a=\partial_\mu V^a-{{\omega_\mu}^a}_b V^b.
                                                       \label{lg:covvec1}
\eea

For completeness we now give the transformation of $\omega_{\mu ab}$
(which are the analogues of the gauge fields of Yang-Mills theory)
under an infinitesimal local LT, parametrized by
\bea
{\Lambda^a}_b(x)=\delta{^a_b}+{\theta^a}_b(x).         \label{lg:inflLT}
\eea
Eq.\ (\ref{lg:Lorentz}) then gives the condition
\bea
\theta_{ba}=-\theta_{ab}.          \label{lg:thetanti}
\eea
where $\theta_{ab}\equiv\eta_{ac}{\theta^c}_b$.
Using Eq.\ (\ref{lg:connconn}), Eq.\ (\ref{lg:inflLT}) and 
${(\Lambda^{-1})^a}_b=\delta{^a_b}-{{\theta^a}_b}$
in Eq.\ (\ref{lg:connvprime}), we obtain
\bea
\omega^\prime_{\mu ab}=\omega_{\mu ab}+\partial_\mu \theta_{ab}
                       -\omega_{\mu ac}{\theta^c}_b
                       +\omega_{\mu bc}{\theta^c}_a.
                                                  \label{lg:omegaprime}
\eea
One can check that the same transformation is obtained on using
Eq.\ (\ref{lg:conn}),
\bea
D(\Lambda(x))=1-\frac{i}{2}S_{ab}\theta^{ab}(x),\label{lg:infD}
\eea
and the fact that $S_{ab}$ satisfy the Lorentz algebra
\bea
[S_{ab},S_{cd}]=i(\eta_{bc}S_{ad}-\eta_{ac}S_{bd}
                  -\eta_{bd}S_{ac}+\eta_{ad}S_{bc})
\eea
in Eq.\ (\ref{lg:connprime}). (It is necessary to make 
$\omega^\prime_{\mu ab}$ manifestly antisymmetric in $a$ and $b$ for having
a complete agreement with Eq.\ (\ref{lg:omegaprime}).) Note that Eq.\ (\ref{lg:infD}) is
consistent with Eq.\ (\ref{lg:inflLT}), since the analogue of 
the former, viz.,
${\Lambda^a}_b(x)=\delta{^a_b}-\frac{i}{2}{[J_{cd}]^a}_b
\theta^{cd}(x)$ leads to the latter on use of 
Eq.\ (\ref{lg:genvec}).

Let us now return to Eq.\ (\ref{lg:covvec1}). From this equation, it is 
straightforward to write down the
covariant derivative of the vierbein:
\bea
{\mathscr D}_\mu {e^a}_\nu
              = \partial_\mu {e^a}_\nu-\Gamma^\lambda_{\mu\nu}{e^a}_\lambda
              -{{\omega_\mu}^a}_b {e^b}_\nu            \label{lg:viercov}
\eea
Here we have added the usual affine connection term for the covariant 
derivative of a coordinate vector. Since the vierbein is 
like the square-root of the curved space metric, it is
natural to demand that its covariant derivative be zero. Put in another way,
\bea
{\mathscr D}_\mu {e^a}_\nu =0                             \label{lg:viercov1}    
\eea
ensures that the covariant derivative of the metric is zero, since
Eq.\ (\ref{lg:ortho1}) holds
and the covariant derivative obeys the Leibniz rule. The spin connections
can now be determined in terms of the vierbeins by making use of the above
condition.

Actually, instead of using the condition (\ref{lg:viercov1}), it is more
convenient to use
the weaker condition
\bea
{\mathscr D}_\mu e_{a\nu} - {\mathscr D}_\nu e_{a\mu} =0.    \label{lg:mu-nu}
\eea
The antisymmetrization on the L.H.S. of Eq.\ (\ref{lg:mu-nu}) ensures that the 
Christoffel symbol 
does not appear when we use Eq.\ (\ref{lg:viercov}) in it:
\bea
\partial_\mu e_{a\nu}- \partial_\nu e_{a\mu}
          -\omega_{\mu a b} {e^b}_\nu +\omega_{\nu a b} {e^b}_\mu=0.
                                                        \label{lg:viervier}          
\eea
Now we multiply with ${e_c}^\mu {e_d}^\nu {e^a}_\lambda$ to get
\bea
{e_c}^\mu {e_d}^\nu {e^a}_\lambda 
                   (\partial_\mu e_{a\nu}- \partial_\nu e_{a\mu})
          +\omega_{\mu d a}{e_c}^\mu {e^a}_\lambda 
          -\omega_{\nu c a}{e_d}^\nu {e^a}_\lambda=0. \label{lg:vierviervier}
\eea
Here we used the vierbein orthonormality relation 
Eq.\ (\ref{lg:ortho2}). Next we use 
Eq.\ (\ref{lg:viervier}) to replace $\omega_{\mu d a} {e^a}_\lambda$ 
and $\omega_{\nu c a} {e^a}_\lambda$ in Eq.\ (\ref{lg:vierviervier}). 
After using Eq.\ (\ref{lg:ortho2}) again, this gives
\bea
\omega_{\mu ab}=\frac{1}{2}
      \Big[{e_a}^\nu(\partial_\nu e_{b\mu}-\partial_\mu e_{b\nu})
      -{e_b}^\nu(\partial_\nu e_{a\mu}-\partial_\mu e_{a\nu})
      +{e_a}^\nu{e_b}^\lambda{e^c}_\mu
       (\partial_\nu e_{c\lambda}-\partial_\lambda e_{c\nu})\Big].
                                                          \label{lg:connvier}
\eea

Finally we note that the Lagrangian given in Eq.\ (\ref{lg:curveD})
accounts only for the gravitational interaction of the Dirac fermion.
Electromagnetic interaction is introduced in the usual way by further
demanding invariance under $U(1)_{\rm EM}$ gauge transformation:
\bea
{\mathscr L}{_{\rm G,EM}^{(f)}}=
          \sqrt{-\tt g}\Big[\Big(\frac{i}{2}\bar\psi\gamma^a
              {e_a}^\mu\Big(\partial_\mu+ie{\cal Q}_f A_\mu
            +\frac{i}{4}
          \sigma^{bc}\omega_{\mu bc}\Big)\psi
          +{\rm h.c.}\Big)-m_f\bar\psi\psi\Big]    \label{lg:curveDEM} 
\eea
where ${\cal Q}_f$ is the charge of the fermion.
Since $A_\mu$ is a local Lorentz scalar and a coordinate vector, the new
term maintains the symmetries we already had.
\section{Interaction vertices}\label{lg:vertices}
QED in curved space is governed by the Lagrangians $\mathscr L
{^{(\gamma)}_{\rm G}}$ and $\mathscr L{^{(f)}_{\rm G,EM}}$ given by
Eqs.\ (\ref{lg:photon1}) and (\ref{lg:curveDEM}) respectively. In
this section, we expand these Lagrangians to $O(\kappa)$ in the
approximation of linearized gravity, and obtain the Feynman rules
for the lowest order interaction of the graviton with photons and 
fermions.
\subsection{Photon-photon-graviton vertex}\label{lg:ppgv}
On use of Eqs.\ (\ref{lg:sqrt-g}) and (\ref{lg:^munu}) in
Eq.\ (\ref{lg:photon1}), the $O(\kappa)$ term in 
$\mathscr L{^{(\gamma)}_{\rm G}}$ turns out to be
\bea
\mathscr L_{AAh}=-\kappa h^{\lambda\rho}(x)\widehat 
T{^{(A)}_{\lambda\rho}}(x)
\eea
with the energy-momentum tensor operator for the photon given by
\bea
\widehat T{^{(A)}_{\lambda\rho}}=
\frac{1}{4} \eta_{\lambda\rho} F_{\alpha\beta} F^{\alpha\beta} +
F_{\lambda\alpha}{F^\alpha}_\rho.
\eea
Writing in terms of the photon field,
\bea
\widehat T{^{(A)}_{\lambda\rho}}
           &=&\frac{1}{2}\eta_{\lambda\rho}
           \Big[(\partial_\alpha A_\beta) (\partial^\alpha A^\beta)
           -(\partial_\alpha A_\beta) (\partial^\beta A^\alpha)\Big]
           +(\partial_\lambda A_\alpha)(\partial^\alpha A_\rho)
           +(\partial_\rho A_\alpha)(\partial^\alpha A_\lambda)
                       \nonumber\\
           &&-(\partial_\alpha A_\lambda) (\partial^\alpha A_\rho)
           -(\partial_\lambda A_\alpha) (\partial_\rho A^\alpha)
                                       \label{lg:emphoton}
\eea

\begin{figure} 
\begin{center}
\begin{picture}(180,60)(0,0)
\Photon(50,30)(100,0){-2}{7}
\ArrowLine(96,8)(86,14)
\Text(105,0)[l]{$A_\mu(k)$}
\Photon(50,30)(100,60){2}{7}
\ArrowLine(86,46)(96,52)
\Text(105,60)[l]{$A_\nu(k')$}
\Photon(0,30)(50,30){2}{7}
\Photon(0,30)(50,30){-2}{7}
\Text(25,35)[b]{$h_{\lambda\rho}$}
\Text(140,30)[l]{$= -i\kappa C_{\mu\nu\lambda\rho}(k,k')$}
\end{picture}
\end{center}

\caption{Feynman rule for the coupling of the photon with
the graviton. $C_{\mu\nu\lambda\rho} (k,k')$ is given in Eq.\ 
(\ref{Cmunulambdarho}). \label{lg:f:AAh}
}
\end{figure}
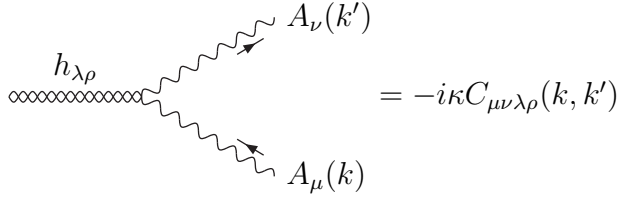 

Let us now consider the vertex given in
Fig.\ \ref{lg:f:AAh}.
Each term of $\mathscr L_{AAh}$ gives two contributions to this vertex,
depending on which of the two photon field operators in the term destroys 
the photon and which creates it. Let us take, as an example, the
$(\partial_\alpha A_\beta) (\partial^\alpha A^\beta)$ part in Eq.\
(\ref{lg:emphoton}). When $A_\beta$ destroys the photon of momentum $k$
and $A^\beta$ creates the photon of momentum $k^\prime$, the contribution
is $(-ik_\alpha)\eta_{\beta\mu}(i k^{\prime\,\alpha})\delta{^\beta_\nu}$;
while for the reverse case, it is $(ik^\prime_\alpha)\eta_{\beta\nu}
(-ik^\alpha)\delta^\beta_\mu$. Adding up, we find that the total contribution
to the vertex from the $-\kappa h^{\lambda\rho} \frac{1}{2}\eta_{\lambda\rho}
(\partial_\alpha A_\beta) (\partial^\alpha A^\beta)$ term of $\mathscr L_{AAh}$
is $-i\kappa\eta_{\lambda\rho}\eta_{\mu\nu} k\cdot k^\prime$. Carrying out a 
similar exercise 
for the other terms of $\mathscr L_{AAh}$, we finally obtain
\cite{lg:berends, lg:photgrav} 
\begin{eqnarray}
C_{\mu\nu\lambda\rho} (k,k') &=& \eta_{\lambda\rho} ( \eta_{\mu\nu} k
\cdot k' - k'_\mu k_\nu ) - \eta_{\mu\nu} (k_\lambda k'_\rho +
k'_\lambda k_\rho ) 
+ k_\nu (\eta_{\lambda\mu} k'_\rho +
\eta_{\rho\mu} k'_\lambda)\nonumber\\
&&+ k'_\mu (\eta_{\lambda\nu} k_\rho +
\eta_{\rho\nu} k_\lambda) 
- k \cdot k' (\eta_{\lambda\mu} \eta_{\rho\nu} +
\eta_{\lambda\nu} \eta_{\rho\mu}).                     \label{Cmunulambdarho}
\end{eqnarray}
\subsection{Fermion-fermion-graviton vertex}\label{lg:ffgv}
This vertex emerges from the Lagrangian 
$\mathscr L{^{(f)}_{\rm G,EM}}$ given by Eq.\ (\ref{lg:curveDEM}), which
involves the vierbein. In the weak field approximation,
the vierbein can be written as
\bea
{e^a}_\mu(x)=\delta{^a_\mu}+\kappa{\phi^a}_\mu(x).     \label{lg:vierwk1}
\eea
Here and below, we drop terms of higher order than what is necessary to
determine the vierbein to $O(\kappa)$. Note that $\delta{^a_\mu}$, and
hence ${\phi^a}_\mu$, are neither local Lorentz vectors nor coordinate vectors,
but only flat space tensors of rank two. Consequently, {\it all} tensors in an 
expression should
be regarded only as flat space tensors {\it after} we use the approximation of
Eq.\ (\ref{lg:vierwk1}) to perform an expansion of the expression in powers of $\kappa$. The 
distinction between greek and latin
indices would then be dropped, and they would be raised and lowered
with the flat space metric. 

The vierbein is determined from the orthonormality relation given by 
Eq.\ (\ref{lg:ortho1}) or Eq.\ (\ref{lg:ortho2}). Thus, use of Eqs.\ (\ref{lg:vierwk1})
and (\ref{lg:hdef}) in Eq.\ (\ref{lg:ortho1}) leads to
\bea
\phi_{\mu\nu}+\phi_{\nu\mu}=2h_{\mu\nu}.                  \label{lg:phi+phi}
\eea 
Eq.\ (\ref{lg:phi+phi}) does not determine the antisymmetric part of $\phi_{\mu\nu}$.
This freedom corresponds to the fact that the curved space action for the Dirac 
field is invariant under local LT: a new vierbein, given by Eq.\ (\ref{lg:viertrans}),
is as good as the old one. The ambiguity is removed by imposing the condition
\bea
\phi_{\mu\nu}=\phi_{\nu\mu}.
       \label{lg:phisymm}
\eea
This is called the Lorentz symmetric gauge \cite{lg:symmgauge}. Let us now
dwell a little on this condition.

The allowed local LT's in the weak field approximation are only those which sustain 
the closeness of the vierbein to
the unit matrix. So we consider the local LT 
\bea
{\Lambda^a}_b(x)=\delta{^a_b} +\kappa{\theta^a}_b(x).
\eea
Using this and Eq.\ (\ref{lg:vierwk1}) in Eq.\ (\ref{lg:viertrans}), we arrive at the
transformation
\bea
\phi{^\prime_{a\mu}}=\phi_{a\mu}+\theta_{a\mu}.  \label{lg:phitrans}
\eea
Because $\theta_{a\mu}$ is antisymmetric (see Eq.\ (\ref{lg:thetanti})), we
can always choose it to cancel out the antisymmetric part of $\phi_{a\mu}$
and implement the gauge condition given by Eq.\ (\ref{lg:phisymm}). Another
point to note is that the gauge condition does not allow any further
transformation like (\ref{lg:phitrans}), and so it does
break the local LT invariance. Finally we remark that the harmonic gauge condition,
which was implemented earlier using general coordinate transformation invariance,
is not disturbed by the implementation of the Lorentz symmetric gauge condition,
since $g_{\mu\nu}$ is a local Lorentz scalar.

Let us now return to Eq.\ (\ref{lg:vierwk1}). Since Eqs.\ (\ref{lg:phi+phi}) and 
(\ref{lg:phisymm}) give $\phi_{\mu\nu}=h_{\mu\nu}$, we have
\bea
{e^a}_\mu(x)=\delta{^a_\mu}+\kappa{h^a}_\mu(x).     \label{lg:vierwk2}
\eea
This implies that $e_{a\mu}=\eta_{a\mu}+\kappa h_{a\mu}$. And use of 
Eq.\ (\ref{lg:^munu}) in ${e_a}^\mu=g^{\mu\nu}e_{a\nu}$ gives
\bea
{e_a}^\mu=\delta{_a^\mu}-\kappa{h_a}^\mu.     \label{lg:vierwk3}
\eea

Now we are in a position to expand the Lagrangian 
$\mathscr L{^{(f)}_{\rm G,EM}}$ to $O(\kappa)$. First consider the terms in
Eq.\ (\ref{lg:curveDEM}) involving the spin connection:
\bea
-\frac{1}{8}\sqrt{-\tt g}{e_a}^\mu \omega_{\mu bc}
(\bar\psi\gamma^a \sigma^{bc}\psi+{\rm h.c.}).       \label{lg:Lvier}
\eea
On using the weak field expressions for the vierbeins in Eq.\ (\ref{lg:connvier}),
we obtain
\bea
\sqrt{-\tt g}{e_a}^\mu \omega_{\mu bc}=\kappa(\partial_c h_{ba}-\partial_b h_{ca}).
                                                        \label{lg:Lvier1}
\eea
Also,
\bea
\bar\psi\gamma^a \sigma^{bc}\psi+{\rm h.c.}
         &=&\bar\psi[\gamma^a,\sigma^{bc}]_+\psi \nonumber\\
         &=&\frac{i}{2}\bar\psi\Big((\gamma^a\gamma^b\gamma^c-\gamma^c\gamma^b\gamma^a)
        + (\gamma^b\gamma^c\gamma^a-\gamma^a\gamma^c\gamma^b)\Big)\psi
         \nonumber\\         
         &=&2\epsilon^{abcd}\bar\psi\gamma_d\gamma_5\psi,        \label{lg:Lvier2}    
\eea 
using
\bea
\gamma^a\gamma^b\gamma^c=\eta^{ab}\gamma^c+\eta^{bc}\gamma^a-\eta^{ca}\gamma^b
                  -i\epsilon^{abcd}\gamma_d\gamma_5.
\eea
The product of the expressions given by Eq.\ (\ref{lg:Lvier1}) and (\ref{lg:Lvier2})
vanish because the graviton field is symmetric. Thus the expression 
(\ref{lg:Lvier}) vanishes at $O(\kappa)$.

The other terms in Eq.\ (\ref{lg:curveDEM}) not involving $A_\mu$ are 
\bea          
\sqrt{-\tt g}\Big[\Big(\frac{i}{2}\bar\psi\gamma^a{e_a}^\mu\partial_\mu\psi
          +{\rm h.c.}\Big)-m_f\bar\psi\psi\Big].
\eea 
Using Eqs.\ (\ref{lg:sqrt-g}) and (\ref{lg:vierwk3}), the $O(\kappa)$ term in the
above expression turns out to be 
\bea
\mathscr L_{ffh}=-\kappa h^{\lambda\rho}(x)\widehat 
T{^{(f)}_{\lambda\rho}}(x)
\eea
with the energy-momentum tensor operator for the fermion given by
\bea
\widehat T{^{(f)}_{\lambda\rho}}(x)=
\Big(\frac{i}{4} \bar\psi(x)[\gamma_\lambda\partial_\rho
+\gamma_\rho\partial_\lambda]\psi(x)+{\rm h.c.}\Big)
-\eta_{\lambda\rho}\mathscr L{^{(f)}_0}(x).      \label{lg:emfermi} 
\eea
(see Eq.\ (\ref{lg:flatD}) for $\mathscr L{^{(f)}_0}$).
Here we have explicitly
symmetrized the term within square brackets in Eq.\ (\ref{lg:emfermi}), making use
of the fact that it comes multiplied with $h^{\lambda\rho}$. Writing out in full,       
\bea
\widehat T{^{(f)}_{\lambda\rho}}&=&\frac{i}{4}\Big(
\bar\psi\gamma_\lambda\partial_\rho\psi
+\bar\psi\gamma_\rho\partial_\lambda\psi
-(\partial_\rho\bar\psi)\gamma_\lambda\psi
-(\partial_\lambda\bar\psi)\gamma_\rho\psi\Big)    \nonumber\\
&&-\eta_{\lambda\rho}\Big(\frac{i}{2}
(\bar\psi\gamma^\mu\partial_\mu\psi
-(\partial_\mu\bar\psi)\gamma^\mu\psi)
-m_f\bar\psi\psi\Big).
\eea

\begin{figure} 
\begin{center}
\begin{picture}(100,60)(-30,0)
\Photon(0,30)(0,60){2}{5}
\Photon(0,30)(0,60){-2}{5}
\Text(5,50)[l]{$h_{\lambda\rho}$}
\ArrowLine(-30,30)(0,30)
\Text(-15,15)[b]{$f(p)$}
\ArrowLine(0,30)(30,30)
\Text(15,15)[b]{$f(p')$}
\Text(40,30)[l]{$=-i\kappa V_{\lambda\rho}(p,p')$}
\end{picture}
\end{center}

\caption{Feynman rule for the tree-level gravitational
vertex of a fermion. $V_{\lambda\rho}(p,p')$ is given in Eq.\ 
(\ref{lg:Vlamrho}).  \label{lg:f:ffh}}
\end{figure} 

Let us now consider the vertex given in
Fig.\ \ref{lg:f:ffh}. Each term of
$\mathscr L_{ffh}$ gives a contribution to this vertex, with $\psi$ destroying
the fermion and $\bar\psi$ creating it. We thus obtain 
\cite{lg:symmgauge, lg:choi, gravnu}
\bea
V_{\lambda\rho}(p,p^\prime)=\frac{1}{4}[\gamma_\lambda(p+p^\prime)_\rho
+\gamma_\rho(p+p^\prime)_\lambda]
-\frac{1}{2}\eta_{\lambda\rho}[(\rlap/p-m)+(\rlap/p^\prime-m)].
                                                           \label{lg:Vlamrho}
\eea
\subsection{Fermion-fermion-photon-graviton vertex}
The term involving $A_\mu$ in Eq.\ (\ref{lg:curveDEM}) is 
\bea
-e{\cal Q}_f \sqrt{-\tt g}{e_a}^\mu \bar\psi\gamma^a\psi A_\mu.
\eea
Using Eqs.\ (\ref{lg:sqrt-g}) and (\ref{lg:vierwk3}), the $O(\kappa)$ term in 
this expression turns out to be \cite{lg:photgrav, lg:choi}
\bea
\mathscr L_{ffAh}=-e\kappa{\cal Q}_f a_{\mu\nu\lambda\rho}h^{\lambda\rho}
    \bar\psi\gamma^\nu\psi A^\mu,              \label{lg:LffAh}
\eea
with
\bea
a_{\mu\nu\lambda\rho}=\eta_{\mu\nu}\eta_{\lambda\rho}
-\frac{1}{2}(\eta_{\mu\lambda}\eta_{\nu\rho}
+\eta_{\mu\rho}\eta_{\nu\lambda}).                                \label{amunulambdarho}
\eea 
Here we have explicitly symmetrized $a_{\mu\nu\lambda\rho}$ in $\lambda$ 
and $\rho$. The Lagrangian of Eq.\ (\ref{lg:LffAh}) gives rise to the
fermion-fermion-photon-graviton vertex with the vertex factor
$-ie\kappa{\cal Q}_f a_{\mu\nu\lambda\rho}\gamma^\nu$.
\subsection{Important features of the vertex factors}
All the three vertex factors deduced by us are seen to be symmetric under the exchange of
$\lambda$ and $\rho$. This is because they represent the coupling to $h_{\lambda\rho}$,
which is a symmetric tensor. We now
proceed to discuss the other important features of these vertex factors.

The photon-photon-graviton vertex factor $-i\kappa C_{\mu\nu\lambda\rho}(k,k')$, with
$C_{\mu\nu\lambda\rho}(k,k')$ given by Eq.\ (\ref{Cmunulambdarho}), 
is also symmetric under the 
simultaneous exchange
of $\mu$ and $\nu$, and, $k$ and $k'$. This reflects the Bose symmetry under
the interchange of the two photons. The vertex factor also satisfies electromagnetic
gauge invariance:
\bea
k^\mu C_{\mu\nu\lambda\rho}(k,k')=0=k'^\nu C_{\mu\nu\lambda\rho}(k,k')\,.   \label{egic}
\eea
It also satisfies gravitational gauge invariance:
\bea
(k-k')^\lambda\epsilon^\mu(k)\epsilon^\nu(k')C_{\mu\nu\lambda\rho}(k,k')=0
=(k-k')^\rho\epsilon^\mu(k)\epsilon^\nu(k')C_{\mu\nu\lambda\rho}(k,k')\,,      \label{ggic}
\eea
$k-k'$ being the four-momentum of the graviton. In writing Eq.\ (\ref{ggic}), we
have taken the polarization vectors of the photons to be real. To prove Eq.\ (\ref{ggic}),
one has to use the on-shell conditions $k^2=0=k'^2$ and $k\cdot\epsilon(k)=0
=k'\cdot\epsilon(k')$ for the photons.

It may be noted that to prove a given kind of gauge invariance, the particles
carrying the corresponding charge must be put on-shell. Since neither the photon nor
the graviton carries electric charge, it was not necessary to put them on-shell
in Eq.\ (\ref{egic}). On the other hand, the photon has gravitational coupling, i.e. carries
``gravitational charge" (this is in fact true for any particle, as all
particles possess energy). So the photons were put on-shell in Eq.\ (\ref{ggic}).

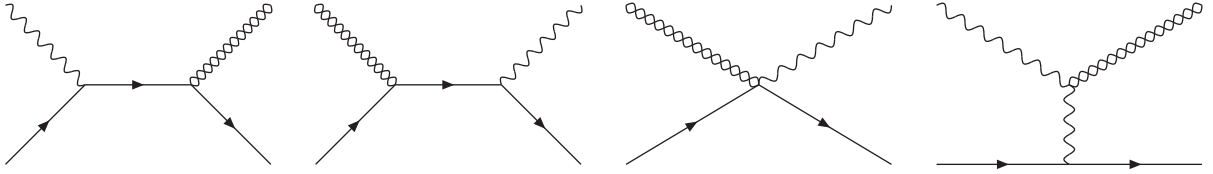
\begin{figure} 
\begin{center}
\begin{picture}(90,60)(0,0)
\ArrowLine(0,0)(30,30)
\ArrowLine(30,30)(70,30)
\ArrowLine(70,30)(100,0)
\Photon(0,60)(30,30){2}{6}
\Photon(70,30)(100,60){2}{6}
\Photon(70,30)(100,60){-2}{6}
\end{picture}
\qquad
\begin{picture}(90,60)(0,0)
\ArrowLine(0,0)(30,30)
\ArrowLine(30,30)(70,30)
\ArrowLine(70,30)(100,0)
\Photon(70,30)(100,60){2}{6}
\Photon(0,60)(30,30){2}{6}
\Photon(0,60)(30,30){-2}{6}
\end{picture}
\qquad
\begin{picture}(90,60)(0,0)
\ArrowLine(0,0)(50,30)
\ArrowLine(50,30)(100,0)
\Photon(50,30)(100,60){2}{7}
\Photon(0,60)(50,30){2}{7}
\Photon(0,60)(50,30){-2}{7}
\end{picture}
\qquad
\begin{picture}(90,60)(0,0)
\ArrowLine(0,0)(50,0)
\ArrowLine(50,0)(100,0)
\Photon(50,0)(50,30){2}{4}
\Photon(0,60)(50,30){2}{7}
\Photon(50,30)(100,60){2}{7}
\Photon(50,30)(100,60){-2}{7}
\end{picture}
\end{center}

\caption{Diagrams of $O(e\kappa)$ with two fermions, one photon and one graviton in the
external lines. These are the tree-level diagrams for the photoproduction of
gravitons off electrons.
\label{f:photprod}
}
\end{figure} 

We now turn to the fermion-fermion-graviton vertex factor $-i\kappa V_{\lambda\rho}(p,p')$,
with $V_{\lambda\rho}(p,p')$ given by Eq.\ (\ref{lg:Vlamrho}). Here again one can
check gravitational gauge invariance:
\bea
(p-p')^\lambda \bar u(\vec P\,')V_{\lambda\rho}(p,p')u(\vec P)=0=
(p-p')^\rho \bar u(\vec P\,')V_{\lambda\rho}(p,p')u(\vec P)
\eea
with the electrons put on-shell for reasons stated just now.

Finally we consider the fermion-fermion-photon-graviton vertex factor $ie\kappa
a_{\mu\nu\lambda\rho}\gamma^\nu$ with $a_{\mu\nu\lambda\rho}$ given by
Eq.\ (\ref{amunulambdarho}). The checking of 
gauge invariance is now less trivial. This is because there are three other diagrams of
$O(e\kappa)$ with the same external lines. These four diagrams, given in Fig.\ 
\ref{f:photprod},
represent the photoproduction of gravitons off electrons at tree-level
(note that the third diagram from the left
is just the vertex under consideration). When all the four diagrams are taken together, it 
is possible to prove  electromagnetic and gravitational
gauge invariance \cite{lg:photgrav}. 

\def\re{\mathop{\rm Re}\nolimits}
\def\im{\mathop{\rm Im}\nolimits}

\chapter{Charged leptons in thermal QED}\label{qed}
\section{Dispersion relation and inertial mass}\label{qed:dis}
We saw in Chapter \ref{th} that in the real-time formulation of
thermal field theory, there are four free-field propagators. Similarly
there are four full or exact propagators with the Heisenberg picture
fields replacing the interaction picture fields in the definitions of
the free-field propagators.
There are also four self-energies $\Sigma_{ab}$ with
$-i\Sigma_{ab}$ ($a,b=1,2$) denoting the 1PI self-energy diagram in
which the momentum enters at a $b$-type vertex and leaves at an
$a$-type vertex. (We are now dealing with the specific case of
fermions.) Thus the free-field propagator, the full or exact
propagator and the self-energy are all $2\times 2$ matrices
\cite{landsman, nieves}. The poles of the
full propagator for a fermion of momentum $p_\mu$ occur at the poles
of the function
\bea
S'_f(p)=[\rlap/p -m_f-\Sigma_f(p)]^{-1} 
\label{qed:pole}
\eea
where $\Sigma_f(p)$ is related to the components of the self-energy matrix.
Let us define
\bea
\re\Sigma_f\equiv\frac{1}{2}(\Sigma_f+\gamma^0\Sigma{_f^\dagger}\gamma^0)
\eea
and
\bea
\im \Sigma_f\equiv\frac{1}{2i}(\Sigma_f-\gamma^0\Sigma{_f^\dagger}\gamma^0),
\eea
and likewise for $\Sigma_{ab}$. Then the aforementioned relations are 
\cite{landsman, nieves}
\bea
\re\Sigma_f(p)=\re\Sigma_{11}(p),                       \label{qed:Real}           
\eea
\bea
\im \Sigma_f(p)=\frac{\im \Sigma_{11}(p)}{1-2\eta_f(p)},      \label{qed:Im}  
\eea
with $\eta_f(p)$ given by Eq.\ (\ref{etaf}).

Considering the analogy of Eq.\ (\ref{qed:pole}) to the vacuum case, we shall call 
$\Sigma_f(p)$ the {\it self-energy of the fermion}.
It has the general form
\bea
\Sigma_f(p)=a\rlap/p+b\rlap/v+c                                \label{qed:form}
\eea
at one loop in an isotropic medium with four-velocity $v_\mu$.
Here $a$, $b$ and $c$ are Lorentz-invariant functions of the
 two Lorentz scalars
\bea
\omega\equiv p\cdot v,
\eea
\bea
P\equiv [(p\cdot v)^2-p^2]^{1/2}.                        \label{qed:Pdef}
\eea
$\omega$ and $P$ are the Lorentz-invariant energy and three-momentum
 respectively, which satisfy
\bea
p^2=\omega^2-{P}^2.
\eea
Eq.\ (\ref{qed:form}) can contain an additional term proportional to
$\sigma^{\mu\nu}p_\mu v_\nu$ in the 
more general case.  However, such
a term does not appear at the level of the one-loop calculations
\cite{qed:weldon} that we are considering in this work, and
therefore we omit it.

We shall be interested only in the dispersive part of the self-energy.
This means that the corresponding term $\bar\psi\Sigma_f\psi$ of the
Lagrangian in momentum space is hermitian, which translates to the condition
\bea
\Sigma_f(p)=\gamma^0 \Sigma{_f^\dagger}(p)\gamma^0.        \label{qed:cond}
\eea
 Thus, the dispersive part of the
self-energy is given by $\re\Sigma_f$. On the other hand, when one is
also interested in the absorption of the fermion in the medium, 
the Lagrangian must no longer be hermitian, and so one has to
consider the entire $\Sigma_f$.
For the relation between $\im \Sigma_f$ and the damping rate of the fermion,
see, for example, Refs. \cite{qed:damping} and \cite{qed:indra}. We shall
deal only with $\re\Sigma_f$ and {\it by $\Sigma_f$, we shall henceforth
 refer only to $\re \Sigma_f$.} Clearly, the Lorentz scalars $a$, $b$ and $c$ of 
Eq.\ (\ref{qed:form}) are real in our case. 

Now $S'_f(p)$ of Eq.\ (\ref{qed:pole}) can be written in the form
\bea
S'_f(p)=N(p)[D(p)]^{-1}                                 \label{qed:N/D}
\eea
with
\bea
N(p)=(1-a)\rlap/p -b\rlap/v+(m_f+c),                    \label{qed:N}
\eea
\bea
D(p) = [(1 - a)p -bv]^2 - (m_f + c)^2 .                 \label{qed:D}
\eea
The poles of $S'_f(p)$ are given by $D(p)=0$. At $O(e^2)$ (i.e., on keeping 
only the terms independent of or linear in $a$, $b$ and $c$), this leads to
\bea
\omega^2&=&{P}^2+m{_f^2}+2[a(\omega^2-{P}^2)+b\omega+m_f c]    \\
        &=&{P}^2+m{_f^2}+\frac{1}{2}\Tr[(\rlap/p+m_f)\Sigma_f]\,.
                                                              \label{disdis}
\eea
Taking the square root, we obtain the pair of solutions
\bea
\omega_\pm=\pm\sqrt{{P}^2+m{_f^2}}\pm\frac{1}{4 \sqrt{{P}^2+m{_f^2}}}
                  \Tr[(\rlap/p+m_f)\Sigma_f].       \label{qed:sol}
\eea
These are implicit equations for $\omega$ as function of $P$, since the
R.H.S. contains $a$, $b$ and $c$ which are again functions of $\omega$.
We now write
\bea
\omega_\pm({P})=\pm E_{f, \bar f}({P})                           \label{qed:sol1} 
\eea
Eq.\ (\ref{qed:sol}) can then be used to solve for $E_f({P})$ and
$E_{\bar f}({P})$. These are the dispersion relations
corresponding to the fermion and the antifermion respectively. In
perturbation theory, the justification for Eq.\ (\ref{qed:sol1}) is
that it reduces to the usual dispersion relations at the tree-level.
(See also Secs.\ \ref{qed:C} and \ref{qed:noC} on charge conjugation. 
For justification at a non-perturbative level, see the last paragraph
of this section.) The inertial masses are defined as
\bea
M_{f,\bar f}\equiv E_{f, \bar f}(0)=\pm\omega_\pm(0).        \label{qed:inn}
\eea
Eq.\ (\ref{qed:sol}) then gives
\bea
M_{f,\bar f}=m_f+\frac{1}{4m_f}\Bigg\{\Tr[(\rlap/p+m_f)\Sigma_f]\Bigg\}
              _{{\omega=\pm m_f}\atop{P=0}},
\eea
using the tree-level values of $\omega$ in the $O(e^2)$ part.

{\it We shall perform all calculations
in the rest frame of the medium}, in which $v^\mu$ has the components
given by Eq.\ (\ref{v}),
and in that frame, we define the components of $p^\mu$ by writing
\begin{eqnarray}
p^\mu = (p^0,\vec P) .
\end{eqnarray}
(Eq.\ (\ref{qed:Pdef}) shows that the magnitude of the rest-frame three-momentum $\vec P$
is indeed equal to the Lorentz-invariant three-momentum $P$. So our notations are
consistent.)
Then $a,b,c$ are functions of the variables $p^0$ and $P$,
which we will indicate by writing them as $a(p^0,P)$, and similarly
for the other ones, when we need to show it explicitly. 
In the same way,
$N(p)$ and $D(p)$ of Eqs.\ (\ref{qed:N}) and (\ref{qed:D}) will be written as
$N(p^0,\vec P)$ and $D(p^0,P)$. Since the condition $\omega=\pm m_f$
now reduces to $p^0=\pm m_f$, we have
\bea
M_{f,\bar f}=m_f+\frac{1}{4}\Bigg\{\Tr[(1\pm\gamma^0)\Sigma_f]\Bigg\}
                         _{p^\mu=(\pm m_f,\vec 0)}
\eea
Defining
\bea
{\cal E}_{f,\bar f}(p^0,\vec P)
          & = &\frac{1}{4}\Tr[(\gamma^0\pm 1)\Sigma_f]    \label{calEoper}\\
          & = &ap^0+b\pm c,                                    \label{calEdef}
\eea
the inertial masses to $O(e^2)$ are given by
\bea
M_f & = & m_f + {\cal E}_f(m_f,\vec 0), \nonumber\\
M_{\bar f} & = & m_f - {\cal E}_{\bar f} (-m_f,\vec 0).  \label{Moper}
\eea
Eq.\ (\ref{calEoper}) is a useful formula that allows us
to extract the matter-induced corrections to the inertial
mass directly from the one-loop expression for $\Sigma_f$.
As we will see later, the wavefunction renormalization factor is
determined in terms of the same quantities ${\cal E}_f$ and ${\cal
E}_{\bar f}$.

It may be noted $CPT$ symmetry, if present, ensures that $E_f(P)=E_{\bar f}(P)$,
and consequently, $M_f=M_{\bar f}$. However, normal matter is $CPT$-asymmetric,
and, as we shall see from explicit calculations, $M_f$ no longer equals
$M_{\bar f}$.

The antifermion dispersion relation of Eq.\ (\ref{qed:sol1}) and the hole dispersion
relation of Ref.\ \cite{weldon89} (see also Ref.\ \cite{physica}) are based
on the same concept, namely, the removal of a particle from the thermal background. 
However, Ref.\ \cite{weldon89} considers a different situation: high-temperature
massless QED (or QCD) at zero chemical potential. 
In this situation, while one can
decompose the propagator as $D(p)=(\omega-\omega_+)(\omega-\omega_-)$
such that $\omega_\pm=\pm{P}$ at the tree-level, there are some important 
differences 
from our case. Firstly, one gets two dispersion relation from $\omega_+$ itself.
One relation, of course, is obtained by writing $\omega_+=E_f({P})$.
The other is obtained by writing $\omega_+=-E_h({P})$. The form of
$\omega_+$ is such that there is an $E_h({P})$ which is positive for all $P$. This is the
hole state, also called plasmino \cite{plasmino}. Secondly, $\omega_-$ does not give dispersion 
relations distinct from those obtained from $\omega_+$, since the medium is $CPT$ symmetric.
\section{Spinor and wavefunction renormalization factor}\label{qed:UandZ}
Consider the momentum-space Dirac equation
\bea
\Big(p^0\gamma^0-\vec P\cdot\vec\gamma-m_f-
                     \Sigma_f(p^0,\vec P)\Big)\xi(\vec P) =0         \label{xi}
\eea
in a medium. Requiring Eq.\ (\ref{xi}) to have non-trivial solutions yields the
condition $D(p^0,P)=0$ (see Eq.\ (\ref{qed:D})). This equation, as we have seen, 
possesses the two 
solutions $p_0=E_f(P)$ and $p_0=-E_{\bar f}(P)$. Now $\xi(\vec P)$ for
$p_0=E_f(P)$ is the fermion spinor $U_s(\vec P)$. On the
other hand, $\xi(\vec P)$ for $p_0=-E_{\bar f}(P)$ is the antifermion spinor
at three-momentum $-\vec P$.

Thus, the fermion spinor $U_s(\vec P)$ in a medium satisfies the equation 
\bea
\Big(\rlap/p - m_f - \Sigma_f(p) \Big) U_s(\vec P) = 0     \label{U_s}
\eea
with $p^\mu = (E_f(P),\vec P)$. We choose the
normalization
\bea
U{^\dagger_s}(\vec P)U_s(\vec P)=1\,.               \label{Unorm}
\eea
The hermitian conjugate of Eq.\ (\ref{U_s}) is
\bea
\overline U_s(\vec P)\Big(\rlap/p - m_f - \Sigma_f(p) \Big)=0      \label{Ubar}
\eea
where the condition (\ref{qed:cond}) was used. We now use the general form of Eq.\ 
(\ref{qed:form}). Multiplying Eq.\ (\ref{U_s}) from the left with $\overline U_s(\vec P)
\gamma_\mu$ and Eq.\ (\ref{Ubar}) from the right with $\gamma_\mu U_s(\vec P)$, and adding
the two resulting equations, one then arrives at the identity
\bea
\overline U_s(\vec P) \gamma_\mu U_s(\vec P) = \left[ {(1-a)p_\mu -bv_\mu \over
m_f  +c} \right] \; \overline U_s(\vec P) U_s(\vec P).                         \label{qed:id}
\eea
The time-component of Eq.\ (\ref{qed:id}), together with the normalization adopted in Eq.\ 
(\ref{Unorm}), give
\begin{eqnarray}
\overline U_s(\vec P) U_s(\vec P) = {m_f+c \over (1-a)E_f - b},          \label{UbarU}
\end{eqnarray}
which, when fed back into Eq.\ (\ref{qed:id}), yields
\begin{eqnarray}
\overline U_s(\vec P) \gamma_\mu U_s(\vec P) = {(1-a)p_\mu -bv_\mu \over
(1-a)E_f - b} .
\end{eqnarray}
For future reference, we note that in particular, in the frame specified by 
Eq.\ (\ref{v}),
\bea
\left[\overline U_s(\vec P)U_s(\vec P)\right]_{\vec P = 0}  =  1
                                                               \label{GordonUP1}
\eea
(use Eq.\ (\ref{calEdef}) and the first of Eqs.\ (\ref{Moper}), or use
Eq.\ (\ref{Ucon}) below),
and
\bea 
\left[\overline U_s(\vec P)\gamma_\mu U_s(\vec P)\right]_{\vec P = 0} & = & v_\mu .
                                                               \label{GordonUP2}
\eea

From Eq.\ (\ref{U_s}), the explicit form of $U_s(\vec P)$ can easily be worked out
in the rest frame of the medium, using any particular representation of the
gamma matrices. The condition for Eq.\ (\ref{U_s}) to possess non-trivial
solution is given by
\bea
[(1-a)E_f-b]^2=(1-a)^2P^2+(m_f+c)^2                             \label{Ucon}
\eea
which, of course, is nothing but the equation $D(p_0, P)=0$ 
with $p_0=E_f$. The $U$-spinors in the Dirac-Pauli representation 
are then given by
\bea
U_{\pm}(\vec P)=\sqrt{\frac{(1-a)E_f-b+m_f+c}{2[(1-a)E_f-b]}}
\left(
\begin{array}{c}
\chi_\pm\\
\frac{\displaystyle (1-a)\vec\sigma\cdot\vec P}
{\displaystyle (1-a)E_f-b+m_f+c}\chi_\pm
\end{array}
\right)                                                            \label{Upm}
\eea                                                            
with the prefactor chosen in accord with the normalization condition of
Eq.\ (\ref{Unorm}). Here
\bea
\chi_+=\left({1\atop 0}\right)~,~~~ \chi_-=\left({0\atop 1}\right)\,. 
\eea
The solutions in Eq.\ (\ref{Upm}) can be used to find that
\bea
\sum\limits_s U_s(\vec P) \overline U_s(\vec P) = \frac{N(E_f,\vec P) }
{2\left[(1 - a)E_f - b\right]}\,.                                  \label{spinsum}
\eea
This relation is representation-independent.

Eq.\ (\ref{spinsum}) is useful in finding out the fermion wavefunction renormalization factor
$Z_f(P)$. In a medium $Z_f$ is defined through the behaviour of the propagator at the
one-particle pole $p_0=E_f(P)$ \cite{weldon89}:
\bea
S'_f(p)\bigg|_{\rm 1-particle} \approx
{Z_f(P) \sum\limits_s U_s(\vec P)\overline U_s(\vec P) \over p_0 - E_f}\,.          \label{qed:Sfull}
\eea
(See App. \ref{appz} for a proof of this relation in vacuum.) Note that
we have explicitly indicated the fact that $Z_f$ depends on $\vec P$, but only on the
magnitude of it, in an isotropic medium. This will be checked below. 
Near the pole considered in Eq.\ (\ref{qed:Sfull}),
Eq.\ ({\ref{qed:N/D}) reduces to
\bea
S'_f(p) \approx
\frac{N(E_f,\vec P)}{(p_0 - E_f)
\left(\frac{\textstyle\partial D}{\textstyle\partial p_0}
\right)_{p^0 = E_f}} \,.                                             \label{Sp_pole}
\eea
The requirement that the residues of these two expressions coincide, then yields
\bea
Z_f(P) = \left\{2 \Big[(1 - a)E_f -b \Big]
\left(\frac{\textstyle\partial D}{\textstyle\partial p_0}
\right)^{-1}\right\}_{p^0 = E_f}\, ,                           \label{Zf1}
\eea
where we have used Eq.\ (\ref{spinsum}). 
Note that since $a$, $b$, $c$ and (hence) $D$ depend on $\vec P$ only through $P$, so
also does $Z_f$. 
Now use
\bea
D(p_0,P)=(1-2a)(p{_0^2}-P^2)-2bp_0-m{_f^2}-2m_f c
\eea
to $O(e^2)$. Then, to this order, 
\bea
Z_f(P)=1+\Bigg[a+\Bigg(p_0-\frac{P^2}{p_0}\Bigg)\frac{\partial a}{\partial p_0}
      +\frac{\partial b}{\partial p_0}+\frac{m_f}{p_0}
       \frac{\partial c}{\partial p_0}\Bigg]_{p_0=E_f}\,.              \label{Zf2}
\eea
For the particular case $\vec P=0$ in which
we shall be interested, Eq.\ (\ref{Zf2}) reduces to
\bea 
Z_f = 1 + \zeta_f \,,                                          \label{Zf}
\eea
where
\bea
\zeta_f = \frac{\partial{\cal E}_f}{\partial p^0}
\Bigg|_{p^\mu = (m_f,\vec 0)} \,,                               \label{zf}
\eea
with ${\cal E}_f$ given by Eq.\ (\ref{calEoper}) or Eq.\ (\ref{calEdef}).
{}From now on whenever we omit the dependence of $Z_f$ on $P$ in an equation, it is to
be understood as the quantity evaluated at $P = 0$.

In this context, we note that in vacuum one can use the result ${Z_f}^{-1}=
1-(d\Sigma_f/d\rlap/p)_{\rlap/p=m}$ just as well as Eq.\ (\ref{Zf1}) in order
to determine the wavefunction renormalization factor. But since in a medium
$a$, $b$ and $c$ depend on $p\cdot v$, one cannot define $d\Sigma_f/d\rlap/p$
unambiguously, and so Eq.\ (\ref{Zf1}) is the result to be
used in the present case. 

We now consider the case of the antiparticles. 
We have already defined the antifermion spinor in the beginning of this section (see
after Eq.\ (\ref{xi})).
Thus, the antifermion spinor $V_s(\vec P)$
satisfies the equation
\bea
\Big(\rlap/p + m_f + \Sigma_f(-p) \Big) V_s(\vec P) = 0 \,,               \label{V_s}
\eea
where $p^\mu = (E_{\bar f}(P),\vec P)$. The
normalization is chosen to be
\begin{eqnarray}
V^\dagger_s(\vec P)V_s(\vec P) = 1 \,.                                       \label{Vnorm}
\end{eqnarray}
The analogy of Eq.\ (\ref{qed:id})
in the present case is
\bea
\overline V_s(\vec P) \gamma_\mu V_s(\vec P) = -\left[ {(1-a(-p))p_\mu + b(-p)
                  v_\mu \over m_f +c(-p)} \right] \; 
                  \overline V_s(\vec P) V_s(\vec P) \,.                      \label{idV}
\eea
From Eqs.\ (\ref{idV}) and (\ref{Vnorm}), one can obtain the expressions for
$\overline V_s(\vec P)V_s(\vec P)$ and $\overline V_s(\vec P)\gamma^\mu V_s(\vec P)$ just as in
the case of the $U$-spinors. It may be noted that since $a(p)$ depends on $\vec P$
only through $P$, the quantity $a(-p)$ essentially translates to $a(p)$ evaluated at
$p_0=-E_{\bar f}$. Similar remarks hold for $b$ and $c$.

The condition for Eq.\ (\ref{V_s}) to possess non-trivial
solution is given by
\bea
[(1-a(-p))E_{\bar f}+b(-p)]^2=(1-a(-p))^2P^2+(m_f+c(-p))^2       \label{Vcon}
\eea
which is the equation $D(p_0, P)=0$ with $p_0=-E_{\bar f}$. The 
normalized $V$-spinors 
in the Dirac-Pauli representation 
are then given by
\bea
V_{\pm}(\vec P)&=&\sqrt{\frac{(1-a(-p))E_{\bar f}+b(-p)+m_f+c(-p)}
            {2[(1-a(-p))E_{\bar f}+b(-p)]}} \nonumber\\
&\times &
\left(
\begin{array}{c}
\frac{\displaystyle (1-a(-p))\vec\sigma\cdot\vec P}
{\displaystyle (1-a(-p))E_{\bar f}+b(-p)+m_f+c(-p)}\chi\,'_\pm\\
\chi\,'_\pm
\end{array}
\right)\,.                                                         \label{Vpm}          
\eea 
Here
\bea
\chi\,'_+=-\left({0\atop 1}\right)~,~~~ \chi\,'_-=\left({1\atop 0}\right)\,. 
\eea
(The choice of $\chi\,'_\pm$
is motivated by consideration of charge conjugation;
this will be explained at the end of Sec.\ \ref{qed:C}.)
Eq.\ (\ref{Vpm}) leads to                                                           
\bea
\sum\limits_s V_s(\vec P) \overline V_s(\vec P) = -\frac{N(-E_{\bar f},-\vec P) }
{2\left[(1 - a(-p))E_{\bar f} + b(-p)\right]}\,.                 \label{VVbar}
\eea

$Z_{\bar f}(P)$ is defined through the relation
\bea
S'_f(p)\bigg|_{\rm 1-antiparticle} \approx
\frac{Z_{\bar f}( P) \sum\limits_s V_s(-\vec P)\overline 
V_s(-\vec P)}{p_0 + E_{\bar f}}\,.          \label{SfullV}
\eea
(See App. \ref{appz} for a proof of this relation in vacuum.) 
Near the pole considered in
Eq.\ (\ref{SfullV}),  Eq.\ (\ref{qed:N/D}) reduces to
\bea
S'_f(p) \approx
\frac{N(-E_{\bar f},\vec P)}{(p_0 + E_{\bar f})
\left(\frac{\textstyle\partial D}{\textstyle\partial p_0}
\right)_{p^0 = -E_{\bar f}}}\,.                       \label{Santi}
\eea
Eqs.\ (\ref{VVbar}), (\ref{SfullV}) and (\ref{Santi}) give
\bea
Z_{\bar f}(P) = -\left\{2 \Big[(1 - a(p))E_{\bar f}+b(p) \Big]
\left(\frac{\textstyle\partial D}{\textstyle\partial p_0}
\right)^{-1}\right\}_{p^0 = -E_{\bar f}}\, .                           \label{Zfbar1}
\eea
Hence we obtain, to $O(e^2)$,
\bea
Z_{\bar f}(P)=1+\Bigg[a+ \Bigg(p_0-\frac{P^2}{p_0}\Bigg)\frac{\partial a}{\partial p_0}
                  +\frac{\partial b}{\partial p_0}+\frac{m_f}{p_0}
              \frac{\partial c}{\partial p_0}\Bigg]_{p_0=-E_{\bar f}}\,.\label{Zfbar2}
\eea
At $P=0$, this reduces to
\begin{eqnarray}
Z_{\bar f} = 1 + \zeta_{\bar f}\,,                                   \label{Zfbar}
\end{eqnarray}
where
\begin{eqnarray}
\zeta_{\bar f} = \frac{\partial{\cal E}_{\bar f}}{\partial p^0}
                  \Bigg|_{p^\mu = (-m_f,\vec 0)} \,.                 \label{zfbar}
\end{eqnarray}

The tree-level spinors $u_s$ and $v_s$ are the limiting cases of the spinors
$U_s$ and $V_s$ when the effects of the medium are neglected.
They satisfy the free Dirac equation in the vacuum, as well as
the relations
\begin{eqnarray}
\bar u_s\gamma_\mu u_s & = & \frac{p_\mu}{m_f}\bar u_s u_s
\label{ugu} \\
\bar u_s u_s & = & {m_f \over E_f} \,
\label{unorm}
\end{eqnarray}
(where $p_0=\sqrt{P^2+m{_f^2}}$), with similar relations for $v_s$ but with the
substitution $p_\mu\rightarrow -p_\mu$ in the above equations.

We conclude this section by pointing out why a medium is expected to
impart new features to the wavefunction renormalization factor which
are absent in vacuum.  Firstly, had the wavefunction renormalization
factor been a function of momentum in vacuum, it would have led to new
kinds of divergence, spoiling the renormalizability of the theory. But
there is no such problem in a medium, since the thermal corrections do
not involve ultraviolet divergences. Secondly, $CPT$ invariance in
vacuum ensures that $Z_f=Z_{\bar f}$. But since a medium can be $CPT$-asymmetric,
this equality may not generally hold.
\section{Effect of charge conjugation invariance}\label{qed:C}
In this section, we consider the consequences of charge 
conjugation invariance for the self-energy, the dispersion relations, the wavefunction
renormalization factors and the antifermion spinor. $C$ invariance is present
when the background contains only photons, and also when it contains, in addition,
charged leptons at zero chemical potential. So these consequences can be
checked in such cases. 

We begin by writing down the effect of the charge conjugation operator
$C$ on a spinor field:
\bea
\psi_C(x)\equiv {C}\psi(x){C}
         ={\sf C}{\gamma_0}^{\rm T}\psi^*(x)\,.                  \label{qed:psiC}
\eea
(Note that $C^2=1$, so that $C^{-1}=C$.)
$C$ invariance of the free Dirac Lagrangian is ensured by the relation
\bea
{\sf C}^{-1}\gamma^\mu {\sf C}=-{\gamma_\mu}^{\rm T}\,           \label{qed:CR}
\eea
where the matrix $\sf C$ is unitary:
\bea
{\sf C}^{-1}={\sf C}^\dagger\,.                                   \label{qed:CU}
\eea
Using $(\psi_C)_C=\psi$, one can then show that
the matrix $\sf C$ is also antisymmetric:
\bea
{\sf C}^{\rm T}=-{\sf C}\,.                                        \label{qed:CA}
\eea

It will be useful to first go through the demonstration of the $C$ invariance of
the tree-level Dirac action
\bea
\mathscr A_f
   =\int d^4x \bar\psi(x)(i\gamma^\mu\partial_\mu-m_f)\psi(x)\,.     \label{qed:A}
\eea
Let us define the Fourier transform of the spinor field by
\bea
\psi(x)=\int\frac{d^4p}{(2\pi)^4}e^{-ip\cdot x}\psi(p)\,.            \label{qed:psiF}
\eea
Putting this in Eq.\ (\ref{qed:A}), we obtain
\bea
\mathscr A_f=\int\frac{d^4p}{(2\pi)^4}\bar\psi(p)(\rlap/p-m_f)\psi(p)\,.
                                                                     \label{qed:AM}
\eea
Now from Eqs.\ (\ref{qed:psiC}) and (\ref{qed:psiF}),
\bea
\psi_C(x)=\int\frac{d^4p}{(2\pi)^4}e^{ip\cdot x}{\sf C}
              {\gamma_0}^{\rm T}\psi^*(p)\,,                     \label{qed:psiC1}                                                                      
\eea
and so,
\bea
\bar\psi_C(x)={\psi_C}^\dagger(x)\gamma_0
          =-\int\frac{d^4p}{(2\pi)^4}e^{-ip\cdot x}\psi^{\rm T}(p){\sf C}^{-1}\,,
                                                                 \label{qed:psiC2}
\eea
where we used Eqs.\ (\ref{qed:CR}) and (\ref{qed:CU}). Substitution in
Eq.\ (\ref{qed:A}) with $\psi\rightarrow\psi_C$ gives
\bea
\mathscr A_{fC}=-\int\frac{d^4p}{(2\pi)^4}\psi^{\rm T}(p){\sf C}^{-1}(-\rlap/p-m_f)
                {\sf C} {\gamma_0}^{\rm T}\psi^*(p)\,.            \label{qed:AC}
\eea
Note that we have $-\rlap/p-m_f$ in the place of
$\rlap/p-m_f$ of Eq.\ (\ref{qed:AM}), due to the presence of $e^{ip\cdot x}$
in Eq.\ (\ref{qed:psiC1}) in the place of $e^{-ip\cdot x}$ in Eq.\ (\ref{qed:psiF}).
A little rearrangement in Eq.\ (\ref{qed:AC}) leads to
\bea
\mathscr A_{fC}=-\int\frac{d^4p}{(2\pi)^4}\bar\psi(p)[{\sf C}^{-1}(\rlap/p+m_f)
                {\sf C}\,]^{\rm T} \psi(p)\,,
\eea
which, on using Eq.\ (\ref{qed:CR}), is found to equal $\mathscr A_f$, as 
it should.         

Now consider the full Dirac action
\bea
\mathscr A{^\prime_f}=\int\frac{d^4p}{(2\pi)^4}\bar\psi(p)\Big(\rlap/p-m_f
           -\Sigma_f(p)\Big)\psi(p)\,,                    \label{fullDiracaction}
\eea
Following the derivation of Eq.\ (\ref{qed:AC}), we can write down
\bea
\mathscr A{^\prime_{fC}}=-\int\frac{d^4p}{(2\pi)^4}\psi^{\rm T}(p){\sf C}^{-1}
                \Big(-\rlap/p-m_f-\Sigma_f(-p)\Big)
                {\sf C} {\gamma_0}^{\rm T}\psi^*(p)\,. 
\eea
Rearranging as in the case $A_{fC}$ and demanding that
\bea
\mathscr A{^\prime_f}=\mathscr A{^\prime_{fC}}\,,                \label{AfAfC}
\eea
we arrive at the condition for $C$ invariance:
\bea
\Sigma_f(p)=[{\sf C}^{-1}\Sigma_f(-p){\sf C}]^{\rm T}\,.       \label{qed:Ccon1}
\eea
Using the general form of $\Sigma_f(p)$ given in Eq.\ (\ref{qed:form}), and
also using Eq.\ (\ref{qed:CR}), it follows that the Lorentz scalars $a$, $b$ and $c$ 
then satisfy
\bea
a(-p)=a(p)~,~~~b(-p)=-b(p)~,~~~c(-p)=c(p)~.                    \label{qed:Ccon2}
\eea

Substitution of these conditions in Eq.\ (\ref{Vcon}) and comparison of the 
resulting equation with Eq.\ (\ref{Ucon}) show that $E_f$ and $E_{\bar f}$ 
satisfy identical equations under $C$ invariance, i.e., the fermion and
antifermion dispersion relations are then the same.

Turning to the wavefunction renormalization factor, we rewrite Eq.\ (\ref{Zfbar2}) as
\bea
Z_{\bar f}(P)=1+\Bigg[a(-p)
             +\Bigg(p_0-\frac{P^2}{p_0}\Bigg)\frac{\partial a(-p)}{\partial p_0}
                  -\frac{\partial b(-p)}{\partial p_0}+\frac{m_f}{p_0}
              \frac{\partial c(-p)}{\partial p_0}\Bigg] _{p_0=E_{\bar f}}\,.
\eea
Substitution of the conditions of Eq.\ (\ref{qed:Ccon2}) and comparison with
Eq.\ (\ref{Zf2}) then shows that $Z_{\bar f}(P)=Z_f(P)$ (on using
$E_{\bar f}=E_f$). So $C$ invariance also leads to the equality
of the fermion and antifermion wavefunction renormalization factors.

Such symmetries between the properties of the particle and the antiparticle
are, in general, the consequence of $CPT$ invariance. However, the form of the 
self-energy as given by Eq.\ (\ref{qed:form}) already implies invariance
of the action under parity and time reversal. So the additional 
imposition of $C$ invariance
through Eq.\ (\ref{qed:Ccon2}) is, in fact, equivalent to the requirement of $CPT$
invariance.

Finally we consider the antifermion spinor. At the tree-level, the fermion 
spinor is related to the antifermion spinor by
\bea
v_s(\vec P)={\sf C}{\gamma_0}^{\rm T}u{_s^*}(\vec P)\,.                      \label{qed:vdef}
\eea
To verify this, start with the equation $(\rlap/p-m_f)u_s(\vec P)=0$. Then $v_s(\vec P)$,
as defined above, satisfies
\bea
\Big(p^\mu {\sf C}{\gamma_0}^{\rm T}{\gamma_\mu}^*
        \big({{\sf C}{\gamma_0}^{\rm T}}\big)^{-1}-m_f\Big)v_s(\vec P)=0\,.\label{qed:vv}
\eea
This, on using the hermitian conjugation property of the gamma matrices 
and then Eq.\ 
(\ref{qed:CR}), reduces to the familiar equation $(\rlap/p+m_f)v_s(\vec P)=0$.

Let us now try to generalize this exercise in the presence of radiative 
corrections. That is, let us start with Eq.\ (\ref{U_s}), and write 
\bea
V_s(\vec P)={\sf C}{\gamma_0}^{\rm T}U{_s^*}(\vec P)\,.                      \label{qed:Vdef}
\eea
Then one can find an equation for $V_s(\vec P)$ just as we found Eq.\ (\ref{qed:vv}).
Demanding that this equation for $V_s(\vec P)$ be the same as 
Eq.\ (\ref{V_s}) leads, on using $E_f=E_{\bar f}$, to the condition
\bea
\Sigma_f(p)={\sf C}{\gamma_0}^{\rm T}{\Sigma_f}^*(-p)
         {\gamma_0}^{\rm T}{\sf C}^{-1}\,.                    \label{qed:Ccon3}
\eea                                    
Use Eq.\ (\ref{qed:cond}) and then Eq.\ (\ref{qed:CA}) in Eq.\ (\ref{qed:Ccon3}).
This reduces it to Eq.\ (\ref{qed:Ccon1}). We therefore conclude that the tree-level 
definition (\ref{qed:vdef}) of the antifermion spinor can be extended
to (\ref{qed:Vdef}) in the presence of $C$ invariance. In the Dirac-Pauli
representation, one can choose $\sf C=i\gamma^2\gamma^0$ (thereby fixing the
overall phase in $\sf C$ which is left arbitrary by Eqs.\ (\ref{qed:CR}),
(\ref{qed:CU}) and (\ref{qed:CA})), while the explicit solutions for the spinors
have already been given in Eqs.\ (\ref{Upm}) and (\ref{Vpm}).                      
One can then verify Eq.\ (\ref{qed:Vdef}), making use of the conditions of Eq.\
(\ref{qed:Ccon2}) together with $E_f=E_{\bar f}$.
\section{Relations when background is not $C$ invariant}\label{qed:noC}
Let us now consider the case when the background contains charged leptons at
non-zero chemical potential. This breaks the invariance under charge conjugation,
even though the underlying Lagrangian, being that of QED, is $C$ invariant.

Recall from Sec.\ \ref{th:TE} that the effect of the medium is introduced through 
the operator $J$ given by Eq.\ (\ref{TE3}), which occurs in the density operator.
The operator $J$ is made up of $H$ and $Q$, which transform under charge conjugation as
\bea
CHC=H,~~~CQC=-Q\,.                                                   \label{CHC}
\eea
The relations (\ref{CHC}) can be explicitly checked at the tree-level, as follows.
Considering the contribution from the charged leptons only, the expressions for $H$ 
and $Q$ are those given in Eq.\ (\ref{HQ}). Let us now consider
Eq.\ (\ref{qed:psiC}) and substitute
the expansion (\ref{psiFour}) into it. Using Eq.\ (\ref{qed:vdef}), and 
$u_s(\vec P)={\sf C}{\gamma_0}^{\rm T}v{_s^*}(\vec P)$ (following from Eq.\ 
(\ref{qed:vdef}) and the properties of the matrix $\sf C$), we obtain
\bea
Cc_s(\vec P)C=d_s(\vec P),~~Cd_s(\vec P)C=c_s(\vec P)\,,              
\eea
so that 
\bea
Cc_s^\dagger(\vec P)c_s(\vec P)C=d_s^\dagger(\vec P)d_s(\vec P)\,,
~~Cd_s^\dagger(\vec P)d_s(\vec P)C=c_s^\dagger(\vec P)c_s(\vec P)\,.     \label{ccdd}
\eea
(For these steps, it is convenient to use $C^\dagger=C$, which follows from
$C^2=1$ and the fact that $C$ is a unitary operator.) Use of Eqs.\ (\ref{ccdd})
in Eqs.\ (\ref{HQ}) at once leads to Eqs.\ (\ref{CHC}).

From Eqs.\ (\ref{TE3}) and (\ref{CHC}), it is clear that $CJC$ does not equal $J$
when $\mu\ne 0$. Note, however, that $J$ stays the same if charge conjugation is 
followed by a reversal of the sign of $\mu$. Consequently, while the relation in Eq.\ 
(\ref{AfAfC}) breaks down, the relation 
\bea
\mathscr A{^\prime_f}=[\mathscr A{^\prime_{fC}}]\,_{\mu\rightarrow -\mu}   
\eea
holds. Eq.\ (\ref{qed:Ccon2}) is then modified to 
\bea
[a(-p)]\,_{\mu\rightarrow -\mu}=a(p)\,,~
[b(-p)]\,_{\mu\rightarrow -\mu}=-b(p)\,,~
[c(-p)]\,_{\mu\rightarrow -\mu}=c(p)\,.                                 \label{abcmu}
\eea
Retracing the subsequent steps of Sec.\ \ref{qed:C}, one then arrives at
\bea
[E_{\bar f}(P)]_{\mu\rightarrow -\mu}&=&E_f(P)\,,                       \label{Emu}\\
\left[Z_{\bar f}(P)\right]_{\mu\rightarrow -\mu}&=&Z_f(P)\,,            \label{Zmu}\\
\left[V_s(\vec P)\right]\,_{\mu\rightarrow -\mu}&=&{\sf C}{\gamma_0}
^{\rm T}U{_s^*}(\vec P)\,. 
\eea 
\section{Calculations at $O(e^2)$}\label{qed:calcu}
\subsection{Self-energy}\label{qed:Self}
\begin{figure}
\begin{center}
%
%
\begin{picture}(180,130)(-90,-30)
\Text(0,-30)[c]{\large\bf (A)}
\ArrowLine(80,0)(40,0)
\Text(60,-10)[c]{$\ell(p)$}
\ArrowLine(40,0)(-40,0)
\Text(0,-10)[c]{$\ell(p+k)$}
\ArrowLine(-40,0)(-80,0)
\Text(-60,-10)[cr]{$\ell(p)$}
\PhotonArc(0,0)(40,0,180){4}{6.5}
\Text(0,50)[cb]{$\gamma(k)$}
\end{picture}
%
%
\begin{picture}(100,100)(-50,-30)
\Text(0,-30)[c]{\large\bf (B)}
\ArrowLine(40,0)(0,0)
\Text(35,-10)[cr]{$\ell(p)$}
\ArrowLine(0,0)(-40,0)
\Text(-35,-10)[cl]{$\ell(p)$}
\Photon(0,0)(0,35){2}{6}
\Text(-4,20)[r]{$\gamma$}
\ArrowArc(0,55)(20,-90,270)
\Text(0,85)[b]{$f(k)$}
\end{picture}
\caption{\sf One-loop diagrams for the self-energy of a charged
lepton $\ell$ in a medium.
\label{f:selfenergy}}
\end{center}
\end{figure}
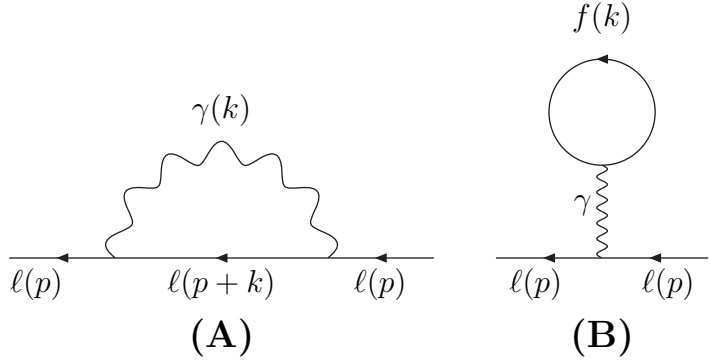
The self-energy diagrams are shown in Fig. \ref{f:selfenergy}. In view of
Eq.\ (\ref{qed:Real}), we need to concern ourselves only with the 1-type
vertex. (Actually, the upper vertex of Fig. \ref{f:selfenergy}B can still
be of 2-type, but we shall see that there is no contribution in this case.)
The propagators for the internal lines are therefore the ``11" propagators
of Chap. \ref{th}. For a fermion, the propagator is thus given by
Eqs.\ (\ref{S}), (\ref{S0}), (\ref{S'}) and (\ref{etaf}), while for
the photon, the propagator in the Feynman gauge is given by
Eqs.\ (\ref{D}), (\ref{D0}), (\ref{D'}) and (\ref{etab}).

The contribution for Fig. \ref{f:selfenergy}B vanishes because the
photon tadpole is zero. There are several reasons for the vanishing
of the photon tadpole in vacuum, but, as we shall see, only one of these 
arguments continue to hold in normal matter. First consider Lorentz invariance.
The photon tadpole must equal a four-vector, but there being no
four-vector available in vacuum (the photon line of the tadpole carries zero 
momentum), the tadpole has to vanish.  However in a medium, we do have a four-vector,
namely, the four-velocity $v^\mu$ of the medium, and the tadpole can be proportional
to it. Next consider invariance under the action of
the charge conjugation operator $C$. Using ${C}|0\rangle=0$ and
${C}A_\mu{C}=-A_\mu$, it is clear that $\langle 0|A_\mu|0\rangle$
vanishes. But in a medium, the expectation value is with respect to the thermal
state which need not be $C$-invariant (normal matter, e.g., contains electrons
but no positrons) and in that case this argument also is invalidated. Finally one 
must realize that the photon tadpole, since it
has no other external lines, represents the coupling of the photon to the net current
in vacuum or in the medium, as the case may be. This net current is zero not only
in vacuum, but also in an electrically neutral medium, thereby ensuring the vanishing
of the photon tadpole.

As an illustration of the formal reasoning presented above,
let us now explicitly evaluate the photon tadpole contained in Fig. \ref{f:selfenergy}B.
The tadpole is given by
\bea
T^\mu=\sum\limits_f \int {d^4k \over (2\pi)^4} \Tr \Big[ ie{\cal Q}_f 
      \gamma^\mu iS_f(k) \Big] \,.                                    \label{qed:T1}                           
\eea
Here the sum is over all species of fermions in the
medium, the charge of each species being denoted by ${\cal Q}_f$ with
the convention that ${\cal Q}_e = -1$. The vacuum part of 
Eq.\ (\ref{qed:T1}), coming from
$iS_{Ff}(p)$ of Eq.\ (\ref{S}), vanishes separately for each fermion in the loop. 
On the other hand, the thermal part, coming from $S_{Tf}(p)$ of Eq.\ (\ref{S}),
equals
\bea
T^\mu=-i4e\sum\limits_f{\cal Q}_f\int\frac{d^4k}{(2\pi)^3}\delta(k^2-m{_f^2})\eta_f(k)k^\mu\,.
\eea
Let us write $T^\mu=v^\mu T$ where $T$ is a scalar, so that $T=v_\mu T^\mu$. 
It is easy to
evaluate $T$ by going over to the rest frame given by Eq.\ (\ref{v}), and 
performing the $k_0$ integration. We thus obtain
\begin{equation}
T^\mu=-iev^\mu\sum_f {\cal Q}_f\left [2\int\frac{d^3K}{(2\pi)^3}
       [f_f(E_K)-f_{\bar f}(E_K)] \right]               \label{qed:T} 
\end{equation}
The distribution functions are given by Eq.\ (\ref{distfunctions}).
Here, as elsewhere, we write the components of a four-vector as
$k^\mu = (k^0,\vec K)$, and use the notation
$E_K \equiv \sqrt{K^2 + m_f^2}$. 
Since the number densities are given by
\begin{eqnarray}
n_{f,\bar f} = 2 \int {d^3K \over (2\pi)^3} \; f_{f,\bar f}(E)\,,
\label{qedreffor2type}
\end{eqnarray}
Eq.\ (\ref{qed:T}) shows that 
$T^\mu$ is proportional to the total charge density of the medium, and
indeed vanishes for an electrically neutral medium such as normal matter.

At this point, we consider the case when the upper vertex in Fig. 
\ref{f:selfenergy}B is a 2-type vertex. Because the thermal part of the ``22"
propagator is the same as that of the ``11" propagator, and the 2-type vertex
differs from the 1-type vertex by a sign, the photon tadpole in this case is
just the negative of the value given in Eq.\ (\ref{qed:T}), and therefore again
vanishes in a neutral medium.          

However, in anticipation of the calculation of the gravitational interaction
of the fermion in Chap.\ \ref{gr}, we mention here that in the presence of the 
gravitational potential, the diagram of Fig.\ \ref{f:selfenergy}B is not zero
by itself. This is because the condition of the vanishing of the photon tadpole
(which is equivalent to require that the medium be electrically neutral)
involves other diagrams. This will be discussed in detail in Sec.\ 
\ref{s:Ztype}.

Let us now return to our present calculation of the self-energy (in the absence 
of gravitational potential), for which we need to  consider only 
Fig. \ref{f:selfenergy}A. Thus the charged lepton self-energy is given by
\begin{eqnarray}
-i\Sigma_\ell(p) = (ie)^2 \int {d^4k\over (2\pi)^4} \; \gamma^\mu
iS_{\ell}(p+k)\gamma^\nu iD_{\mu\nu}(k)  \,.
\label{Sigmap}
\end{eqnarray}
When Eqs.\ (\ref{S}) and (\ref{D}) are substituted into Eq.\
(\ref{Sigmap}), four terms are produced.  Since we are interested in
the background induced contributions only, we disregard the term
involving both $S_{F\ell}$ and $\Delta_F$. Among the other three, the
one involving both $S_{T\ell}$ and $\Delta_T$ contributes only to the
absorptive part of the self-energy 
(see Eq.\ (\ref{qed:Im})), and does not concern us here. 
The contributions to the dispersive part of the self-energy
(see Eq.\ (\ref{qed:Real})) arises from the remaining two terms, which
can be written in the form
\begin{eqnarray}
\label{Sigma1,2}
\Sigma'_\ell(p) = \Sigma'_{\ell1}(p) + \Sigma'_{\ell2}(p) \,,
\end{eqnarray}
where
\begin{eqnarray}
\label{Sigma1}
\Sigma'_{\ell1}(p) & = &
2e^2 \int {d^4k \over (2\pi)^3}\;\delta(k^2) \eta_\gamma(k)
{\rlap/p + \rlap/ k - 2m_\ell \over p^2 + 2k\cdot p - m^2_\ell}\,,\\
\label{Sigma2}
\Sigma'_{\ell2}(p) & = &
-\, 2e^2 \int {d^4k \over (2\pi)^3} \;
\delta(k^2 - m_\ell^2) \eta_\ell(k) \;
{\rlap/ k - 2m_\ell \over p^2 - 2k\cdot p + m_\ell^2} \,.
\end{eqnarray}

We shall now write
\bea
\Sigma'_{\ell1}(p) =a_1(p)\rlap/p+b_1(p)\rlap/v+c_1(p)       \label{CCC}
\eea
and check the conditions of $C$ invariance stated in Eq.\ (\ref{qed:Ccon2}).
Eq.\ (\ref{CCC}) can be inverted to express $a_1(p)$, $b_1(p)$ and $c_1(p)$
in terms of $\Tr[\rlap/p\Sigma'_{\ell 1}(p)]$,
$\Tr[\rlap/v\Sigma'_{\ell 1}(p)]$ and $\Tr[\Sigma'_{\ell 1}(p)]$.
Then using the expression of Eq.\ (\ref{Sigma1}), we obtain
\bea
a_1(p)&=&\frac{2e^2}{(p\cdot v)^2-p^2}\int {d^4k \over (2\pi)^3}\;
       \delta(k^2) \eta_\gamma(k)
       \frac{(p\cdot v)(k\cdot v+p\cdot v)-(k\cdot p+p^2)}
       {p^2 + 2k\cdot p - m^2_\ell}\,,\nonumber\\
b_1(p)&=&\frac{2e^2}{(p\cdot v)^2-p^2}\int {d^4k \over (2\pi)^3}\;
       \delta(k^2) \eta_\gamma(k)
       \frac{(p\cdot v)(k\cdot p)-p^2k\cdot v}
       {p^2 + 2k\cdot p - m^2_\ell}\,,\nonumber\\
c_1(p)&=&-4e^2m_\ell\int {d^4k \over (2\pi)^3}\;
       \delta(k^2) \eta_\gamma(k)
       \frac{1}{p^2 + 2k\cdot p - m^2_\ell}\,.
\eea
Changing $p$ to $-p$, and, simultaneously, $k$ to $-k$, in each of these 
expressions, one can readily check the properties given by 
Eq.\ (\ref{qed:Ccon2}). 

Let us next write
\bea
\Sigma'_{\ell2}(p) =a_2(p)\rlap/p+b_2(p)\rlap/v+c_2(p)\,.   
\eea
We then have
\bea
a_2(p)&=&\frac{2e^2}{(p\cdot v)^2-p^2}\int {d^4k \over (2\pi)^3}\;
       \delta(k^2-m^2_\ell) \eta_\ell(k)
       \frac{k\cdot p-(p\cdot v)(k\cdot v)}
       {p^2 - 2k\cdot p + m^2_\ell}\,,\nonumber\\
b_2(p)&=&\frac{2e^2}{(p\cdot v)^2-p^2}\int {d^4k \over (2\pi)^3}\;
       \delta(k^2-m^2_\ell) \eta_\ell(k)
       \frac{p^2k\cdot v-(p\cdot v)(k\cdot p)}
       {p^2 - 2k\cdot p + m^2_\ell}\,,\nonumber\\
c_2(p)&=&4e^2m_\ell\int {d^4k \over (2\pi)^3}\;
       \delta(k^2-m^2_\ell) \eta_\ell(k)
       \frac{1}{p^2 - 2k\cdot p + m^2_\ell}\,.
\eea
To check the conditions of Eq.\ (\ref{abcmu}), we first have to change 
$\mu_\ell$ to $-\mu_\ell$. Eq.\ (\ref{etaf}) shows that this is equivalent
to changing $\eta_\ell(k)$ to $\eta_\ell(-k)$. After making this change, one
only has to change $p$ to $-p$, and, simultaneously, $k$ to $-k$, and the
conditions check out.
\subsection{Calculation of inertial mass }\label{calin}
Using Eq.\ (\ref{calEoper}), and according to the decomposition given
in Eq.\ (\ref{Sigma1,2}), we write
\begin{eqnarray}
\label{Edecomp}
{\cal E}_\ell = {\cal E}_{\ell1} + {\cal E}_{\ell2}\,,
\end{eqnarray}
where
\begin{eqnarray}
\label{E1}
{\cal E}_{\ell1} & = & 2e^2 \int\frac{d^4k}{(2\pi)^3}
\delta(k^2) \eta_\gamma(k)\frac{p_0 + k _0- 2m_\ell}
{p^2 + 2k\cdot p - m_\ell^2} \,,\\
\label{E2}
{\cal E}_{\ell2} & = & -2e^2 \int\frac{d^4k}{(2\pi)^3}
\delta(k^2 - m_\ell^2) \eta_\ell(k)\frac{k_0 - 2m_\ell}
{p^2 - 2k\cdot p + m_\ell^2} \,.
\end{eqnarray}
We can make a similar decomposition of ${\cal E}_{\bar\ell}$. The
quantities ${\cal E}_{\bar\ell1}$ and ${\cal E}_{\bar\ell2}$ are
obtained from ${\cal E}_{\ell1}$ and ${\cal E}_{\ell2}$ by replacing
$m_\ell$ by $-m_\ell$.

The inertial mass is determined by applying Eq.\ (\ref{Moper}), and,
according to the decomposition given in Eq.\ (\ref{Edecomp}), we
write it as
\begin{eqnarray}
\label{Melldef}
M_\ell = m_\ell + m_{\ell 1} + m_{\ell 2} \,,
\end{eqnarray}
and similarly for the anti-leptons.  Substituting $p^\mu =
(m_\ell,\vec 0)$ in Eq.\ (\ref{E1}), and using the fact that the terms
in the integrand that are odd in $k$ yield zero, we obtain
\begin{eqnarray}
m_{\ell 1} \equiv {\cal E}_{\ell 1}(m_\ell,\vec 0) & =  &
\frac{e^2}{m_\ell}\int\frac{d^4k}{(2\pi)^3}\delta(k^2)\eta_\gamma(k)
\nonumber\\
\label{m1}
& = & \frac{e^2T^2}{12m_\ell} \,.
\end{eqnarray}
This is the contribution to the inertial mass from the photons in the
background, in agreement with the result quoted in Eq.\ (\ref{DHR:M}),
and it is non-zero for any the charged lepton propagating through the
medium.  In a similar fashion we find
\begin{eqnarray}
m_{\bar\ell 1} \equiv -{\cal E}_{\bar\ell 1}(-m_\ell,\vec 0) =
\frac{e^2T^2}{12m_\ell} \,,                                   \label{mbar1}
\end{eqnarray}
and therefore the photon contribution for the anti-particle is the same
as for the corresponding particle.

In order to justify the usage of the term ``inertial mass,'' we now deduce
the $O(e^2)$ dispersion relation when only photons are present in the 
background. For this purpose, Eq.\ (\ref{Sigma1}) has to be used in Eq.\
(\ref{disdis}). Since the tree-level dispersion relation 
$p^2=\omega^2-{P}^2=m{^2_\ell}$ can be used in $O(e^2)$ terms,
we obtain 
\bea
\Tr[(\rlap/p+m_\ell)\Sigma'_{\ell 1}]=\frac{e^2T^2}{3}\,,
                                                           \label{qeddd}
\eea
after a term odd in the integration variable drops out, leading to
the same integral as in Eq.\ (\ref{m1}). Consequently, Eq.\ (\ref{disdis}) 
reads
\bea
\omega^2={P}^2+m{_\ell^2}+\frac{e^2T^2}{6}\,,
\eea
and the dispersion relations are 
\bea
E_\ell({P})=E_{\bar\ell}({P})=\sqrt{{P}^2+m{^2_{\ell T}}}          \label{ddd}
\eea
with
\bea
m_{\ell T}=m_\ell+\frac{e^2T^2}{12m_\ell}\,.                \label{mlT}
\eea 
Thus the dispersion relations in photon background are of the same form 
as in vacuum, and reduce to $E_\ell=E_{\bar\ell}=m_{\ell T}+
{P}^2/2m_{\ell T}$ in the non-relativistic limit. From this, it is clear that
$m_{\ell T}$ is the inertial mass. Now, from Eq.\ (\ref{ddd}), $m_{\ell T}$
also equals $E_\ell(0)$ or $E_{\bar\ell}(0)$, thereby justifying our 
definition of inertial mass given in Eq.\ (\ref{qed:inn}). Note that Eq.\  
(\ref{mlT}) is in agreement with  the results given in Eqs.\ (\ref{m1})
and (\ref{mbar1}).

When the medium also contains fermions,
one finds that the dispersion relations no longer possess the
form given in Eq.\ (\ref{ddd}). This is because, unlike in Eq.\ (\ref{qeddd}),
$\Tr[(\rlap/p+m_\ell)\Sigma'_{\ell 2}]$ is a function of momentum.
We have chosen to call $E_\ell(0)$ 
and $E_{\bar\ell}(0)$ the inertial masses even in this case, by a
generalization of the usual application of the term. Sometimes these
quantities are called the phase-space masses \cite{DHR84}.

In passing, we remark that the first equality in Eq.\ (\ref{ddd}) is as
expected from $C$ invariance, which is present in the case of a photon 
background.

We now turn to the fermion background term given in Eq.\ (\ref{E2}),
and find its effect on the inertial mass.
In a background that contains electrons but not
the other charged leptons, the distribution functions for the muon and
the tau vanish. As a result,
\begin{eqnarray}
m_{\mu 2} = m_{\tau 2} = m_{\bar\mu 2} = m_{\bar\tau 2} = 0 \,.
\end{eqnarray}
For the electron, we obtain
\begin{eqnarray}
m_{e2}\equiv {\cal E}_{e2}(m_e,\vec 0) =
\frac{e^2}{m_e}\int\frac{d^4k}{(2\pi)^3}\delta(k^2 - m_e^2)
\eta_e(k)\left[\frac{k_0 - 2m_e}{k_0 - m_e}\right]
\end{eqnarray}
Performing the integrations over $k_0$ and the angular variables,
we obtain
\begin{eqnarray}
\label{me2}
\label{E+e2}
m_{e2} =
\frac{e^2}{2\pi^2 m_e} \int_0^\infty dK \frac{K^2}{2E_K}
\left[\left({E_K - 2m_e \over
E_K - m_e} \right) f_e(E_K) + \left( {E_K + 2m_e \over
E_K + m_e} \right) f_{\bar e} (E_K) \right] \,.
\end{eqnarray}
Here $E_K$ is given by Eq.\ (\ref{EK}) (with $m$ replaced by $m_e$) and
the distribution functions by Eq.\ (\ref{distfunctions}).
Similarly,
\begin{eqnarray}
m_{\bar e 2} & \equiv & -{\cal E}_{\bar e2}(-m_e,\vec 0) \nonumber\\
& = &
\frac{e^2}{2\pi^2 m_e} \int_0^\infty dK \frac{K^2}{2E_K}
\left[\left({E_K + 2m_e \over
E_K + m_e} \right) f_e(E_K) + \left( {E_K - 2m_e \over
E_K - m_e} \right) f_{\bar e} (E_K) \right]\,.\nonumber\\*
\label{mbare2}
\end{eqnarray}

The integration over $K$ can be performed only when the
distribution functions are specified, and we will consider some
examples in Sec.\ \ref{s:cases}. Here we only note 
the general result of Eq.\ (\ref{Emu}), obtained by charge conjugation,
is borne out at $P=0$ by Eqs.\ (\ref{me2}) and (\ref{mbare2}): the inertial
mass corrections for the particle and the antiparticle  are the same
only after we implement $\mu_e\rightarrow -\mu_e$, or equivalently,
\bea
f_e\leftrightarrow f_{\bar e}\,,                    \label{antiprescription2}
\eea
in one of them.
\subsection{Calculation of wavefunction renormalization factor}\label{calZ}
We decompose
\begin{eqnarray}
\zeta_\ell = \zeta_{\ell 1} + \zeta_{\ell 2}
\end{eqnarray}
with a similar decomposition for the anti-leptons, where, from Eq.\ (\ref{zf}),
\begin{eqnarray}
\zeta_{\ell i} =
\frac{\partial{\cal E}_{\ell i}}{\partial p^0} \Bigg|_{
p^\mu = (m_\ell,\vec 0)} \qquad \mbox{for $i = 1,2$},
\end{eqnarray}
and, from Eq.\ (\ref{zfbar}),
\begin{eqnarray}
\zeta_{\bar\ell i} =
\frac{\partial{\cal E}_{\bar\ell i}}{\partial p^0}\Bigg|_{
p^\mu = (-m_\ell,\vec 0)} \qquad \mbox{for $i = 1,2$}.
\end{eqnarray}
From Eq.\ (\ref{E1}), we obtain
\begin{eqnarray}
\frac{\partial{\cal E}_{\ell 1}}{\partial p^0} \Bigg|_{
p^\mu = (m_\ell,\vec 0)}
=
\frac{\partial{\cal E}_{\bar\ell 1}}{\partial p^0}\Bigg|_{
p^\mu = (-m_\ell,\vec 0)}
= -\frac{e^2}{m_\ell^2} \int\frac{d^3K}{(2\pi)^3}
\frac{f_\gamma(K)}{K}
\left( 1 - \frac{m^2_\ell}{K^2}\right) \,,
\end{eqnarray}
which implies
\begin{eqnarray}
\label{Zpsi1}
\zeta_{\ell 1}  = \zeta_{\bar\ell 1} =
-\, {e^2T^2\over 12m_\ell^2} + {e^2 \over 2\pi^2} \int_0^\infty
{dK\over K} \; f_\gamma(K) \,,
\end{eqnarray}
the photon distribution function being given by Eq.\ (\ref{fgamma}).
Notice that the integral in Eq.\ (\ref{Zpsi1}) is infrared divergent.
This divergence will turn out to be of importance in Chap. \ref{gr}, since
there it will cancel a similarly divergent term in the gravitational vertex
contribution to the gravitational mass (see Eq.\ (\ref{A1final})).

Since the electron background terms do not contribute to the
self-energy of the muon or the tau, it follows that
\begin{eqnarray}
\zeta_{\mu 2} = \zeta_{\tau2}
= \zeta_{\bar\mu 2} = \zeta_{\bar\tau 2} = 0 \,.
\end{eqnarray}
For the electron, Eq.\ (\ref{E2}) implies
\begin{eqnarray}
\frac{\partial{\cal E}_{e2}}{\partial p^0}\Bigg|_{
p^\mu = (m_e,\vec 0)} & = & - {{\cal E}_{e2} (m_e,\vec 0) \over m_e}
\nonumber\\
\frac{\partial{\cal E}_{\bar e2}}{\partial p^0}\Bigg|_{
p^\mu = (-m_e,\vec 0)} & = & {{\cal E}_{\bar e 2}(-m_e,\vec 0) \over
m_e} \,,
\end{eqnarray}
which yield
\begin{eqnarray}
\label{ZU2}
\label{Le2}
\zeta_{e2} & = & - \frac{m_{e2}}{m_e}\nonumber\\
\zeta_{\bar e 2} & = & - \frac{m_{\bar e 2}}{m_e} \,,
\end{eqnarray}
with $m_{e2}$ and $m_{\bar e 2}$ given in Eqs.\ (\ref{me2}) and
(\ref{mbare2}) respectively.
The result (\ref{Zpsi1}) shows that $Z_e=Z_{\bar e}$ in a photon background,
while the result (\ref{ZU2}) shows that $Z_e$ transforms to $Z_{\bar e}$ 
under $f_e\leftrightarrow f_{\bar e}$ (so that $m_{e2}$ becomes $m_{\bar e 2}$) 
as expected from considerations of charge conjugation.

 \chapter{Gravitational couplings of charged leptons in a medium}\label{gr}
 \section{Introduction}\label{gr:intro}
 The underlying Lagrangian for the calculations of this chapter is that of
 the weak gravitational field version of QED in curved space, which
 was discussed in Chapter \ref{lg}. Thus we shall use the graviton
 couplings of Sec.\ \ref{lg:vertices}, in addition to, of course,
 the flat space QED coupling. Since the coupling constant $\kappa$
 is of mass dimension $-1$, the theory we are considering is 
 non-renormalizable. But this will not be a problem, since our calculations 
 will be confined to the lowest order in $\kappa$, and the radiative
 corrections to be considered will involve an expansion in the electromagnetic
 coupling constant $e$. All our calculations will be valid at
 energies low compared to $\kappa^{-1}\approx m_{\rm Planck}$, as stated in
 Sec.\ \ref{lg:intro}. (This is comparable, to some extent, to tree-level calculations
 using the four-Fermi interaction, which are valid at energies much less
 than $m_W$.)
 \section{Gravitational mass}\label{gr:mass}
 The gravitational mass is a measure of the strength of the coupling of
 the fermion to the graviton.  It can be determined in terms of the
 fermion's vertex function for the gravitational interaction. For this
 purpose, we are going to consider the $S$-matrix element of quantum field 
 theory for the case of scattering of a slowly moving fermion, by a weak
 and static external gravitational field
 (these conditions define the Newtonian limit of gravitational interaction).
 Since our aim is to calculate gravitational mass in a medium, we must 
 consider this scattering to take place in the presence of a thermal background.

 So, let us first consider the relevant part
 \bea
 \mathscr L_{\rm eff}=-\kappa h^{\lambda\rho}(x)
 \widehat T_{\lambda\rho}(x)                                        \label{gr:S0}
 \eea
 of the interaction Lagrangian.
 Here $\widehat T_{\lambda\rho}(x)$ is the total energy-momentum tensor 
 operator, involving quantized fermion and photon fields, such that
 $\mathscr L_{\rm eff}$ represents the possible interactions of these fields
 with the classical field $h_{\lambda\rho}(x)$ to all orders in $e$.
 The $S$-matrix, to lowest order in $\kappa$, is then
 \bea
 S=1-i\kappa\int d^4x\, h^{\lambda\rho}(x)\widehat T_{\lambda\rho}(x).
					  \label{gr:S1}
 \eea
 Eq.\ (\ref{gr:S1}) follows from the well-known $S$-matrix expansion
 \bea
 S=T[\exp(i\int d^4x\,\mathscr L_{int})]   \label{gr:S2}
 \eea
 where $T$ denotes time-ordering.
 The standard derivation of this formula assumes that the interaction
 $\mathscr L_{int}$ does not contain derivatives of fields, so that
 the commutation relations of the interacting theory are the same as those
 in the free-field theory, and also $\mathscr H_{int}=-\mathscr L_{int}$.
 In the present case, $\mathscr L_{\rm eff}$ does contain derivatives
 through $\widehat T_{\lambda\rho}$ (see, for example, Eqs.\ 
 (\ref{lg:emphoton}) and (\ref{lg:emfermi})). However the formula 
 (\ref{gr:S2}) can be shown to be generally valid, irrespective of
 $\mathscr L_{int}$ containing derivatives or not, provided one uses
 the commutation relations of the free theory in calculating the $S$-matrix
 elements \cite{gr:mandl}. This justifies our use of Eq.\ (\ref{gr:S2}),
 and also the derivation of the interaction vertices in Secs.\ 
 \ref{lg:ppgv} and \ref{lg:ffgv}.

 Returning to Eq.\ (\ref{gr:S1}), we now consider the $S$-matrix element
 between incoming and outgoing fermion states:
 \bea 
 S_{ff}&\equiv &\langle f(\vec P\,',s) |S | f(\vec P,s) \rangle
			\label{gr:S3'}\\
     &=& -i\kappa                    
 \langle f(\vec P\,',s) |\widehat T_{\lambda\rho} (0) | f(\vec P,s) \rangle
	      \int d^4x\, e^{-iq\cdot x}h^{\lambda\rho}(x)\,,
					       \label{gr:S3}
 \eea
 where we have used the relation
 \bea
 \widehat T_{\lambda\rho} (x) 
 =e^{i\widehat P.x} \widehat T_{\lambda\rho} (0) 
  e^{-i\widehat P.x},                         \label{gr:S3''}
 \eea
 $\widehat P_\mu$ being the momentum operator, and also defined
 \bea
 q=p-p'\,.                                              \label{qdef}
 \eea
 Here $q_0=E_f-E'_f$, each of the energies satisfying the dispersion
 relation of charged leptons in thermal QED. 

 At zeroth order in $e$, the matrix element of Eq.\ (\ref{gr:S3})
 equals ${\bar u}_s(p')V_{\lambda\rho}(p,p')u_s(p)$
 with $V_{\lambda\rho}(p,p')$ given by Eq.\ (\ref{lg:Vlamrho}). When higher
 orders in $e$ are taken into account, we write 
 \bea
 \langle f(\vec P\,',s) |\widehat T_{\lambda\rho} (0) | f(\vec P,s) \rangle
 = \sqrt{Z_f(P) Z_f(P')} \;
 \overline U_{s}(\vec P\,')\Gamma_{\lambda\rho}(p,p') U_s(\vec P)            \label{gr:S4}
 \end{eqnarray}
 following the LSZ reduction formula of vacuum field theory. 
 Here $\Gamma_{\lambda\rho}
 (p,p')$ is the 1PI vertex function: it is the sum of all connected,
 amputated diagrams contributing to the fermion-fermion-graviton vertex.
 Though the vertex function can be defined in general for off-shell momenta
 (see Eq.\ (\ref{vertexdef})), the arguments of $\Gamma_{\lambda\rho}$ in Eq.\ (\ref{gr:S4})
 are on-shell momenta,
 and satisfy the dispersion relation of the charged lepton.
 It may be noted that the replacement of the tree-level spinor $u_s(p)$
 with $\sqrt{Z_f(P)}U_s(\vec P)$ in going beyond the lowest order, 
 as done in Eq.\ (\ref{gr:S4}),
 also gives the residue of the one-particle pole of the dressed propagator 
 in Eq.\ (\ref{qed:Sfull})
 \cite{gr:LSZ}.

 Using Eq.\ (\ref{gr:S4}) in Eq.\ (\ref{gr:S3}), and also taking $h^{\lambda\rho}$ to
 be independent of time, we obtain
 \begin{eqnarray}
 S_{ff} = -i \kappa (2\pi)\delta(E_f - E'_f) Z_f(P)
 \overline U_s(\vec P\,') \Gamma_{\lambda\rho}(p,p') U_s(\vec P)
 h^{\lambda\rho}(\vec Q).                             \label{gr:S5}
 \end{eqnarray}
 Here we have denoted the three-vector part of $q^\mu$ by upper case letter,
 and also defined the Fourier transform by
 \begin{eqnarray}
 f(\vec x) = \int\frac{d^3Q} {(2\pi)^3}
 f (\vec Q) e^{-i\vec Q\cdot\vec x}                                \label{gr:FT}
 \end{eqnarray}
 for any function $f(\vec x)$. Note that $E_f=E'_f$ implies that $P=P'$, so that the
 wavefunction renormalization factors for the initial and the final states are the same.

 Now we relate $h^{\lambda\rho}$ to an external gravitational potential
 $\phi^{\rm ext}$. First consider Poisson's equation
 \bea
 \nabla^2\phi^{\rm ext}(\vec x)=4\pi G\rho^{\rm ext}(\vec x),
 \eea
 the static mass density $\rho^{\rm ext}(\vec x)$ being the source of the
 gravitational potential. In momentum space, this becomes
 \bea
 -2Q^2\phi^{\rm ext}(\vec Q)=\kappa^2\rho^{\rm ext}(\vec Q).      \label{gr:S6}
 \eea
 Next, the energy-momentum tensor corresponding to $\rho^{\rm ext}$ is
 \bea
 T_{\lambda\rho}=v_\lambda v_\rho \rho^{\rm ext}                  \label{gr:S7}
 \eea
 where $v_\lambda$ is the four-velocity of the mass density. This 
 $T_{\lambda\rho}$ determines the desired $h_{\lambda\rho}$ through
 the linearized Einstein field equations, which, in the harmonic gauge, read
 \bea
 \Box h_{\lambda\rho}=-\kappa(T_{\lambda\rho}-\frac{1}{2}\eta_{\lambda\rho}
		       {T^\mu}_\mu).                     \label{gr:S8}
 \eea
 (To obtain Eq.\ (\ref{gr:S8}), one starts with the field equations 
 in the form
 \bea
 R_{\lambda\rho}=-8\pi G(T_{\lambda\rho}-\frac{1}{2}g_{\lambda\rho}
		       {T^\mu}_\mu),                   
 \eea
 and evaluates each side to the lowest order in $\kappa$. Thus, one 
 puts $R_{\lambda\rho}$ correct to $O(\kappa)$ (from Eqs.\ (\ref{lg:Ricci})
 and (\ref{lg:Chris})) on the L.H.S. Since the R.H.S. is already of $O(\kappa^2)$,
 one can put $g_{\lambda\rho}=\eta_{\lambda\rho}$ on this side. Then the
 gauge condition (\ref{lg:gauge1}) is used.) After noting that $\Box$ becomes
 $-\nabla^2$ for the static case, we write Eq.\ (\ref{gr:S8}) in momentum space,
 and use first Eq.\ (\ref{gr:S7}) and then Eq.\ (\ref{gr:S6}) in it. 
 In this way, we finally arrive at
 \begin{eqnarray}
 \kappa h^{\lambda\rho}(\vec Q) = 
 (2v^\lambda v^\rho - \eta^{\lambda\rho}) \phi^{\rm ext} (\vec Q)    \label{gr:S9}
 \end{eqnarray}
 It may be noted that in the frame in which the mass distribution is at rest,
 i.e. for 
 $v^\lambda=(1,\vec 0)$, Eq.\ (\ref{gr:S9}) immediately leads to
 $g_{00}=1+2\phi^{\rm ext}$, which can also be deduced by demanding that
 the geodesic equation reduces to Newton's equation of motion in the appropriate limit
 \cite{gr:g00}. Another important point is that we shall take $v^\lambda$
 equal to the four-velocity of the medium, which occurs in the thermal propagators.
 This means that we shall take the medium to be at rest relative to the mass
 distribution producing the external gravitational field (a special case being
 that the medium itself is the source of the gravitational field). 
 Eqs.\ (\ref{gr:S5}) and (\ref{gr:S9}) constitute our end-result for the
 $S$-matrix element.

 We now define the gravitational mass of the fermion. The mass density operator
 for the fermion, $\widehat\rho_f(t,\vec x)$, is determined by writing another effective
 Lagrangian
 \begin{eqnarray}
 {\mathscr L'}_{\rm eff} = -\widehat\rho_f(t,\vec x)\phi^{\rm ext}(\vec x)
								       \label{gr:S10}
 \end{eqnarray}
 such that it reproduces $S_{ff}$ as given by Eqs.\ (\ref{gr:S5}) and (\ref{gr:S9}).
 The gravitational mass $M'_f$ is then defined  through the matrix element 
 of the mass operator (obtained by integrating $\widehat\rho_f$ at any particular
 instant $t_\rho$ over all space) in the zero-momentum limit as follows:
 \bea
 \lim_{\vec P\rightarrow 0}
 \langle f(\vec P\,',s)|\int d^3x\,\widehat\rho_f(t_\rho,\vec x) |f(\vec P,s)\rangle=
	   (2\pi)^3 [\delta^{(3)}(\vec P - \vec P\,')] _
	   {\vec P\rightarrow 0} M'_f \,.                                \label{gr:S11}
 \eea
 A number of remarks on this definition are in order. Firstly, 
 the normalization of one-particle states adopted by us, as given by Eq.\ 
 (\ref{statenorm}),
 is consistent with the definition given in Eq.\ (\ref{gr:S11}). 
 Secondly, the mass operator is
 independent of $\vec x$ (as it involves integration over all space) and so carries
 zero momentum. Consequently, on the L.H.S. of Eq.\ (\ref{gr:S11}) we have $P=\vec P\,'$.
 The delta function on the R.H.S. embodies this fact (see also the derivation
 of Eq.\ (\ref{gr:S13}) below.)
 Thirdly, we shall see that $M'_f$ is independent of the choice of $t_\rho$.

 The remaining task is to find an expression for the gravitational mass. Eqs.\ (\ref{gr:S2}),
 (\ref{gr:S3'}) and (\ref{gr:S10}) give, to lowest order in interaction,
 \begin{eqnarray}
 S_{ff} =
 -i2\pi\delta(E'_f - E_f) \phi^{\rm ext} (\vec P - \vec P\,')
 \langle f(\vec P\,',s) |\widehat\rho_f(t_\rho,\vec 0) |f(\vec P,s)\rangle\,,      \label{gr:S12}
 \end{eqnarray}
 where we used
 \bea
 \widehat \rho_f(t,\vec x) 
 =e^{i\widehat H(t-t_\rho)-i\widehat{\vec P}\cdot\vec x} 
 \widehat \rho_f(t_\rho,\vec 0) 
 e^{-i\widehat H(t-t_\rho)+i\widehat{\vec P}\cdot\vec x}\,,                \label{gr:S12'}
 \eea
 and the definition of Fourier transform given in Eq.\ (\ref{gr:FT}).
 Again Eq.\ (\ref{gr:S11}) gives (on using Eq.\ (\ref{gr:S12'}) with $t=t_\rho$)
 \begin{eqnarray}
 M'_f = \lim_{\vec P\rightarrow 0}\Big [
 \langle f(\vec P\,',s) |\widehat\rho_f(t_\rho,\vec 0) |f(\vec P,s)\rangle
 \Big]_{\vec P\,'=\vec P}\, .                        \label{gr:S13}
 \end{eqnarray}
 Now $\vec P\,'=\vec P$ implies that $E'_f=E_f$. So the matrix element 
 occurring on the R.H.S. of Eq.\ 
 (\ref{gr:S13}) can be determined by using Eq.\ (\ref{gr:S12}) on one hand, and
 Eqs.\ (\ref{gr:S5}) and (\ref{gr:S9}) on the other hand. We are thus led to
 \bea
 M'_f &=& (2v^\lambda v^\rho - \eta^{\lambda\rho})
 \lim_{\vec P\rightarrow 0}
 Z_f(P)\Bigg\{ \overline U_s(\vec P\,') \Gamma_{\lambda\rho}(p,p') U_s(\vec P)
 \Bigg\}
 _{{E'_f=E_f}\atop{\vec P\,'\rightarrow P}}\,.                       \label{gr:S14}
 \eea
 Note that in the last step we have relaxed the equality $\vec P\,'=\vec P$ to a 
 limiting procedure, while retaining the weaker condition $E'_f=E_f$. 
 The significance of
 this exercise will be explained in detail at the end of Sec.\ \ref{gr:op}.

 We end this section with the comment that our aim was
 only to get the expression for the gravitational mass in a medium 
 \cite{gr:charge}. Calculation
 of scattering cross-section, which also involves modification of phase space factors
 at finite temperature, does not concern us here.
 \section{Operational definition at $O(e^2)$}\label{gr:op}
 We write the complete 1PI vertex function at one-loop
 in the form
 \begin{eqnarray}
 \Gamma_{\lambda\rho} = V_{\lambda\rho} + \Gamma'_{\lambda\rho} \,.
 \label{V+G}
 \end{eqnarray}
 Since $V_{\lambda\rho}$  denotes the tree-level vertex function,
 $\Gamma'_{\lambda\rho}$ is the $O(e^2)$ contribution. Then, using
 Eqs.\ (\ref{V+G}) and (\ref{Zf}), the formula given by Eq.\ (\ref{gr:S14})
 can be rewritten at $O(e^2)$ in the form
 \begin{eqnarray}
 \label{M'f}
 M'_f &=& (2v^\lambda v^\rho - \eta^{\lambda\rho})
 \times \lim_{\vec P\rightarrow 0}\Bigg\{ 
 \overline U_s(\vec P) V_{\lambda\rho}(p,p)U_s(\vec P)
 + \zeta_f \bar u_s(\vec P)V_{\lambda\rho}(p,p)u_s(\vec P)
 \nonumber\\*
 &&+\bigg[\bar u_s(\vec P\,')\Gamma'_{\lambda\rho}(p,p')u_s(\vec P)
 \bigg]_{{E'_f=E_f}\atop{\vec P\,'\rightarrow P}}      
 \Bigg\} \,.
 \end{eqnarray}
 Since $\zeta_f$ and $\Gamma'_{\lambda\rho}$ are $O(e^2)$, in any term
 that contains either of these factors we have substituted the tree-level
 expressions for the other quantities.  Now, the terms
 involving $V_{\lambda\rho}$ can be evaluated immediately with the help
 of the identities given in Eqs.\ (\ref{GordonUP1}) and (\ref{GordonUP2}). 
 Remembering
 that $E_f(0)=M_f$, we finally obtain the operational definition to
 $O(e^2)$
 \begin{eqnarray}
 \label{operationalM'}
 M'_f = 3M_f - 2m_f + \zeta_f m_f
 + (2v^\lambda v^\rho - \eta^{\lambda\rho})
 \lim_{\vec P\rightarrow 0}\left\{\left[
 \vphantom{\frac{1}{2}}
 \overline u_s(\vec P')\Gamma'_{\lambda\rho}(p,p')u_s(\vec P)
 \right]_{{E'_f=E_f}\atop{\vec P\,'\rightarrow P}}\right\} \,,
 \end{eqnarray}
 where we can set $E_f = \sqrt{P^2 + m_f^2}$ in the last term.

 We now turn to the derivation of an operational definition at $O(e^2)$
 for the gravitational mass of the antilepton. We first aim to write down
 an expression for $\langle \bar f(\vec P,s) |S | \bar f(\vec P\,',s) \rangle$
 from Eq.\ (\ref{gr:S5}) by crossing. To this end, let us note the 
 action of the field operators $\psi$ and $\bar\psi$, contained,
 through $\mathscr L_{\rm eff}$, in $S$. In the $S$-matrix element
 of Eq.\ (\ref{gr:S5}), $\psi$ kills $f(\vec P,s)$ and gives
 $\sqrt{Z_f(P)}U_s(\vec P)e^{-ip\cdot x}$, while in
 $\langle \bar f(\vec P,s) |S | \bar f(\vec P\,',s) \rangle$,
 $\psi$ creates $\bar f(\vec P, s)$ and gives
 $\sqrt{Z_{\bar f}(P)}V_s(\vec P)e^{ip\cdot x}$ (see the tree-level
 Fourier expansion for $\psi$ given in Eq.\ (\ref{psiFour})). This
 does not only identify the external leg factors, but also shows that
 the incoming momentum $p$ in the amputated vertex corresponding to
 the first case gets replaced in the second case with an outgoing
 momentum $p$, equivalent to an incoming momentum $-p$.
 Similarly, one can go through the action of $\bar\psi$. 
 Note also that the graviton comes out
 with three-momentum $\vec P\,' - \vec P$ in 
 $\langle \bar f(\vec P,s) |S | \bar f(\vec P\,',s) \rangle$,
 while the
 fermion exchange rule gives an extra minus sign. We therefore have
 \bea
 \langle \bar f(\vec P,s) |S | \bar f(\vec P\,',s) \rangle
 &=& (-1)(-i \kappa)2\pi
 \delta(E_{\bar f} - E'_{\bar f})\nonumber\\*
 &&\times Z_{\bar f}(P)\;
 \overline V_s(\vec P') \Gamma_{\lambda\rho}(-p,-p') V_s(\vec P)
 h^{\lambda\rho}(\vec P\,' - \vec P) \,.
							 \label{cross}
 \eea
 Note that $p$ and $p'$,
 of which $\Gamma_{\lambda\rho}$ is a function of in Eq.\ (\ref{cross}),
 are on-shell, and satisfy the antifermion dispersion relation.

 Next, we simply interchange $\vec P$ and $\vec P\,'$ in Eq.\ (\ref{cross})
 to obtain the antiparticle equation corresponding to Eq.\ (\ref{gr:S5}):
 \begin{eqnarray}
 \label{Sbarff}
 S_{\bar f\bar f}&\equiv&
 \langle \bar f(\vec P\,',s) |S | \bar f(\vec P,s) \rangle
 \nonumber\\*
  &=& (-1)(-i \kappa)2\pi
 \delta(E_{\bar f} - E'_{\bar f}) Z_{\bar f}(P)\;
 \overline V_s(\vec P) \Gamma_{\lambda\rho}(-p',-p) V_s(\vec P')
 h^{\lambda\rho}(\vec P - \vec P') 
 \end{eqnarray}
 This leads to an equation that is analogous to Eq.\ (\ref{M'f}),
 but with an extra minus sign in front and some obvious changes
 in the corresponding symbols, which in turn lead to the $O(e^2)$ formula
 \begin{eqnarray}
 \label{opMbar'}
 M'_{\bar f}& =&  3M_{\bar f} - 2m_f + \zeta_{\bar f} m_f
 \nonumber\\*
 &&- (2v^\lambda v^\rho - \eta^{\lambda\rho})
 \lim_{\vec P\rightarrow 0}\left\{\left[
 \vphantom{\frac{1}{2}}
 \overline v_s(\vec P)\Gamma'_{\lambda\rho}(-p',-p)v_s(\vec P')
 \right]_{{E'_{\bar f}=E_{\bar f}}\atop{\vec P\,'\rightarrow P}}\right\}\,.
 \end{eqnarray}
 Using the usual relation between the free particle and antiparticle spinors
 given in Eq.\ (\ref{qed:vdef}), and the properties of the charge conjugation
 matrix $\sf C$ given by Eqs.\ (\ref{qed:CR}), (\ref{qed:CU})
 and (\ref{qed:CA}), the spinor matrix element that appears in Eq.\ 
 (\ref{opMbar'}) can be rewritten in the form
 \begin{eqnarray}
 \label{Crelation}
 \overline v_s(\vec P)\Gamma'_{\lambda\rho}(-p',-p)v_s(\vec P') =
 -\overline u_s(\vec P'){\sf C}\,\Gamma'^{\rm T}_{\lambda\rho}(-p',-p)
 {\sf C}^{-1}u_s(\vec P) \,.
 \end{eqnarray}
 Finally, using Eq.\ (\ref{vertexC}) (which actually holds at all orders
 and even for off-shell momenta) in Eq.\ (\ref{Crelation}), and
 substituting the relation in Eq. (\ref{opMbar'}), we arrive at
 \begin{eqnarray}
 \label{CoperationalMbar'}
 M'_{\bar f} &=& 3M_{\bar f} - 2m_f + \zeta_{\bar f} m_f
 \nonumber\\*
 &&+ (2v^\lambda v^\rho - \eta^{\lambda\rho})
 \lim_{\vec P\rightarrow 0}\left\{\left[
 \vphantom{\frac{1}{2}}
 \left[\overline u_s(\vec P\,')\Gamma'_{\lambda\rho}(p,p')u_s(\vec P)
 \right]_{\mu\rightarrow -\mu}
 \right]_{{E'_f=E_f}\atop{\vec P\,'\rightarrow P}}
 \right\} \,.
 \end{eqnarray}

 From Eqs.\ (\ref{operationalM'}) and (\ref{CoperationalMbar'}),
 it follows that $M'_{\bar f}$ and $M'_f$ are related by
 \bea
 [M'_{\bar f}]\,_{\mu\rightarrow -\mu}=M'_f\,.                 \label{M'mu}
 \eea
 Since the gravitational mass of the antiparticle should equal that of the 
 particle for $C$ symmetric Lagrangian and medium, 
 Eq.\ (\ref{M'mu}) is expected from the discussion of
 Sec.\ \ref{qed:noC} for $C$ symmetric Lagrangian and $C$ violating 
 background. The consequence of Eq.\ (\ref{M'mu}) is that we can just
 obtain $M'_{\bar f}$ from $M'_f$ by implementing the prescription
 of Eq.\ (\ref{antiprescription2}) in the $O(e^2)$ correction.

 We take the opportunity to emphasize the following point. In the
 calculations that follow, we will find expressions for the various
 contributions to $\Gamma'_{\lambda\rho}(p,p')$, which are given as
 integrals over the propagators and thermal distribution functions. In
 general, such expressions do not have a unique limiting value as we
 let $p'\rightarrow p$ in an arbitrary way
 \cite{zeromomprob}. More importantly, some of the integrals involve 
 ill-defined expressions if
 the limit is not taken properly.  In our case, the precise order in
 which the various limits must be taken has been dictated by the
 physical issue at hand. Thus, since we are interested in the
 interaction of the particle with a static gravitational potential, the
 quantity that enters is $\Gamma'_{\lambda\rho}(p,p')$ evaluated for
 $E'_f=E_f$.  Next we set $\vec P' = \vec P$ since we
 want the gravitational field to vary slowly over a macroscopic region. 
 Finally we set $\vec
 P\rightarrow 0$ to obtain the coupling at zero momentum, which
 determines the gravitational mass. This justifies the somewhat
 cumbersome notation regarding the limits in Eq.\
 (\ref{operationalM'}), but it is meant to indicate precisely what we
 have just explained.
 We will see that this prescription allowed us to evaluate all the
 integrals involved (including those
 that superficially seem to be ill-defined)
 in a unique and well-defined way, without having to introduce 
 any special regularization technique. Our result for the case
 of the photon background agrees with the result of DHR \cite{DHR84},
 who used a special regularization technique, while our result for
 the case of the electron background agrees with the answer obtained 
 by the use of a regularization technique similar to that of DHR
 (see Appendix \ref{applambda}).
 \section{Gravitational vertex}\label{s:indgrav}
 \subsection{Irreducible diagrams}
\begin{figure}[btp]
 \begin{center}
 %
 %
 \begin{picture}(100,170)(-50,-30)
 \Text(0,-30)[cb]{\large\bf (A)}
 \ArrowLine(40,0)(0,0)
 \Text(35,-10)[cr]{$\ell(p)$}
 \ArrowLine(0,0)(-40,0)
 \Text(-35,-10)[cl]{$\ell(p')$}
 \Photon(0,0)(0,40){2}{6}
 \Text(-4,20)[r]{$\gamma$}
 \ArrowArc(0,60)(20,90,270)
 \ArrowArc(0,60)(20,-90,90)
 \Text(23,60)[l]{$f(k)$}
 \Text(-23,60)[r]{$f(k-q)$}
 \Photon(0,80)(0,120){2}{6}
 \Photon(0,80)(0,120){-2}{6}
 \end{picture}
 %
 %
 \begin{picture}(100,170)(-50,-30)
 \Text(0,-30)[cb]{\large\bf (B)}
 \ArrowLine(40,0)(0,0)
 \Text(35,-10)[cr]{$\ell(p)$}
 \ArrowLine(0,0)(-40,0)
 \Text(-35,-10)[cl]{$\ell(p')$}
 \Photon(0,0)(0,40){2}{6}
 \Text(-4,20)[r]{$\gamma$}
 \ArrowArc(0,60)(20,-90,270)
 \Text(0,85)[b]{$f(k)$}
 \Photon(0,40)(40,30){2}{6}
 \Photon(0,40)(40,30){-2}{6}
 \end{picture}
 %
 %
 \begin{picture}(100,170)(-50,-30)
 \Text(0,-30)[cb]{\large\bf (C)}
 \ArrowLine(40,0)(0,0)
 \Text(35,-10)[cr]{$\ell(p)$}
 \ArrowLine(0,0)(-40,0)
 \Text(-35,-10)[cl]{$\ell(p')$}
 \Photon(0,0)(0,40){2}{6}
 \Text(-4,20)[r]{$\gamma$}
 \ArrowArc(0,60)(20,-90,270)
 \Text(0,85)[b]{$f(k)$}
 \Photon(0,20)(40,20){2}{6}
 \Photon(0,20)(40,20){-2}{6}
 \end{picture}
 %
 %
 \begin{picture}(100,170)(-50,-30)
 \Text(0,-30)[cb]{\large\bf (D)}
 \ArrowLine(40,0)(0,0)
 \Text(35,-10)[cr]{$\ell(p)$}
 \ArrowLine(0,0)(-40,0)
 \Text(-35,-10)[cl]{$\ell(p')$}
 \Photon(0,0)(0,40){2}{6}
 \Text(-4,20)[r]{$\gamma$}
 \ArrowArc(0,60)(20,-90,270)
 \Text(0,85)[b]{$f(k)$}
 \Photon(0,0)(40,20){2}{6}
 \Photon(0,0)(40,20){-2}{6}
 \end{picture}
 \end{center}
 \caption{\sf 
 Tadpole diagrams for the one-loop gravitational vertex of charged leptons.
 These diagrams contribute to the vertex of any charged lepton in a
 background of electrons and nucleons.
 \label{f:Ztype}}
 \end{figure}
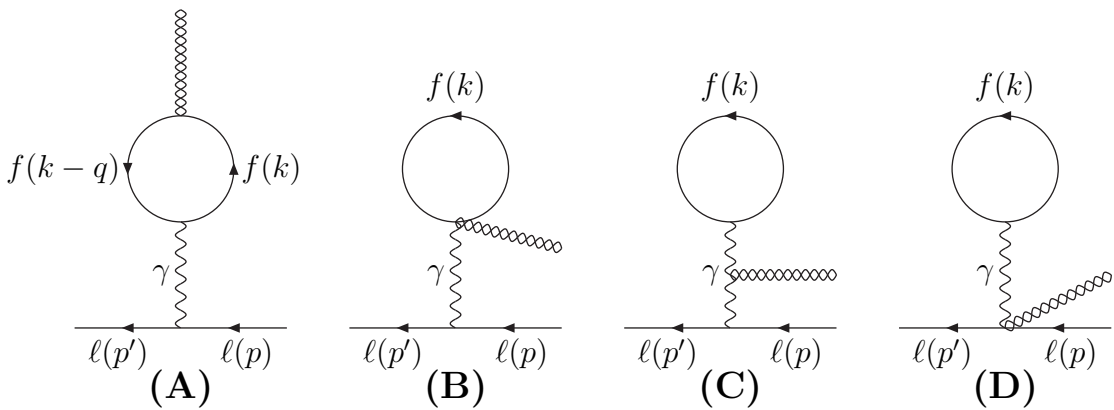

 The irreducible one-loop diagrams for the vertex function are given in
 Figs.~\ref{f:Wtype} and \ref{f:Ztype}. The four-vector $q^\mu$, which
 has been defined in Eq.\ (\ref{qdef}), gives the momentum of the
 outgoing graviton.
 %
 %
 \begin{figure}[tbp]
 \begin{center}
 %
 %
 \begin{picture}(180,150)(-90,-65)
 \Text(35,-35)[ct]{\large\bf (A)}
 \ArrowLine(80,0)(40,0)
 \Text(60,-10)[c]{$\ell(p)$}
 \ArrowLine(40,0)(0,0)
 \Text(20,-10)[c]{$\ell(k)$}
 \ArrowLine(0,0)(-40,0)
 \Text(-20,-10)[c]{$\ell(k-q)$}
 \ArrowLine(-40,0)(-80,0)
 \Text(-60,-10)[c]{$\ell(p')$}
 \Photon(0,0)(0,-45){2}{4}
 \Photon(0,0)(0,-45){-2}{4}
 \Text(0,-50)[l]{$q$}
 \PhotonArc(0,0)(40,0,180){4}{7.5}
 \Text(0,50)[c]{$\gamma$}
 \end{picture}
 %
 %
 \begin{picture}(180,130)(-90,-65)
 \Text(0,-35)[ct]{\large\bf (B)}
 \ArrowLine(80,0)(40,0)
 \Text(60,-10)[c]{$\ell(p)$}
 \ArrowLine(40,0)(-40,0)
 \Text(0,-10)[c]{$\ell(p-k)$}
 \ArrowLine(-40,0)(-80,0)
 \Text(-60,-10)[cr]{$\ell(p')$}
 \Photon(0,44)(0,80){2}{4}
 \Photon(0,44)(0,80){-2}{4}
 \Text(0,85)[bl]{$q$}
 \PhotonArc(0,0)(40,0,180){4}{6.5}
 \Text(40,40)[c]{$\gamma$}
 \Text(-40,40)[c]{$\gamma$}
 \end{picture}
 %
 %
 \begin{picture}(180,130)(-90,-65)
 \Text(0,-35)[ct]{\large\bf (C)}
 \ArrowLine(80,0)(40,0)
 \Text(60,-10)[c]{$\ell(p)$}
 \ArrowLine(40,0)(-40,0)
 \Text(0,-10)[c]{$\ell(k)$}
 \ArrowLine(-40,0)(-80,0)
 \Text(-60,-10)[c]{$\ell(p')$}
 \PhotonArc(0,0)(40,0,180){4}{7.5}
 \Text(0,50)[c]{$\gamma$}
 \Photon(40,0)(40,-50){2}{4}
 \Photon(40,0)(40,-50){-2}{4}
 \Text(40,-55)[l]{$q$}
 \end{picture}
 %
 %
 %
 \begin{picture}(180,130)(-90,-65)
 \Text(0,-35)[ct]{\large\bf (D)}
 \ArrowLine(80,0)(40,0)
 \Text(60,-10)[c]{$\ell(p)$}
 \ArrowLine(40,0)(-40,0)
 \Text(0,-10)[c]{$\ell(k)$}
 \ArrowLine(-40,0)(-80,0)
 \Text(-60,-10)[c]{$\ell(p')$}
 \PhotonArc(0,0)(40,0,180){4}{7.5}
 \Text(0,50)[c]{$\gamma$}
 \Photon(-40,0)(-40,-50){2}{4}
 \Photon(-40,0)(-40,-50){-2}{4}
 \Text(-40,-55)[r]{$q$}
 \end{picture}
 \caption{\sf
 Bubble diagrams for the one-loop gravitational vertex of charged leptons.
 These diagrams contribute in a background of photons and electrons.
 The braided line represents the graviton.
 \label{f:Wtype}}
 \end{center}
 \end{figure}
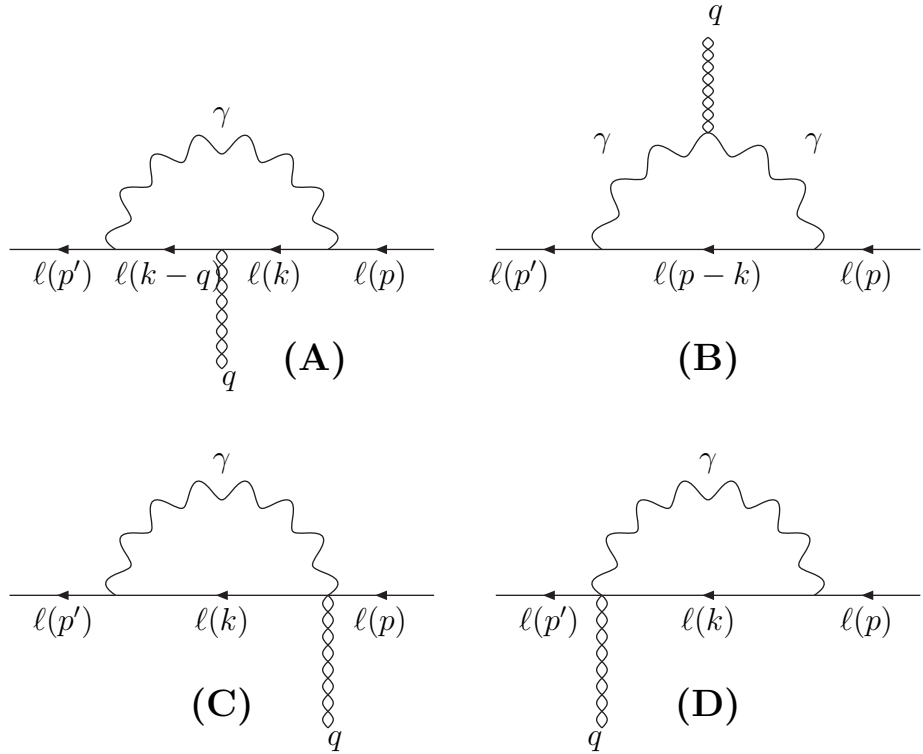

 %
 %

 We calculate only the terms that contribute to the
 dispersive part of the vertex function. To determine the
 condition for this, let us return to Eq.\ (\ref{gr:S0}), and demand that
 $\mathscr L_{\rm eff}$ be hermitian. This means $\widehat T{_{\lambda\rho}
 ^\dagger}(x) =\widehat T_{\lambda\rho}(x)$. Next consider Eq.\ (\ref{gr:S4}).
 Exchange of $p$ and $p'$ in this equation, followed by complex conjugation,
 gives
 \bea
 \langle f(\vec P,s) |\widehat T_{\lambda\rho} (0) | f(\vec P\,',s) \rangle^*
 = \sqrt{Z_f(P) Z_f(P')} \;
 \overline U_{s}(\vec P\,')\gamma_0\Gamma^\dagger_{\lambda\rho}(p',p)
 \gamma_0 U_s(\vec P)\,.                                        \label{gr:S4'}
 \eea
 But since $\widehat T_{\lambda\rho}$ is a hermitian operator, the left
 hand sides
 of Eqs.\ (\ref{gr:S4}) and (\ref{gr:S4'}) are equal. So equating the 
 right hand sides of these equations, we are led to the desired condition
 \begin{eqnarray}
 \Gamma_{\lambda\rho} (p,p') = \gamma_0 \Gamma_{\lambda\rho}^\dagger
 (p',p) \gamma_0                      \label{grdisper}
 \end{eqnarray}
 for the dispersive part of the vertex function.
 The absorptive part, on the other hand, corresponds to antihermitian
 $\mathscr L_{\rm eff}$ and satisfies $\Gamma_{\lambda\rho} (p,p') = 
 -\gamma_0 \Gamma_{\lambda\rho}^\dagger(p',p) \gamma_0$. It contributes
 to fermion damping, with which we are not concerned in the present
 work.

 When the formulas given in Eqs.\ (\ref{S}) and (\ref{D}) for the
 propagators are substituted in the expressions corresponding to the
 diagrams, we obtain terms of different kind.  One of them is
 independent of the background medium, in which we are not interested.
 Those involving two factors of the thermal part of the propagators
 contribute to the absorptive part of the vertex, while those involving
 three factors of the thermal part vanish because of the various
 $\delta$-functions appearing in it. Thus, the background induced
 contribution to the dispersive part of the vertex
 contains the thermal part of only one of the
 propagators, and they are the only kind of term that we retain.

 We have omitted the one-particle reducible diagrams in which the
 graviton line comes out from one of the external fermion legs, because
 they do not contribute to $\Gamma_{\lambda\rho}$. The proper way to
 take them into account in the calculation of the amplitude for any
 given process, is to choose the external spinor to be
 the solution of the effective Dirac equation for the propagating
 fermion mode in the medium (instead of the spinor representing the
 free-particle solution of the equation in vacuum), and multiply it with 
 the square-root of the wavefunction renormalization factor, as
 discussed in Sec.~\ref{gr:mass}.
 \subsection{Bubble diagrams given in Fig.\ \ref{f:Wtype}}
 \subsubsection*{Diagram \ref{f:Wtype}A}
 The amplitude of the diagram in Fig.~\ref{f:Wtype}A can be written as
 \begin{eqnarray}
 -i\kappa \Gamma^{(A)}_{\lambda\rho} (p,p') &=& \int {d^4k \over
 (2\pi)^4} \;
 ie\gamma_\alpha \, iS_\ell(k') \, (-i\kappa) V_{\lambda\rho} (k,k')
 \, iS_\ell(k) \,
 ie\gamma_\beta \,iD^{\alpha\beta} (k-p) \,
 \end{eqnarray}
 where
 \begin{eqnarray}
 k' \equiv k-q \,.
 \end{eqnarray}
 As already explained, to determine the contribution to the dispersive
 part of the vertex function we need to retain the terms that contain
 the thermal part of only one of the propagators.  Any of them contains
 some combination of the form
 \begin{eqnarray}
 \Lambda_{\lambda\rho} (k_1,k_2) \equiv \gamma_\alpha (\rlap/k_2+m_\ell)
 V_{\lambda\rho} (k_1,k_2) (\rlap/k_1+m_\ell) \gamma^\alpha \,.
 \end{eqnarray}
 After some straightforward algebra, this can be written as
 \begin{eqnarray}
 \Lambda_{\lambda\rho} (k_1,k_2) &=& -{1\over 2}
 \Big[ (k_1+k_2)_\rho (\rlap/k_1 \gamma_\lambda \rlap/k_2 +
 m^2_\ell \gamma_\lambda)
 + (k_1+k_2)_\lambda (\rlap/k_1 \gamma_\rho \rlap/k_2 + m^2_\ell
 \gamma_\rho) \Big] \nonumber\\*
 && + \eta_{\lambda\rho} \Big[ (k_1^2-m^2_\ell) (\rlap/k_2 -2m_\ell) +
 (k_2^2-m^2_\ell) (\rlap/k_1 - 2m_\ell) \Big] \nonumber\\*
 && + 2m_\ell (k_1+k_2)_\lambda (k_1+k_2)_\rho \,.
 \label{Lambda}
 \end{eqnarray}

 For the sake of convenience, we divide the total contribution into two
 parts
 \begin{eqnarray}
 \label{Gammadecomp}
 \Gamma'^{(A)}_{\lambda\rho} (p,p') = \Gamma'^{(A1)}_{\lambda\rho}
 (p,p') + \Gamma'^{(A2)}_{\lambda\rho} (p,p') \,,
 \end{eqnarray}
 where $\Gamma'^{(A1)}_{\lambda\rho}$ contains the distribution
 function of the photon and therefore contributes to the gravitational
 vertex for all charged leptons, and $\Gamma'^{(A2)}_{\lambda\rho}$
 contains the distribution function of the electrons and contributes
 only to the vertex for the electrons. Changing the
 integration variable from $k$ to $k+p$, we obtain
 \begin{eqnarray}
 \Gamma'^{(A1)}_{\lambda\rho} (p,p') &=& -e^2 \int {d^4k \over
 (2\pi)^3} \;
 {\delta(k^2) \eta_\gamma(k) \over [(k+p')^2-m^2_\ell][(k+p)^2-m^2_\ell]}
 \Lambda_{\lambda\rho} (k+p,k+p') \,
 \label{GamA1}
 \end{eqnarray}
 and similarly,
 \begin{eqnarray}
 \Gamma'^{(A2)}_{\lambda\rho} (p,p') &=& e^2
 \int {d^4k \over (2\pi)^3} \; \delta(k^2-m^2_\ell) \eta_\ell(k)
 \nonumber\\* && \times
 \Bigg( {\Lambda_{\lambda\rho} (k,k-q)
 \over [(k-q)^2-m_\ell^2](k-p)^2} +
 {\Lambda_{\lambda\rho} (k+q,k) \over [(k+q)^2-m_\ell^2](k-p')^2}
 \Bigg) \,.
 \label{A2}
 \end{eqnarray}
 %
 \subsubsection*{Diagram \ref{f:Wtype}B}
 For this diagram
 \begin{eqnarray}
 -i\kappa \Gamma^{(B)}_{\lambda\rho} (p,p') = \int {d^4k \over
 (2\pi)^4} \;
 ie\gamma^\alpha \, iS_\ell(p-k) \, ie\gamma^\beta
 (-i\kappa) C_{\mu\nu\lambda\rho} (k,k') \,
 iD^{\nu\alpha} (k) iD^{\mu\beta} (k')\,,
 \end{eqnarray}
 and we decompose it in analogy with Eq.\ (\ref{Gammadecomp}).
 The part that contains the photon distribution function is
 \begin{eqnarray}
 \Gamma'^{(B1)}_{\lambda\rho} (p,p')
 = e^2 \int {d^4k \over (2\pi)^4} \;
 \gamma^\nu S_{F\ell}(p-k) \gamma^\mu
 C_{\mu\nu\lambda\rho} (k,k') \,
 \Big[ \Delta_F (k) \Delta_T(k') + \Delta_F (k') \Delta_T(k) \Big] \,.
 \end{eqnarray}
 Making a change of the integration variable in one of the terms, this
 can be written as
 \begin{eqnarray}
 \Gamma'^{(B1)}_{\lambda\rho} (p,p')
 &=& e^2 \int {d^4k \over (2\pi)^3} \; \delta(k^2) \eta_\gamma(k)
 \Bigg[ { \gamma^\nu (\rlap/p'- \rlap/k + m_\ell) \gamma^\mu
 C_{\mu\nu\lambda\rho} (k+q,k) \over
 [(p'-k)^2-m_\ell^2] (k+q)^2}
 \nonumber\\*
 &&+ {\gamma^\nu (\rlap/p- \rlap/k +m_\ell) \gamma^\mu
 C_{\mu\nu\lambda\rho} (k,k-q) \over [(p-k)^2-m_\ell^2] (k-q)^2} \Bigg]
 \,, \quad
 \label{GamB1}
 \end{eqnarray}    
 while
 \begin{eqnarray}
 \Gamma'^{(B2)}_{\lambda\rho} (p,p') &=& e^2 \int {d^4k \over (2\pi)^4} \;
 \gamma^\nu S_{T\ell}(k) \gamma^\mu
 C_{\mu\nu\lambda\rho} (p-k,p'-k) \,
 \Delta_F (p-k) \Delta_F (p'-k)
 \nonumber\\*
 &=& -e^2 \int {d^4k \over (2\pi)^3} \; \delta(k^2-m_\ell^2) \eta_\ell(k)
 \nonumber\\*
 &&\times\gamma^\nu (\rlap/k + m_\ell) \gamma^\mu \;
 {C_{\mu\nu\lambda\rho} (p-k,p'-k) \over (p-k)^2 (p'-k)^2}
 \label{vertex.B2}
 \end{eqnarray}
 gives the lepton background part.
 \subsubsection*{Diagrams \ref{f:Wtype}C and \ref{f:Wtype}D}
 For these two diagrams the manipulations are similar and, omitting
 the details, the results are
 \begin{eqnarray}
 \label{GammaC1D1}
 \Gamma'^{(C1+D1)}_{\lambda\rho} (p,p') &=& -\, e^2
 a_{\mu\nu\lambda\rho}
 \int {d^4k \over (2\pi)^3} \;
 \delta(k^2) \eta_\gamma(k) 
 \nonumber\\*
 &&\times\left[
 {\gamma^\mu (\rlap/k + \rlap/p' + m_\ell) \gamma^\nu \over(k+p')^2
 -m_\ell^2} +
 {\gamma^\nu (\rlap/k + \rlap/p+m_\ell) \gamma^\mu \over (k+p)^2
 -m_\ell^2} \right] \,,
 \end{eqnarray}
 and
 \begin{eqnarray}
 \label{GammaC2D2}
 \Gamma'^{(C2+D2)}_{\lambda\rho} (p,p') &=& e^2 a_{\mu\nu\lambda\rho}
 \int {d^4k \over (2\pi)^3} \;
 \delta(k^2-m^2_\ell) \eta_\ell(k) 
 \nonumber\\*
 &&\times\left[
 {\gamma^\mu (\rlap/k +m_\ell) \gamma^\nu \over (k-p')^2}
 + {\gamma^\nu (\rlap/k +m_\ell) \gamma^\mu \over (k-p)^2}
 \right] \,.
 \end{eqnarray}
 %
 \subsection{Tadpole diagrams given in Fig.~\ref{f:Ztype}}\label{s:Ztype}
 \subsubsection*{The question of the photon tadpole}
 The calculation of the diagrams of Fig.\ \ref{f:Ztype} leads to an immediate
 problem. While the contributions from the diagrams \ref{f:Ztype}B,
 \ref{f:Ztype}C and \ref{f:Ztype}D appear to vanish due to the charge neutrality
 of the medium (see Sec.\ \ref{qed:calcu}), the contribution from the diagram 
 \ref{f:Ztype}A appears to diverge for $q=0$ (due to the 
 divergence of the
 photon propagator in this limit). The following discussion is devoted to the
 setting up of a framework of calculation which leads to the disappearance
 of the aforementioned divergence.

 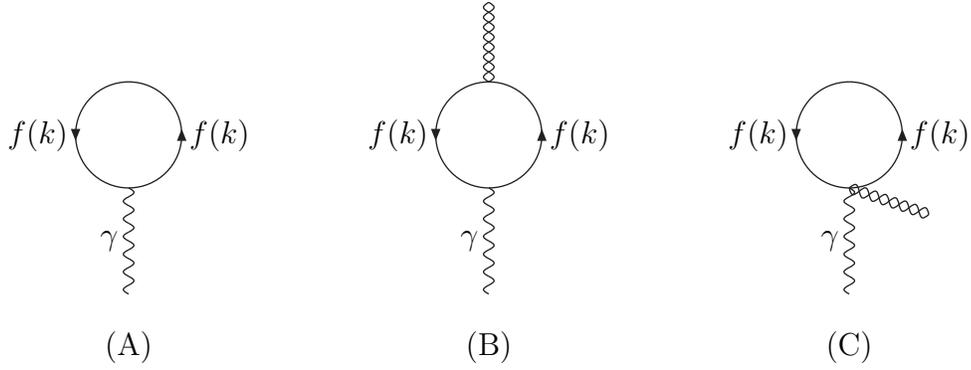
\begin{figure}[bhtp]
 \begin{center}
 \begin{picture}(100,140)(-50,-40)
 \Photon(0,0)(0,40){2}{6}
 \Text(0,-20)[c]{(A)}
 \Text(-4,20)[r]{$\gamma$}
 \ArrowArc(0,60)(20,90,270)
 \ArrowArc(0,60)(20,-90,90)
 \Text(23,60)[l]{$f(k)$}
 \Text(-23,60)[r]{$f(k)$}
 \end{picture}
 \hspace{1cm}
 \begin{picture}(100,140)(-50,-40)
 \Text(0,-20)[c]{(B)}
 \Photon(0,0)(0,40){2}{6}
 \Text(-4,20)[r]{$\gamma$}
 \ArrowArc(0,60)(20,90,270)
 \ArrowArc(0,60)(20,-90,90)
 \Text(23,60)[l]{$f(k)$}
 \Text(-23,60)[r]{$f(k)$}
 \Photon(0,80)(0,110){2}{4}
 \Photon(0,80)(0,110){-2}{4}
 \end{picture}
 \hspace{1cm}
 \begin{picture}(100,140)(-50,-40)
 \Text(0,-20)[c]{(C)}
 \Photon(0,0)(0,40){2}{6}
 \Text(-4,20)[r]{$\gamma$}
 \ArrowArc(0,60)(20,90,270)
 \ArrowArc(0,60)(20,-90,90)
 \Text(23,60)[l]{$f(k)$}
 \Text(-23,60)[r]{$f(k)$}
 \Photon(0,40)(30,30){2}{4}
 \Photon(0,40)(30,30){-2}{4}
 \end{picture}
 \end{center}
 \caption{\small\sf The one-loop diagrams
 that contribute to order $\kappa$ to the
 photon tadpole in a medium, in the presence of
 a static and homogeneous ($q = 0$) gravitational potential. The
 fermion loop involves a sum over all the species of fermions present in
 the medium.}\label{f:tadpole}
 \end{figure}

 Let us recall from the discussion of Sec.\ \ref{qed:calcu} that the electrical
 neutrality of the medium requires the vanishing of the photon tadpole. The 
 one-loop diagrams that contribute to the photon tadpole 
 {\it in the presence of a static and homogeneous gravitational potential}
 are shown in Fig.~\ref{f:tadpole}, where the graviton line represents
 represents the $q=0$ background field. In the absence of the
 background field, only the diagram \ref{f:tadpole}A
 contributes to the photon tadpole. In that case, the
 requirement that the tadpole vanishes yields the familiar condition
 (see Eq.\ (\ref{qed:T}))
 \begin{equation}
 \label{neutralcond}
 {\cal Q}^{(\ref{f:tadpole}A)} \equiv
 \sum_f {\cal Q}_f\left[2\int\frac{d^3K}{(2\pi)^3}[f_f(E_K)-f_{\bar f}(E_K)]
 \right]
 = 0\,.
 \end{equation}
 and the quantity ${\cal Q}^{(\ref{f:tadpole}A)}$ is identified
 with the total charge of the medium.
 However, in the presence of the background field, and
 to the order that we are calculating, we have to take into account
 the contributions of the diagrams \ref{f:tadpole}B and \ref{f:tadpole}C
 to the photon tadpole or, equivalently, to
 the total charge of the system. If we denote them
 by ${\cal Q}^{(\ref{f:tadpole}B)}$ and ${\cal Q}^{(\ref{f:tadpole}C)}$
 respectively, the condition for the vanishing of the photon tadpole is
 \bea
 {\cal Q}^{(\ref{f:tadpole}A)} + {\cal Q}^{(\ref{f:tadpole}B)}
 + {\cal Q}^{(\ref{f:tadpole}C)}=0\,,                         \label{Q+Q+Q}
 \eea
 instead of Eq.\ (\ref{neutralcond}). Physically, this means
 that the number density of the particles are not determined by their free
 distribution functions. The particle distributions
 rearrange themselves in a way that depends on
 the background gravitational field.

 Consider now the problem of finding out the dispersion relation of the fermion
 in an external gravitational field. 
 The dispersion relation is determined by the 
 poles of the function given in Eq.\ (\ref{qed:pole}), with $\Sigma_f(p)$ now
 denoting the fermion self-energy {\it in an external gravitational field}. Thus, 
 the one-loop diagrams contributing to $\Sigma_f(p)$ at $O(\kappa)$ are the
 diagrams of Fig.\ \ref{f:selfenergy}, Fig.\ \ref{f:Wtype} and Fig.\ 
 \ref{f:Ztype}. Let us confine our attention to the tadpole diagrams, i.e.,
 the diagram of Fig.\ \ref{f:selfenergy}B and the diagrams of Fig.\ \ref{f:Ztype},
 and consider the implications of the charge neutrality
 condition of Eq.\ (\ref{Q+Q+Q}) for the sum of the contributions arising from
 these diagrams.

 Firstly, the unadorned tadpole of Fig.\ \ref{f:tadpole}A is now itself of 
 order $\kappa$
 because of the condition (\ref{Q+Q+Q}). Since the
 diagrams \ref{f:Ztype}C and \ref{f:Ztype}D contain an explicit factor of
 $\kappa$ apart from the unadorned tadpole, their contribution is
 actually of order $\kappa^2$ and therefore we can neglect them.
 Secondly, the diagram \ref{f:selfenergy}B cancels the $q$-independent
 contributions from the diagrams \ref{f:Ztype}A and
 \ref{f:Ztype}B. Since the loop in diagram \ref{f:Ztype}B in
 independent of $q$, this diagram is totally cancelled.

 In summary, the only contribution from the various tadpole diagrams
 arises from the $q$-dependent part of the tadpole
 subdiagram of Fig.~\ref{f:Ztype}A (by subdiagram, we mean that the lower
 part, beginning with the photon propagator, is amputated). It seems reasonable 
 to assume that
 this holds true also when we consider the gravitational vertex function
 and its contribution to the gravitational mass. It then turns out that
 the $q$-dependent part of the tadpole
 subdiagram of Fig.~\ref{f:Ztype}A, when multiplied with $2v^\lambda v^\rho
 -\eta^{\lambda\rho}$, and evaluated at $q_0=0$ and $\vec Q\rightarrow 0$,
 is proportional to $v_\alpha Q^2$ (see Eqs.\ (\ref{GamX.5}), (\ref{Aq})
 (\ref{Aq1}) and (\ref{Aqlast})). When multiplied with the
 photon propagator at $q_0=0$, it gives zero for the $\delta(Q^2)$ part
 of the propagator, and cancels the $1/Q^2$
 in the other part.
 This latter contribution will be labeled by the letter `X' in
 order not to confuse it with the contributions of
 Fig.~\ref{f:Wtype}A.
 \subsubsection*{The non-vanishing contribution}\label{s:non0}
 We denote the vertex contribution coming from Fig.~\ref{f:Ztype}A by
 \begin{eqnarray}
 \Gamma^{(X)}_{\lambda\rho} (p,p') &=&
 {e^2 \gamma^\alpha \over q^2} X_{\lambda\rho\alpha} (q) \,,
 \label{Xq}
 \end{eqnarray}
 where $X_{\lambda\rho\alpha} (q)$ is the photon-graviton mixing
 diagram with external momentum $q$
 \begin{eqnarray}
 X_{\lambda\rho\alpha} (q) =
 \sum_f \int {d^4k \over (2\pi)^4} \Tr \Big[
 V_{\lambda\rho} (k,k') iS_f(k) i{\cal Q}_f \gamma_\alpha iS_f(k')
 \Big] \,.
 \label{defX}
 \end{eqnarray}
 Then, taking the above discussion above into account, the quantity
 which will appear in the expression for the gravitational mass is
 given by
 \begin{eqnarray}
 \widetilde\Gamma^{(X)}_{\lambda\rho} (p,p') &=&
 {e^2 \gamma^\alpha \over q^2} \Big[ X_{\lambda\rho\alpha} (q)
 - X_{\lambda\rho\alpha} (0) \Big] \,.
 \label{Xq-X0}
 \end{eqnarray}
 (As mentioned in Sec.\ \ref{qed:calcu}, the sum in Eq.\ (\ref{defX})
 is over all species of fermions in the
 medium, the charge of each species being denoted by ${\cal Q}_f$ with
 the convention that ${\cal Q}_e = -1$.)
 The medium-dependent
 contribution to $X_{\lambda\rho\alpha} (q)$ can be written as
 \begin{eqnarray}
 X_{\lambda\rho\alpha} (q)
= \sum_f {\cal Q}_f
\int {d^4k\over (2\pi)^3}
\; \delta(k^2-m_f^2) \eta_f(k) \left[
{A_{\lambda\rho\alpha}(k,k-q) \over q^2 - 2k\cdot q} +
{A_{\lambda\rho\alpha}(k+q,k) \over q^2 + 2k\cdot q}  \right] \,,
\label{GamX}
\end{eqnarray}
where, for arbitrary 4-momenta $k_1$ and $k_2$, we have defined
\begin{eqnarray}
A_{\lambda\rho\alpha} (k_1,k_2) &=&
\Tr  \Big[
V_{\lambda\rho} (k_1,k_2)
(\rlap/k_1 +m_f) \gamma_\alpha (\rlap/k_2+m_f) \Big]
\nonumber\\*
&=& \Big[
(2k_{1\lambda} k_{1\rho} + k_{1\lambda} k_{2\rho} + k_{2\lambda}
k_{1\rho}) k_{2\alpha}
+ (m_f^2-k_1\cdot k_2) (\eta_{\lambda\alpha} k_{1\rho} +
\eta_{\rho\alpha} k_{1\lambda}) \nonumber\\*
&& - 2 \eta_{\lambda\rho} (k_1^2-m_f^2)
k_{2\alpha} \Big]
+ \Big[ k_1 \leftrightarrow k_2\Big] \,.
\end{eqnarray}
Putting $k^2=m_f^2$, we obtain
\begin{eqnarray}
A_{\lambda\rho\alpha} (k,k-q) &=&
[8k_\lambda k_\rho - 4 (k_\lambda
q_\rho + k_\rho q_\lambda) + 2 q_\lambda q_\rho] k_\alpha
- [4k_\lambda k_\rho - (k_\lambda
q_\rho + k_\rho q_\lambda)] q_\alpha \nonumber\\* &&
+ k\cdot q [\eta_{\lambda\alpha} (2k-q)_\rho + \eta_{\rho\alpha}
(2k-q)_\lambda ]
- 2 \eta_{\lambda\rho} (q^2-2k\cdot q) k_\alpha \,.
\label{A...}
\end{eqnarray}
Since $A_{\lambda\rho\alpha} (k_1,k_2) = A_{\lambda\rho\alpha}
(k_2,k_1)$ by definition, $A_{\lambda\rho\alpha} (k+q,k)$ is
obtained by changing the sign of $q$ in this expression.
\subsection{Additional remarks}
This is an appropriate point to consider the cases when 2-type vertices
can occur in the diagrams contributing to the gravitational vertex. Since
the vertices connected to the external legs can only be of 1-type, a
2-type vertex cannot occur in any of the diagrams of Fig.\ \ref{f:Wtype} or
in the diagram of Fig.\ \ref{f:Ztype}B. However, in Figs.\ \ref{f:Ztype}A, 
\ref{f:Ztype}C and \ref{f:Ztype}D, the vertex where the photon line connects to
the fermion loop can be of 2-type.

Now, due to the reason given in Sec.\ \ref{qed:Self} (see after Eq.\ 
(\ref{qedreffor2type})), the unadorned tadpole with a 2-type vertex is just
the negative of the unadorned tadpole with a 1-type vertex, and hence of $O(\kappa)$
because of the charge neutrality condition. So the diagrams of Figs.\ \ref{f:Ztype}C 
and \ref{f:Ztype}D, even with a 2-type vertex, are of $O(\kappa^2)$, and
can be neglected.

In Fig.\ \ref{f:Ztype}A, the occurrence of the 2-type vertex makes the photon
propagator a ``12" propagator. So the photon propagator is proportional to 
$\delta(q^2)$, which becomes $\delta(Q^2)$ for $q^0=0$. However, the value of 
the graviton momentum is taken to be $\vec Q\rightarrow 0$, and not
$Q=0$ (having $Q=0$ in addition to $q^0=0$ would mean a potential constant in space and time, 
resulting
in zero gravitational field). Consequently, $\delta(Q^2)$ in the ``12" propagator
causes this contribution to vanish. Therefore, 2-type vertices do not play any
role in the present calculation.

Next, we make some comments on the expressions deduced in this section.

Firstly, one can easily check that the expressions given in Eqs.\ 
(\ref{GamA1}), (\ref{A2}), (\ref{GamB1}), (\ref{vertex.B2}), (\ref{GammaC1D1}),
(\ref{GammaC2D2}), and (\ref{Xq}) (coupled with (\ref{GamX})), all satisfy
the dispersive condition of Eq.\ (\ref{grdisper}). (Similarly, it can be checked
that the terms involving the thermal parts of two propagators from
$\Gamma'^{(A)}_{\lambda\rho}$, $\Gamma'^{(B)}_{\lambda\rho}$,
$\Gamma'^{(C+D)}_{\lambda\rho}$ and $\Gamma'^{(X)}_{\lambda\rho}$
are all absorptive.)

Secondly, it can be verified that the different contributions to the 
dispersive part of $\Gamma'_{\lambda\rho}$ also satisfy the charge
conjugation relation given by Eq.\ (\ref{vertexC}).

Thirdly, the complete one-loop vertex function
satisfies the transversality condition, which is implied by the gravitational
gauge invariance of the theory.
This is shown in Appendix~\ref{app:transv}.
\section{Calculation of the gravitational mass}
\label{s:calc}
As seen in Eq.\ (\ref{operationalM'}), there are three types of
$O(e^2)$ correction to the gravitational mass.  One of them is
proportional to the inertial mass that was calculated in
Sec.~\ref{calin}, and another one involves the wave function
renormalization factor derived in Sec.~\ref{calZ}. In this section we
find the contributions from the irreducible one-loop vertex diagrams. Since the
expressions for those already have an explicit factor of $e^2$ outside
the integral, to evaluate them we can use the tree-level values for
the dispersion relation and the spinors associated with the external
lepton.
\subsection{Terms with the photon distribution from Fig.~\ref{f:Wtype}}
We first evaluate those terms obtained in Sec.~\ref{s:indgrav}
that contain the photon distribution function. In fact, if the
temperature of the ambient medium is low ($T\ll m_e$) and the chemical
potential of the background electrons is zero, these are the only
terms that contribute and they are precisely the ones that were
calculated in Ref.\ \cite{DHR84}.
Since we have performed the calculations in a different
way, using 1-particle irreducible diagrams only, the following results
serve as a good checkpoint between the earlier calculations of
Ref.\ \cite{DHR84} and ours.
\subsubsection*{Contribution (A1)}
{}From the formula for the gravitational mass given in
Eq.\ (\ref{operationalM'}),
it follows that we need to calculate the
vertex only for $p=p'$, in which case
\begin{eqnarray}
\Gamma'^{(A1)}_{\lambda\rho} (p,p) &=& -e^2 \int {d^4k \over
(2\pi)^3} \;
{\delta(k^2) \eta_\gamma(k) \over 4(k\cdot p)^2}
\Lambda_{\lambda\rho} (k+p,k+p) \,.
\end{eqnarray}
{}From Eq.\ (\ref{Lambda}) it follows that, for any 4-vector $y^\mu$,
\begin{eqnarray}
\Lambda_{\lambda\rho} (y,y) = -4 y_\lambda y_\rho (\rlap/y - 2m_\ell) +
(y^2-m_\ell^2) \bigg[ (\gamma_\lambda y_\rho + \gamma_\rho y_\lambda) +
2\eta_{\lambda\rho} (\rlap/y - 2m_\ell) \bigg] \,,
\label{Lambdayy}
\end{eqnarray}
which leads to
\begin{eqnarray}
\left[\overline u_s(\vec P)\Gamma'^{(A1)}_{\lambda\rho} (p,p)u_s(\vec P)
\right]_{p^\mu = (m_\ell,\vec 0)} & = &
-  e^2 \int {d^4k \over
(2\pi)^3} \;
{\delta(k^2) \eta_\gamma(k) \over (k\cdot p)^2}
 \overline u_s(\vec P)\Bigg[
-{k\cdot p\over m_\ell} k_\lambda p_\rho\nonumber\\*
&&+ m_\ell (k_\lambda k_\rho + p_\lambda p_\rho)
+ {(k\cdot p)^2 \over m_\ell} \eta_{\lambda\rho}\Bigg]u_s(\vec P) \,,
\end{eqnarray}
where we have used Eq.\ (\ref{ugu}) and omitted the terms odd in
$k$, which integrate to zero.  Using the notation
\begin{eqnarray}
m'_{(A1)} =
(2v^\lambda v^\rho - \eta^{\lambda\rho}) \left[
\overline u_s(\vec P) \Gamma'^{(A1)}_{\lambda\rho}(p,p)
u_s(\vec P)\right]_{p^\mu = (m_\ell,\vec 0)}
\,,
\end{eqnarray}
we obtain
\begin{eqnarray}
m'_{(A1)} &=& e^2 \int {d^3K \over
(2\pi)^3} \; f_\gamma(K) \bigg[ {1\over m_\ell K} - {m_\ell\over K^3} \bigg]
\nonumber\\*
&=& {e^2T^2 \over 12m_\ell} - {e^2 m_\ell \over 2\pi^2} \int_0^\infty {dK\over
K} \; f_\gamma(K) \,.
\label{A1final}
\end{eqnarray}
The remaining integral is infrared divergent, but its contribution
to the gravitational mass is canceled
by a similar term that arises from the wavefunction renormalization,
as we show below.
\subsubsection*{Contribution (B1)}
This term has to be treated carefully because the denominators
in the integrand of Eq.\ (\ref{GamB1})
vanish for $q=0$. Indeed, by using the formulas
\bea
{\mathscr Pr}\Bigg(\frac{1}{x}\Bigg)&=&\lim_{\epsilon\rightarrow 0}\frac{x}{x^2+\epsilon^2}\,,
\nonumber\\*
\delta(x)&=&\lim_{\epsilon\rightarrow 0}\frac{1}{\pi}
     \frac{\epsilon}{x^2+\epsilon^2}\,
                                                                   \label{Pdelta}
\eea
(with $\mathscr Pr$ denoting the principal value), one can show that \cite{bedaque}
\bea
{\mathscr Pr}\Bigg(\frac{1}{k^2-m^2}\Bigg)\delta(k^2-m^2)=\frac{1}{2}\frac{d}{dm^2}
  \delta(k^2-m^2)                                                     \label{Pdel}
\eea
which is ill-defined even for $m\neq 0$, whereas for $m=0$, one can
interpret the R.H.S. of 
Eq.\ (\ref{Pdel}) only as a limiting form with $m\rightarrow 0$.
However, a careful evaluation of the term given in Eq.\ (\ref{GamB1}),
following the procedure indicated in
Eq.\ (\ref{operationalM'}),
shows that the limit exists.
Denoting
\begin{eqnarray}
m'_{(B1)} \equiv
(2v^\lambda v^\rho - \eta^{\lambda\rho})
\lim_{\vec P\rightarrow 0}\left\{\left[
\vphantom{\frac{1}{2}}
\overline u_s(\vec P')\Gamma'^{(B1)}_{\lambda\rho}(p,p')u_s(\vec P)
\right]_{{E'_\ell=E_\ell}\atop{\vec P\,'\rightarrow P}}\right\} \,,
\end{eqnarray}
the result is
\begin{eqnarray}
m'_{(B1)} &=& - {e^2T^2 \over 3m_\ell} \,.
\label{B1final}
\end{eqnarray}
The details of the derivation of this result are given in
Appendix~\ref{app:tough:B1}. This result can also be obtained by means of
a special regularization at $q=0$ (following Eq. (\ref{Pdel})), as shown 
in Appendix \ref{applambda}.

\subsubsection*{Contributions (C1+D1)}
We can proceed as in the evaluation of $m'_{(A1)}$ above. Thus, from
Eq.\ (\ref{GammaC1D1}),
\begin{eqnarray}
\left[\overline u_s(\vec P)\Gamma'^{(C1+D1)}_{\lambda\rho} (p,p)u_s(\vec P)\right]_{
p^\mu=(m_\ell,\vec 0)} &=&
-\, e^2 a_{\mu\nu\lambda\rho}
\int {d^4k \over (2\pi)^3} \;
\frac{\delta(k^2) \eta_\gamma(k)}{k\cdot p}\nonumber\\*
&&\times\left[\overline u_s(\vec P)\gamma^\mu \rlap/k \gamma^\nu u_s(\vec P)\right] \,,
\end{eqnarray}
using the fact that $a_{\mu\nu\lambda\rho}$ is symmetric in the
indices $\mu,\nu$. Then using
\begin{eqnarray}
(2v^\lambda v^\rho - \eta^{\lambda\rho}) a_{\mu\nu\lambda\rho} =
-\eta_{\mu\nu} - 2v_\mu v_\nu
\end{eqnarray}
it follows that
\begin{eqnarray}
\label{C1D1final}
m'_{(C1+D1)} & \equiv  &
(2v^\lambda v^\rho - \eta^{\lambda\rho})
\left[\overline u_s(\vec P)\Gamma'^{(C1+D1)}_{\lambda\rho}(p,p)
u_s(\vec P)\right]_{p^\mu = (m_\ell,\vec 0)} \nonumber\\
& = & 0 \,.
\end{eqnarray}
%
\subsection{Terms with the electron distribution from Fig.~\ref{f:Wtype}}
These terms contribute only to the gravitational vertex involving electrons and
positrons.  The integration over $k_0$ and the angular variables can
be done exactly.  The remaining integral can be evaluated analytically
only for special cases of the distribution functions, some of which we
consider afterwards.
\subsubsection*{Contribution (A2)}
As can be seen from Eq.\ (\ref{A2}),
the denominators of the integrand of this term vanish as
$q\to0$. Consequently, the prescription indicated in
Eq.\ (\ref{operationalM'}) has to be followed carefully in this case.
As we show in detail in Appendix \ref{app:tough:A2},
defining
\begin{eqnarray}
m'_{(A2)} \equiv
(2v^\lambda v^\rho - \eta^{\lambda\rho})
\lim_{\vec P\rightarrow 0}\left\{\left[
\vphantom{\frac{1}{2}}
\overline u_s(\vec P')\Gamma'^{(A2)}_{\lambda\rho}(p,p')u_s(\vec P)
\right]_{{E'_\ell=E_\ell}\atop{\vec P\,'\rightarrow P}}\right\} \,,
\end{eqnarray}
the final result for this term is
\begin{eqnarray}
m'_{(A2)} & = & {e^2 \over m_e} \int {d^3K \over (2\pi)^32E_K}
\left\{
{2E_K^2-m_e^2 \over E_K} \left( {E_K-2m_e \over E_K-m_e}
\; {\partial f_e \over
\partial E_K} + {E_K+2m_e \over E_K+m_e} \; {\partial f_{\bar e} \over
\partial E_K} \right)\right. \nonumber\\*
&& + {2E_K^4 - E_K^3m_e - 5E_K^2m_e^2 + 2E_Km_e^3 - 2m_e^4
\over m_e E_K^2
(E_K-m_e)} f_e \nonumber\\*
&& - \left.{2E_K^4 + E_K^3m_e - 5E_K^2m_e^2 - 2E_Km_e^3 - 2m_e^4
\over m_e E_K^2
(E_K+m_e)} f_{\bar e}
\right\} \,,
\label{A2final}
\end{eqnarray}
where $E_K$ is defined in Eq.\ (\ref{EK}). This result is also
derived in Appendix \ref{applambda} by means of a special regularization
at $q=0$.
\subsubsection*{Contribution (B2)}
{}From Eq.\ (\ref{vertex.B2}) it is seen that
the integrand is not singular in the limit $q\to0$.
Therefore we can evaluate directly
\begin{eqnarray}
\Gamma'^{(B2)}_{\lambda\rho} (p,p)
&=& -e^2 \int {d^4k \over (2\pi)^3} \; \delta(k^2-m_e^2) \eta_e(k)
\gamma^\nu (\rlap/k+m_e) \gamma^\mu \;
{C_{\mu\nu\lambda\rho} (p-k,p-k) \over (p-k)^4} \,
\end{eqnarray}
and the contribution to the gravitational mass is given by
\begin{eqnarray}
m'_{(B2)} =
(2v^\lambda v^\rho - \eta^{\lambda\rho})
\left[\overline u_s(\vec P)\Gamma'^{(B2)}_{\lambda\rho}(p,p) u_s(\vec P)\right]_{
p^\mu = (m_e,\vec 0)} \,.
\end{eqnarray}
In the expression for $C_{\mu\nu\lambda\rho}$, any term having a
factor of $(p-k)_\mu$ or $(p-k)_\nu$ does not contribute to the
integral. This is because, within the spinors, we can write
\begin{eqnarray}
\gamma^\nu (\rlap/k+m_e) \gamma^\mu (p-k)_\mu =
\gamma^\nu (\rlap/k+m_e) (m_e-\rlap/k)
= \gamma^\nu (m_e^2-k^2) \,,
\end{eqnarray}
which vanishes because of the $\delta$-function. The argument is
similar for $(p-k)_\nu$. Thus,
\begin{eqnarray}
&&\hspace{-1cm}
(2v^\lambda v^\rho - \eta^{\lambda\rho}) \overline u_s(\vec P)
\gamma^\nu (\rlap/k+m_e)\gamma^\mu C_{\mu\nu\lambda\rho}(p-k,p-k)
u_s(\vec P) \nonumber\\
&&=
\overline u_s(\vec P)\Big\{
8(\rlap/k-2m_e)  (p\cdot v - k\cdot v)^2 + 4 (m_e - 2k\cdot v \rlap/v)
(p-k)^2 \Big\} u_s(\vec P)\,, \quad
\end{eqnarray}
ignoring all terms which have a factor of $k^2-m_e^2$.
Using Eqs.\ (\ref{ugu}) and (\ref{unorm}), we then obtain
\begin{eqnarray}
m'_{(B2)}
&=& -\, {2e^2\over m_e^2} \int {d^4k \over
(2\pi)^3} \; \delta(k^2-m_e^2) \eta_e(k) \left( k_0 + {m_e^2
\over k_0-m_e}
\right)
\nonumber\\
&=& -\, {e^2\over m_e^2} \int {d^3K\over
(2\pi)^3} \left( f_e - f_{\bar e} \right)
- {e^2\over 2\pi^2} \int dK \; {K^2 \over E_K}
\left[ {f_e(E_K) \over E_K -m_e } - {f_{\bar e}(E_K)
\over E_K +m_e }\right] .
\label{B2final}
\end{eqnarray}
%
\subsubsection*{Contributions (C2+D2)}
Similarly, for this term we can evaluate directly
\begin{eqnarray}
m'_{(C2 + D2)} =
(2v^\lambda v^\rho - \eta^{\lambda\rho})
\left[\overline u_s(\vec P)\Gamma'^{(C2 + D2)}_{\lambda\rho}(p,p) u_s(\vec P)\right]_{
p^\mu = (m_e,\vec 0)} \,,
\end{eqnarray}
with
\begin{eqnarray}
\Gamma'^{(C2+D2)}_{\lambda\rho} (p,p) = 2e^2 a_{\mu\nu\lambda\rho}
\int {d^4k\over
(2\pi)^3} \delta(k^2-m_e^2) \eta_e(k) \; {\gamma^\mu (\rlap/k+m_e)
\gamma^\nu \over (k-p)^2} \,.
\end{eqnarray}
By straight forward algebra
\begin{eqnarray}
(2v^\lambda v^\rho - \eta^{\lambda\rho}) a_{\mu\nu\lambda\rho}
\gamma^\mu (\rlap/k+m_e) \gamma^\nu = 4 \rlap/k - 6m_e - 4 k\cdot v
\rlap/v \,,
\end{eqnarray}
and using Eqs.\ (\ref{ugu}) and (\ref{unorm}),
\begin{eqnarray}
m'_{(C2+D2)} &=& 6e^2 \int {d^4k \over
(2\pi)^3} \; \delta(k^2-m_e^2) \eta_e(k) {1 \over k_0-m_e}
\nonumber\\
&=& {3e^2\over 2\pi^2} \int dK  \; {K^2 \over E_K}
\Bigg[ {f_e(E_K) \over E_K-m_e} - {f_{\bar e} (E_K) \over
E_K+m_e} \Bigg] .
\label{C2D2final}
\end{eqnarray}
%
\subsection{Terms from Fig.~\ref{f:Ztype}}
The contribution to the gravitational mass due to this term is
\begin{eqnarray}
\label{mXdef}
m'_{(X)} \equiv
(2v^\lambda v^\rho - \eta^{\lambda\rho})
\lim_{\vec P\rightarrow 0}\left\{\left[
\vphantom{\frac{1}{2}}
\overline u_s(\vec P') \widetilde\Gamma^{(X)}_{\lambda\rho}(p,p') u_s(\vec P)
\right]_{{q^0 =0}\atop{\vec Q\rightarrow 0}}\right\} \,,
\end{eqnarray}
where, from Eq.\ (\ref{Xq-X0}),
\begin{eqnarray}
\label{XQ-X0}
\widetilde\Gamma^{(X)}_{\lambda\rho} (p,p') \Bigg|_{q^0=0}  =
- \, {e^2 \gamma^\alpha \over Q^2}
\left[ X_{\lambda\rho\alpha} (q) \Big|_{q^0 = 0}
- X_{\lambda\rho\alpha} (0) \right] \,.
\end{eqnarray}
Using the expression for $A_{\lambda\rho\alpha}$ from Eq.\ (\ref{A...})
we obtain
\begin{eqnarray}
(2v^\lambda v^\rho - \eta^{\lambda\rho})
X_{\lambda\rho\alpha} (q) \Big|_{q^0 = 0} &=&
8\sum_f {\cal Q}_f
\int {d^4k\over (2\pi)^3}
\; \delta(k^2-m_f^2) \eta_f(k) k_0 v_\alpha 
\Bigg[
(2k_0^2-m_f^2 - \frac12 Q^2)\nonumber\\*
&&\times\left( {1\over 2 \vec K\cdot \vec Q - Q^2} -
{1\over 2 \vec K\cdot \vec Q + Q^2} \right)\Bigg] \,,
\end{eqnarray}
where we have omitted the terms that vanish by symmetric integration
over $\vec K$, as well as all those terms that are independent
of $q$, because they drop out of
Eq.\ (\ref{XQ-X0}), and in addition
all the terms that are
proportional to $q_\alpha$, because in Eq.\ (\ref{mXdef}) they yield
a factor of $\rlap/q$ which vanishes between spinors.
Performing the integration over $k_0$,
\begin{eqnarray}
(2v^\lambda v^\rho - \eta^{\lambda\rho})
X_{\lambda\rho\alpha} (q) \Big|_{q^0 = 0}
&=&
4v_\alpha\sum_f {\cal Q}_f
\int {d^3K\over (2\pi)^3}
\; \Big( f_f - f_{\bar f} \Big)
(2E_K^2-m_f^2 - \frac12 Q^2)\nonumber\\*
&&\times\left( {1\over 2 \vec K\cdot \vec Q - Q^2} -
{1\over 2 \vec K\cdot \vec Q + Q^2} \right).\;
\label{GamX.4}
\end{eqnarray}
For the term that contains an explicit factor of $Q^2$ in the
numerator we use the angular integration formula of Eq.\
(\ref{angint}), which yields
\begin{eqnarray}
(2v^\lambda v^\rho - \eta^{\lambda\rho})
X_{\lambda\rho\alpha} (q)\Big|_{q^0 = 0}
\!\!&= & v_\alpha\sum_f {\cal Q}_f \Bigg[
{Q^2\over 2\pi^2}
\int dK \; \Big( f_f - f_{\bar f} \Big)
 + \!4 \int \! {d^3K\over (2\pi)^3}
\; \Big( f_f - f_{\bar f} \Big)\nonumber\\*
&&\times (2E_K^2-m_f^2)
\left( {1\over 2 \vec K\cdot \vec Q - Q^2} -
{1\over 2 \vec K\cdot \vec Q + Q^2} \right) \!\!\Bigg]\, .
\label{GamX.5}
\end{eqnarray}
The evaluation of the rest of the integral is presented in
Appendix \ref{app:tough:X}. Substituting the results into
Eq.\ (\ref{XQ-X0}), the contribution of this diagram to the
gravitational mass is found to be given by
\begin{eqnarray}
m'_{(X)}
\label{Xfinal}
&=& - e^2 \sum_f {{\cal Q}_f \over 6\pi^2}
\int dK \Bigg[  \Big( f_f - f_{\bar f} \Big) -
 {2E_K^2-m_f^2\over 2E_K} {\partial \over \partial E_K} \Big( f_f -
f_{\bar f} \Big) \Bigg] \,.
\end{eqnarray}
%
\subsection{Summary}
Starting from Eq.\ (\ref{operationalM'}), the total contribution to
the gravitational mass of charged leptons can be written in the form
\begin{eqnarray}
M'_\ell = m_\ell + m'_{\ell1} + m'_{\ell2} + m'_{(X)} \,,
\label{Mell'def}
\end{eqnarray}
where $m'_{(X)}$ is the contribution from Eq.\ (\ref{Xfinal}), which
is the same for all charged leptons, $m'_{\ell1}$ represents the terms
that contain the photon distribution function, and $m'_{\ell2}$
contains the terms that depend on the electron distribution
function. They are given as follows.

Substituting into Eq.\ (\ref{operationalM'}) the results given in
Eqs.\ (\ref{A1final}), (\ref{B1final}) and (\ref{C1D1final}), and
using the expression for the wave-function normalization and the
correction to the inertial mass given in Eqs.\ (\ref{Zpsi1}) and
(\ref{m1}), we find
\begin{eqnarray}
m'_{\ell1} = - {e^2T^2 \over 12m_\ell} \,,
\end{eqnarray}
in agreement with the DHR result\cite{DHR84}, quoted in Eq.\
(\ref{DHR:Mg}).  Notice that the infrared divergence contained in the
$m'_{(A1)}$ term cancels with a similar one that arises from the wave
function renormalization correction $\zeta_{\ell 1}$.

The terms from the diagrams in Fig.~\ref{f:Wtype} that involve the
fermion distribution function contribute only to the gravitational
mass of the electron, and therefore
\begin{eqnarray}
m'_{\mu2} = m'_{\tau2} = 0 \,.
\end{eqnarray}
The individual contributions of this type to the electron
gravitational mass appear in Eqs.\ (\ref{A2final}), (\ref{B2final})
and (\ref{C2D2final}).
Substituting those results into Eq.\ (\ref{operationalM'}),
and using the results for the
inertial mass and the wave-function normalization factor, given in
Eqs.\ (\ref{me2}) and (\ref{ZU2}) respectively, we obtain
\begin{eqnarray}
m'_{e2} & = & {\displaystyle e^2 \over \displaystyle\pi^2 m_e}
\int_0^\infty dK \; {K^2  \over 2E_K} \Bigg\{
\left( {3\over 2} + {m_e^2 \over E_K^2} - {m_e \over
E_K - m_e} \right) f_e(E_K) \nonumber\\*
& & + \left( {3\over 2} + {m_e^2 \over E_K^2} + {m_e \over
E_K + m_e} \right) f_{\bar e} (E_K) \nonumber\\*
& & +
{2E_K^2-m_e^2 \over 2E_K} \bigg( {E_K-2m_e \over E_K-m_e} \;
{\partial f_e \over
\partial E_K} + {E_K+2m_e \over E_K+m_e} \; {\partial f_{\bar e} \over
\partial E_K} \bigg) \Bigg\} \,.
\label{m'2summary}
\end{eqnarray}

The corresponding formulas for the antileptons are obtained by making
the substitution given in Eq.\ (\ref{antiprescription2}), as explained
after Eq.\ (\ref{M'mu}). Thus, 
\begin{eqnarray}
\label{Mbarell'def}
M'_{\bar\ell} = m_\ell + m'_{\ell1} + m'_{{\bar\ell}2} - m'_{X} \,,
\end{eqnarray}
where
\begin{eqnarray}
m'_{\bar\mu2} = m'_{\bar\tau2} = 0\,,
\end{eqnarray}
while the result for $m'_{{\bar e}2}$ is obtained from
Eq.\ (\ref{m'2summary}}) by making the substitution
$f_e \leftrightarrow f_{\bar e}$.

\section{Results for particular cases}
\label{s:cases}
In contrast with the $m_{\ell 1}$ and $m'_{\ell 1}$ terms,
which depend on the photon momentum distribution, 
$m_{e2}$, $m_{\bar e2}$, $m'_{e2}$, $m'_{\bar e2}$ and $m'_{(X)}$ depend
on the fermion distribution functions and cannot be
evaluated exactly in the general case.
Therefore, for illustration, we consider in detail 
their calculation for the specific
situation in which the background is composed of non-relativistic protons 
and electrons. In this case we can set (for $f = e,p$)
\begin{eqnarray}
f_{\bar f}(E) \approx 0
\end{eqnarray}
and
\begin{eqnarray}
\label{Enonrel}
E_K \approx m_f + \frac{K^2}{2m_f} \,.
\end{eqnarray}
We consider in detail two cases separately,
according to whether the electron gas is classical or degenerate.

\subsection{Classical electron gas and classical proton gas}
In this situation we can set
\begin{eqnarray}
\label{classdist}
f_f (E) = e^{-\beta(E-\mu_f)}
\end{eqnarray}
for both $f = e,p$. This implies the relation
\begin{eqnarray}
\label{classrel}
{\partial f_f\over \partial E} = -\beta f_f \,,
\end{eqnarray}
as well as the integration formula
\begin{eqnarray}
\int dK \; K^{2r} f_f = 2 \pi^{3/2} \Gamma \left(r+\frac12 \right) 
\left( {\beta\over 2m_f} \right)^{1-r} n_f\,,
\label{Kint}
\end{eqnarray}
where $n_f$ is the number density, given by
\begin{eqnarray}
n_f = 2 \int {d^3K \over (2\pi)^3} \; f_f(E) \approx 2 \left( {m_f
\over 2\pi\beta} \right)^{3/2} e^{-\beta(m_f-\mu_f)} \,.
\end{eqnarray}

Let us consider $m_{e2}$ and $m_{\bar e2}$, given in Eqs.\ (\ref{me2})
and (\ref{mbare2}), respectively. Setting $f_{\bar e} = 0$ and using
Eq.\ (\ref{Enonrel}) to expand the co-efficients of $f_e$ in the
integrands in powers of $K$, the remaining integrals are evaluated by
means of Eq.\ (\ref{Kint}) to yield
\begin{eqnarray}
\label{me2mbare2class}
m_{e2} & = & - \frac{e^2n_e}{2m_e T} + O(e^2n_e/m_e^2)\,,
\nonumber\\
m_{\bar e2} & = & \frac{3e^2n_e}{8m_e^2} + 
O(e^2n_e T^2/m_e^4)\,.
\end{eqnarray}
(Note that the subleading terms are small because they are suppressed by powers of $T/m_e$,
and $T < m_e$ for the electrons to be non-relativistic.)
Similarly, from Eq.\ (\ref{Xfinal}) we obtain for this case
\begin{eqnarray}
\label{Xclassical}
m'_{(X)} & = & - {7e^2 \over 24T} \sum_{f = e,p} {{\cal Q}_f n_f \over
m_f} 
+ O(e^2n_e T/m_e^3)\nonumber\\*
& \approx & \frac{7e^2n_e}{24m_e T} \,,
\end{eqnarray}
where we have used the charge-neutrality condition which, neglecting
terms $O(\kappa$), is simply $\sum_f{\cal Q}_f n_f=0$.  Applying the
same procedure in Eq.\ (\ref{m'2summary}), the leading contribution,
in powers of $T/m_e$, comes from the $\partial f_e/\partial E_K$ term
in that equation, and leads to
\begin{eqnarray}
m'_{e2} = {e^2 n_e \over 2T^2} + O(e^2n_e/Tm_e) \,.
\end{eqnarray}
By the substitution indicated in Eq.\ (\ref{antiprescription2}),
the corresponding result for the positron is
\begin{eqnarray}
m'_{\bar e2} = - {3e^2 n_e \over 8m_eT} + O(e^2n_e/m^2_e)\,.
\end{eqnarray}
Therefore, using Eqs.\ (\ref{Melldef}), (\ref{Mell'def}) and
(\ref{Mbarell'def}), the inertial and gravitational masses for charged
leptons $\ell$ other than the electron are obtained as
\begin{eqnarray}
M_{\bar\ell} = M_{\ell} & = & m_\ell + {e^2 T^2 \over 12m_\ell} \,, 
\nonumber\\*
M'_{\ell,\bar\ell} &=& m_\ell - {e^2 T^2 \over 12m_\ell} \pm
\frac{7e^2n_e}{24 m_e T}\,,
\end{eqnarray}
where the upper sign corresponds to the leptons and the lower one to
the anti-leptons. The corresponding formulas for the electron are
\begin{eqnarray}
M_{e} & = & m_e + {e^2 T^2 \over 12m_e} - {e^2 n_e \over 2m_eT}\,,
\nonumber\\*
M'_e & = & m_e - {e^2 T^2 \over 12m_e} + {e^2 n_e \over 2T^2} \,,
\end{eqnarray}
and for the positron they are
\begin{eqnarray}
M_{\bar e} &=& m_e + {e^2 T^2 \over 12m_e} + {3e^2 n_e \over 8m_e^2}  \,,
\nonumber\\* 
M'_{\bar e} &=& m_e - {e^2 T^2 \over 12m_e} - {2e^2 n_e \over 3m_eT} \,.
\end{eqnarray}

We now estimate how large these corrections could be for the
electron. Those due to the photon background
were estimated by DHR \cite{DHR84} and were found to be very
small. This is because the fractional changes in the inertial and gravitational masses 
in that case are
$O(e^2T^2/m{_e^2})$, and $T<m_e$. Therefore, neglecting that contribution,
the fractional changes
are given by
\begin{eqnarray}
\label{fracMe}
\left| {M_e - m_e \over m_e} \right| & = & {e^2 n_e \over 2m_e^2T}\\*
\label{fracM'e}
\left| {M'_e - m_e \over m_e} \right| & = & {e^2 n_e \over 2m_eT^2} \,.
\end{eqnarray}
Although it may seem that the effects are more noticeable as the
temperature decreases, they are bounded by the conditions that the
electron gas is non-degenerate and non-interacting,
which require that
\begin{eqnarray}
T > \frac{n_e^{2/3}}{m_e}
\label{classicalcond}
\end{eqnarray}
and 

\begin{eqnarray}
\label{freecond}
T > {e^2 \over r_{\rm av}} \sim e^2 n_e^{1/3} \,,
\end{eqnarray}
since $r_{\rm av} \sim n_e^{-1/3}$ \cite{ll:pk}.  Using the fact that $n_p=n_e$,
it follows that the corresponding conditions for the proton gas do not
imply further restrictions, because they are automatically satisfied
whenever Eqs.\ ({\ref{classicalcond}) and (\ref{freecond}) hold.

By writing the right-hand side of Eq.\ (\ref{fracMe}) in
the alternative forms 
\begin{eqnarray}
\frac{e^2 n_e}{2m^2_eT} & = & \frac{1}{2}\left(\frac{e^2 n^{1/3}_e}{T}\right)
\left(\frac{n^{1/3}_e}{m_e}\right)^2 \nonumber\\
& = & \left(\frac{n_e^{2/3}}{m_e T}\right)^2
\left(\frac{T}{2m_e}\right)\left(\frac{e^2m_e}{n_e^{1/3}}\right)\,,
\end{eqnarray}
it is seen that
\begin{eqnarray}
\left| {M_e - m_e \over m_e} \right| < \left\{
\begin{array}{ll}
e^4/2 & \mbox{if $n_e^{1/3} < e^2 m_e$} \,, \\[12pt]
T/2m_e & \mbox{if $n_e^{1/3} > e^2 m_e$}\,.
\end{array}
\right.
\end{eqnarray}
Similarly, writing 
\begin{eqnarray}
\frac{e^2n_e}{2m_eT^2} & = & \frac{1}{2} \left(\frac{e^2 n^{1/3}_e}{T}\right)^2
\left(\frac{n^{1/3}_e}{e^2m_e}\right) \nonumber\\
& = & \frac{1}{2}\left(\frac{n_e^{2/3}}{m_e T}\right)^2
\left(\frac{e^2 m_e}{n_e^{1/3}}\right) \,,
\end{eqnarray}
it follows that
\begin{eqnarray}
\left| {M'_e - m_e \over m_e} \right| <  \frac{1}{2}
\end{eqnarray}
in either case.
Therefore, while the fractional correction to the electron's
inertial mass is likely to be small in most situations 
with the conditions that we are presently considering,
the fractional change in the gravitational mass 
could be substantial.  For example, if we use the temperature and
density at the solar core, i.e., $T=1.57\times10^7$\,K,
$n_e=9.5\times10^{25}$\,cm$^{-3}$, we obtain
\begin{eqnarray}
\left| {M_e - m_e \over m_e} \right| &=& 9.8 \times 10^{-5}\,,
\nonumber\\ 
\left| {M'_e - m_e \over m_e} \right| &=& 3.5 \times 10^{-2}\,,
\end{eqnarray}
which shows that the correction to the gravitational mass of the
electron can at least be appreciable in realistic physical
situations.

\subsection{Degenerate electron gas and classical proton gas}
For a degenerate electron gas,
\begin{eqnarray}
T < \frac{n_e^{2/3}}{m_e} \sim \frac{K_F^2}{m_e} \,,
\label{degenerate}
\end{eqnarray}
where $K_F$ is the Fermi momentum of the electron gas. We assume
\begin{eqnarray}
K_F < m_e
\label{NRcondition}
\end{eqnarray}
so that the electrons are non-relativistic. We also assume the electron gas to be
non-interacting. This requires
that the average kinetic energy $\sim K{_F^2}/m_e$ of an electron is larger
than the average Coulomb interaction energy $\sim e^2 n_e^{1/3} \sim e^2 K_F$.
(Unlike the case of the classical gas, $T$ contributes negligibly to the
average kinetic energy in the present case; see Eq.\ (\ref{degenerate}).)
This implies that \cite{lifshitz}
\begin{eqnarray}
\label{Freecondition}
K_F > e^2 m_e \,.
\end{eqnarray}
We now show that, under these conditions, the protons can be treated as a 
non-interacting, non-relativistic, classical gas \cite{raffelt:book}.
The only additional condition we have to impose is the condition for
non-interaction for a classical proton gas:
\begin{eqnarray}
T > e^2 n_p^{1/3} \sim e^2 K_F
\label{pfreecond}
\end{eqnarray}
(using $n_p=n_e$). Since Eqs.\ (\ref{degenerate}) and (\ref{NRcondition})
only imply that $T<K_F$, it is still possible to satisfy Eq.\ (\ref{pfreecond}).
Eqs.\ (\ref{degenerate}) and (\ref{NRcondition})
also imply that $T<m_e$. So the non-relativistic condition for the classical proton gas
is automatically satisfied. Finally, since $K_F<m_e\ll m_p$, Eq.\ (\ref{pfreecond})
gives $T\gg e^2K{_F^2}/m_p$, so that the non-degeneracy condition
\begin{eqnarray}
T > \frac{K_F^2}{m_p} \sim \frac{n_p^{2/3}}{m_p} 
\label{pNDcond}
\end{eqnarray}
is satisfied as well.

Therefore, Eq.\ (\ref{classdist})
applies to the proton, while for the electron
\begin{eqnarray}
\label{fedegen}
f_e = \Theta (K_F-K)
\end{eqnarray}
with
\begin{eqnarray}
K_F = (3\pi^2 n_e)^{1/3} \,,
\end{eqnarray}
which in turn imply the relation
\begin{eqnarray}
\label{derfedegen}
{df_e \over dK} = -\delta (K_F-K)\,.
\end{eqnarray}

We repeat the calculation of the quantities $m_{e2}$, $m_{\bar e2}$,
$m'_{e2}$, $m'_{\bar e2}$ and $m'_{X}$ for this case, neglecting the
terms that are a factor $\sim O(K^2_F/m^2_e)$ smaller than the ones
that we retain. {}From Eqs.\ (\ref{me2}) and (\ref{mbare2}), setting
$f_{\bar e} = 0$ and using Eq.\ (\ref{Enonrel}), we obtain
\begin{eqnarray}
\label{me2degen}
m_{e2} & = & - \frac{e^2K_F}{2\pi^2} \,,\nonumber\\
m_{\bar e2} & = & \frac{e^2 K_F^3}{8\pi^2 m_e^2} \,.
\end{eqnarray}
{}From Eq.\ (\ref{Xfinal}),
\begin{eqnarray}
m'_{(X)} = \frac{e^2}{6\pi^2} 
\int dK \Bigg[ f_e - 
\left(K + \frac{m^2_e}{2K}\right) 
\frac{df_e}{dK}\Bigg] - {7e^2 \over 24T} {n_p \over m_p} \,,
\end{eqnarray}
where we have borrowed the result for the proton contribution from 
Eq.\ (\ref{Xclassical}), while in the electron term we have
expressed $E_K$ in terms of $K$ and used 
\begin{eqnarray}
{d \over dE_K} = \frac{E_K}{K}\frac{d}{dK}
\label{EdE=KdK}
\end{eqnarray}
for any function of $E_K$.  Using Eqs.\ (\ref{fedegen}) and
(\ref{derfedegen}) this finally yields
\begin{eqnarray}
\label{Xdegen}
m'_{(X)} = \frac{e^2 m^2_e}{12\pi^2 K_F} \,. 
\end{eqnarray}
Here we have neglected the proton contribution because it is $\sim
e^2K_F^3/(Tm_p) < e^2K_F$ from Eq.\ (\ref{pNDcond}).  In a similar
fashion, from Eq.\ (\ref{m'2summary}),
\begin{eqnarray}
m'_{e2} = \frac{e^2 m^2_e}{2\pi^2 K_F} \,,
\end{eqnarray}
and by the substitution indicated in Eq.\ (\ref{antiprescription2}),
the corresponding result for the positron is
\begin{eqnarray}
m'_{\bar e2} = - \frac{3e^2 K_F}{8\pi^2} \,.
\end{eqnarray}

Thus, substituting these results into Eqs.\ (\ref{Melldef}),
(\ref{Mell'def}) and (\ref{Mbarell'def}), we obtain the following
expressions for the inertial and gravitational masses, retaining only
the leading terms in powers of $K_F/m_e$.  For the charged leptons
$\ell$ other than the electron,
\begin{eqnarray}
\label{Melldegen}
M_{\bar\ell} = M_\ell & = & m_\ell + {e^2 T^2 \over 12m_\ell}
\nonumber\\*
M'_{\ell,\bar\ell} & = & m_\ell - {e^2 T^2 \over 12m_\ell} \pm 
\frac{e^2 m^2_e}{12\pi^2 K_F} \,,
\end{eqnarray}
with the upper sign corresponding to the leptons and the lower one to
the anti-leptons, while for the electron
\begin{eqnarray}
\label{Medegen}
M_e &=& m_e + {e^2 T^2 \over 12m_e} - {e^2 K_F\over
2\pi^2}  \,, \nonumber\\* 
M'_e &=& m_e - {e^2 T^2 \over 12m_e} 
+ \frac{7e^2 m^2_e}{12\pi^2 K_F} \,,
\end{eqnarray}
and for the positron
\begin{eqnarray}
\label{Mbaredegen}
M_{\bar e} &=& m_e + {e^2 T^2 \over 12m_e} + {e^2 K_F^3\over
8\pi^2m_e^2}  \,, \nonumber\\* 
M'_{\bar e} &=& m_e - {e^2 T^2 \over 12m_e} - \frac{e^2 m_e^2}{12\pi^2 K_F} \,.
\end{eqnarray}

It is interesting to note that Eqs.\ (\ref{degenerate}) and
(\ref{NRcondition}) imply that the photon contributions in Eqs.\
(\ref{Melldegen})-(\ref{Mbaredegen}) are much smaller than the
contribution due to the electron background in each case. In fact,
using Eq.\ (\ref{Freecondition}), we see that the fractional
corrections to the gravitational mass can be as large as about
$7/12\pi^2$ for the electron and $1/12\pi^2$ for the positron and the
other leptons.

\chapter{Some aspects of high-temperature QED}\label{high}
\section{Introduction}\label{high:intro}
So far in this work, we have considered low temperatures and densities.
Such thermal backgrounds ensure the validity of a straightforward
expansion in $e^2$, something we have always assumed. A negative illustration of this 
point is provided by Eqs.\ (\ref{Melldef}) and (\ref{m1}) 
for the inertial mass of the charged lepton 
in photon background. It is clear from these equations that 
if $T$ exceeds $m_\ell$ by
a large amount, the change in inertial mass can no longer be treated as a perturbation.

In this chapter, we turn to an entirely different kind of thermal background, namely,
a background at high temperature. More specifically, we will now consider 
high-temperature QED with exact chiral invariance and at zero chemical potential.
In this case, the medium effects are not, in general, of $O(e^2)$ (see, for example,
Sec.\ {\ref{subtle}). Exact chiral invariance means that the full Dirac action of Eq.\ 
(\ref{fullDiracaction}) is invariant under $\psi\rightarrow e^{i\gamma_5\theta}\psi$
for any real constant $\theta$. So, the electron is massless in vacuum, and the self-energy
cannot have the constant term on the R.H.S. of Eq.\ (\ref{qed:form}). (Chiral symmetry
also rules out the $\sigma^{\mu\nu}p_\mu v_\nu$ term in the self-energy, which, anyway,
is absent at one-loop, beyond which we will not be interested in.) Thus, we now have the 
general form
\bea
\Sigma(k)=a\rlap/k +b\rlap/v                                               \label{eq:1}
\eea
for the self-energy of the electron. 
We shall use $k^\mu$ to denote the electron four-momentum in this 
chapter \cite{notation}. The scalars $a$ and $b$ are then functions of
\bea
\omega\equiv k\cdot v\,,~~ K\equiv [(k.v)^2-k^2]^{1/2}
\eea
which satisfy
\bea
k^2=\omega^2-K^2\,.                                                         \label{eq:k^2}
\eea
 The pole in the full propagator is given by
\bea
f(\omega,K)\equiv(\omega-K)(1-a)-b=0\,.                                \label{eq:27a}
\eea
(This is nothing but Eq.\ (\ref{Ucon}) at zero mass. Since we now take the chemical
potential to be zero, $C$ symmetry is unbroken, and the particle and antiparticle
dispersion relations are the same. So, instead of $E_f$ and $E_{\bar f}$, we just
use $\omega$ to denote the energy.) To leading order in $T$ and at one-loop, it was shown by
Weldon in his seminal work \cite{qed:weldon} that
\bea
a&=&-\frac{e^2T^2}{8K^2}\Bigg[1-\frac{\omega}{2K}
         \ln\Bigg(\frac{\omega+K}{\omega-K}\Bigg)\Bigg]\,,\nonumber\\*
b&=&\frac{e^2T^2}{8K}\Bigg[\frac{\omega}{K}-\frac{1}{2}\Bigg(\frac{\omega^2}{K^2}-1\Bigg)
         \ln\Bigg(\frac{\omega+K}{\omega-K}\Bigg)\Bigg]\,.                 \label{aandb}
\eea
Substitution of Eq.\ (\ref{aandb}) into Eq.\ (\ref{eq:27a}) then gives the electron 
dispersion relation
\bea
\omega-K=\frac{e^2T^2}{8K}\Bigg[1+\frac{1}{2}\Bigg(1-\frac{\omega}{K}\Bigg)
        \ln\Bigg(\frac{\omega+K}{\omega-K}\Bigg)\Bigg]\,.                   \label{correct}
\eea
An important point is that if we write 
$\omega-K=b/(1-a)\approx b$ from Eq.\ (\ref{eq:27a}) by neglecting $a^2$ and $ab$, and then 
substitute Eq.\ (\ref{aandb}), we do not get the
correct answer given in Eq.\ (\ref{correct}). This is an example of what we meant by the 
invalidity of a straightforward expansion in $e^2$ at the beginning of this section. 
 
In the same work, Weldon demonstrated the gauge independence of this leading order dispersion 
relation, by showing that gauge dependence appears in the self-energy only at
subleading order in temperature. However, one expects that even this subleading
order gauge dependence should not show up in the dispersion 
relation. We are therefore going to investigate the gauge independence of the 
dispersion relation after
taking the subleading temperature dependence into account \cite{qed:indra}.
\section {One-loop electron self-energy in a general linear
covariant gauge} \label{sec:selfenergy}
$\re\Sigma$ is defined through the equations
\bea
a(\omega, K)&=&a_R(\omega,K)+ia_I(\omega,K)\,,\\*
b(\omega, K)&=&b_R(\omega,K)+ib_I(\omega,K)\,,\\*
\re\Sigma&=&a_R\rlap/k+b_R\rlap/v\,.
\eea
To obtain the dispersion relation, we need to consider only $\re\Sigma$ and, 
consequently, only the ``11" 
propagators, 
at one-loop \cite{real}. As in Chapter \ref{qed}, {\it we will denote $\re\Sigma$ 
by just $\Sigma$}.
Because of the additional features of masslessness of the fermion, zero chemical potential,
and inclusion of gauge dependence in the photon propagator, it will be convenient
for us to rewrite the ``11" propagators for the purpose of the calculations in this 
chapter. Thus, we are going to use
the free-particle massless fermion 
propagator at zero chemical potential given by
\bea
S(p)&=&\rlap/p\Bigg[\frac{1}{p^2 + i\epsilon} +2\pi i \delta(p^2) f_F(p)\Bigg]\,,
                                                                        \label{eq:8}\\
f_F(p)&=&[e^{|p_0|/T} +1]^{-1}\,,                                            \label{eq:9}
\eea
and the free photon propagator in a general covariant gauge
given by
\cite{kowalski}
\bea
D^{\mu\nu}(p)&=& D{^{\mu\nu}_{\rm{FG}}}(p)+ D{^{\mu\nu}_\xi}(p)\,,\\
D{^{\mu\nu}_{\rm{FG}}}(p)&=&-\eta^{\mu\nu}
      \Bigg[\frac{1}{p^2 + i\epsilon} -2\pi i \delta(p^2) f_B(p)\Bigg]\,,
                                                                         \label{eq:11}\\
 D{^{\mu\nu}_\xi}(p)&=&-\xi p^\mu p^\nu \Bigg[\frac{1}{(p^2 + i\epsilon)^2}
       +2\pi i f_B(p)\frac{ d\delta(p^2)}{dp^2}\Bigg]\,,                  \label{eq:12}\\
f_B(p)&=&[e^{|p_0|/T} -1]^{-1}\,.                                          \label{eq:13}
\eea
Here $\xi$ is the gauge parameter and FG denotes the Feynman gauge ($\xi=0$).
Since {\it the calculation will be performed in the rest-frame of the medium}, 
we have used $p_0$ in the place of $p\cdot v$ in
Eqs.\ (\ref{eq:9}) and (\ref{eq:13}). In this frame, we define the components
of the electron four-momentum by writing
\bea
k^\mu=(k^0,\vec K)\,.
\eea
Note that this choice of frame will not lead to any loss of generality in the proof
of gauge independence, since the dispersion relation is Lorentz invariant.

Let us write
\bea
\Sigma=\Sigma^{T=0} + \Sigma^\prime.                         \label{eq:sigmas}
\eea
We now evaluate the real parts of these functions 
by putting the expressions
for the fermion
and the photon propagators in the expression for the self-energy given by 
Eq.\ (\ref{Sigmap}).
Let us first consider the $T=0$ part
\bea
\Sigma^{T=0}=a^{T=0}(k^2)\rlap/k\,,               \label{eq:sigmazero}
\eea
and write $a^{T=0}=a{^{T=0}_{\rm FG}}+a{^{T=0}_\xi}$. 
Dimensional regularization of $\Sigma{^{T=0}_{\rm FG}}$
in $4-\epsilon^\prime$ dimensions gives
\bea
a{^{T=0}_{\rm FG}}=-\frac{e^2}{(4\pi)^2}\Bigg[\frac{2}{\epsilon^\prime}
               -\gamma-1+\ln(4\pi)-2\int_0^1 dx\:x \ln\Big(x(1-x)k^2\Big)
               +O(\epsilon^\prime)\Bigg]
\eea
where $\gamma$ is the Euler-Mascheroni constant.
On the other hand, one easily obtains
\bea
\Sigma{^{T=0}_{\xi}}=
-i\xi e^2\int\frac{d^4p}{(2\pi)^4}\frac{1}{(p^2+i\epsilon)^2}
      \Bigg[\rlap/p - \frac{p^2 \rlap/k +k^2 \rlap/p}
      {(p+k)^2+i\epsilon}\Bigg].
\eea
The odd $\rlap/p$ term vanishes on integration. Evaluation of the rest
leads to
\bea
a{^{T=0}_{\xi}}=-\frac{\xi e^2}{(4\pi)^2}\Bigg[\frac{2}{\epsilon^\prime}
               -\gamma-1+\ln(4\pi)-\int_0^1 dx\: \ln\Big(x(1-x)k^2\Big)
               +O(\epsilon^\prime)\Bigg].
\eea
Finally adding the counter-term to $a^{T=0}$, as fixed by the
renormalization condition
\bea
a{^{T=0}_{ren}}(k^2=\sigma^2)=0,
\eea
we arrive at
\bea
a{^{T=0}_{ren}}(k^2)=(1+\xi)\frac{e^2}{(4\pi)^2}\ln\frac
                               {k^2}{\sigma^2}.      \label{eq:azero}
\eea
where $\sigma$ is the renormalization scale.
 
We now turn to the $T$-dependent part, and write
\bea
\Sigma'=\Sigma'_{\rm{FG}} + \Sigma'_\xi.
\eea
To obtain $\Sigma'_{\rm FG}$, one has to put Eqs.\ (\ref{eq:8})
and (\ref{eq:11}) in the expression (\ref{Sigmap}) for the self-energy, and
consider the relevant terms.  It is then convenient to change $p$ to $-p-K$ in the 
$f_F$-containing
term. Finally setting $p^2=0$, as allowed by $\delta(p^2)$,
yields
\bea
\Sigma'_{\rm FG}= 2e^2\int\frac{d^4p}{(2\pi)^4}
     [(\rlap/p+\rlap/k)f_B(p) + \rlap/p f_F(p)]
     2\pi\delta(p^2)\re\frac{1}{2p.k+k^2+i\epsilon}\,.    \label{eq:16}
\eea
Putting (\ref{eq:8}) and (\ref{eq:12}) in (\ref{Sigmap}), one arrives at
$\Sigma'_\xi$. Its $f_F$-containing part and
$f_B$-containing part will be indicated explicitly:
\bea
\Sigma'_\xi= \Sigma'_{\xi,F} + \Sigma'_{\xi,B}.                 \label{eq:17}
\eea
In $\Sigma'_{\xi,F}$, changing $p$ to $-p-K$ and then setting
$p^2=0$ (allowed by the delta function) gives us
\bea
\Sigma'_{\xi,F} = \xi e^2\int\frac{d^4 p}{(2\pi)^4}
    [k^2 \rlap/p-2p.k\rlap/k] 2\pi\delta(p^2)f_F(p)
    \re\frac{1}{(2p.k + k^2 +i\epsilon)^2}\,.
                                                               \label{eq:21}
\eea
In $\Sigma'_{\xi,B}$, simplification leads to
\bea
\Sigma'_{\xi,B}=\xi e^2\int\frac{d^4p}{(2\pi)^3}f_B(p)
                  \frac{d\delta(p^2)}{dp^2}
      \Bigg[\rlap/p - (p^2 \rlap/k +k^2 \rlap/p)\re\frac{1}
      {(p+k)^2+i\epsilon}\Bigg].                          \label{eq:simple}
\eea
The $\rlap/p$ term, being odd, drops out on integration. 
To deal with the remaining part, we first write down
\bea
\frac{d\delta(p^2)}{dp^2}=\lim_{\epsilon\rightarrow 0}\frac{1}{2\pi i}
             \Bigg[\frac{1}{(p^2+i\epsilon)^2}
             -\frac{1}{(p^2-i\epsilon)^2}\Bigg]
\eea
and then use the regularization
\bea
\frac{1}{(p^2\pm i\epsilon)^2}=\lim_{\lambda\rightarrow 0}\frac{\partial}
{\partial\lambda^2}\frac{1}{p^2-\lambda^2\pm i\epsilon}           \label{eq:reg}
\eea
so that
\bea
\frac{d\delta(p^2)}{dp^2}=-\lim_{\lambda\rightarrow 0}\frac{\partial}
{\partial\lambda^2}\delta(p^2-\lambda^2).                       \label{eq:delta}
\eea
Let us now make use of Eq.\ (\ref{eq:delta}) in  Eq.\ (\ref{eq:simple}). We then commute the
integration over $p_0$ with the limit and the differentiation
involving $\lambda$ , and set $p^2=\lambda^2$
(allowed by the delta function). This gives
\bea
\Sigma'_{\xi,B} &=& \xi e^2 \int\frac{d^3P}{(2\pi)^3}
      \lim_{\lambda\rightarrow 0} \frac{\partial}{\partial\lambda^2}\int dp_0
      f_B(p)\delta(p^2-\lambda^2)(k^2\rlap/p+\lambda^2\rlap/k)\nonumber\\* 
      &&\times\re\frac{1}{k^2+2p.k+\lambda^2+i\epsilon}.
\eea
After integrating over $p_0$, the operations involving $\lambda$ are carried out.
While this last step is easily performed for the part proportional to $\lambda^2$
(since
\bea
\lim_{\lambda\rightarrow 0}\frac{\partial}{\partial\lambda^2}
(\lambda^2 f(\lambda^2)) = f(0)                 \label{trick}
\eea
with $f$ denoting a function with finite $\partial f/ \partial\lambda^2$
at $\lambda=0$ \cite{footnote3}), a more tedious algebra is to be worked out for the
remaining part, finally giving 
\bea
\Sigma'_{\xi,B} &=& \xi \frac{e^2}{2}\int\frac{d^3P}
       {(2\pi)^3}\frac{1}{e^{\beta P}-1}\frac{1}{P}
      \Bigg[\Bigg(\rlap/k+\frac{k^2\vec{P}.\vec{\gamma}}{2P^2}\Bigg)
      \re\Bigg(\frac{1}{D_+}+\frac{1}{D_-}\Bigg)    \nonumber  \\*
      &&-k^2 P{\gamma}_0 \re\Bigg(\frac{1+k_0/P} {{D_+}^2}
      -\frac{1-k_0/P}{{D_-}^2}\Bigg) 
      +k^2 \vec{P}.\vec{\gamma} \re\Bigg(\frac{1+k_0/P}{{D_+}^2}
      +\frac{1-k_0/P}{{D_-}^2}\Bigg)\Bigg]    \nonumber   \\*
      &&- \xi \frac{e^2\beta}{4}\int\frac{d^3P} {(2\pi)^3}
      \frac{e^{\beta P}}{(e^{\beta P}-1)^2}\frac{1}{P^2}
      \Bigg[k^2P\gamma_0 \re\Bigg(\frac{1}{D_+}-\frac{1}{D_-}\Bigg
       )\nonumber\\* 
      &&-k^2\vec{P}.\vec{\gamma}\re\Bigg(\frac{1}{D_+}+\frac{1}{D_-}\Bigg)
      \Bigg]\,.                                                      \label{eq:23}
\eea
Here
\bea
D{_\pm} = k^2\pm 2Pk_0 - 2\vec P.\vec K + i\epsilon.
\eea
While the expressions (\ref{eq:16}) and (\ref{eq:21}) are the same as those in
Ref.\ \cite{qed:weldon} (except that we have carefully incorporated the $i\epsilon$,
 which will be needed for the calculations in Appendix \ref{app:masseqn}), the
expression (\ref{eq:23}) is a different one. We have found this form of
$\Sigma{^\prime_{\xi,B}}$ convenient for actually carrying out the
integration (this, again, is done in Appendix \ref{app:masseqn}).

Let us now investigate the high-$T$ behaviour of (\ref{eq:16}), (\ref{eq:21})
and (\ref{eq:23}). Following Ref.\ \cite{qed:weldon}, this behaviour can be inferred
from the degree of ultraviolet divergence of the integral in each
expression in the absence of the cut-off of $O(T)$ provided by $f_B$ or $f_F$
or $e^{\beta P}/(e^{\beta P} - 1)^2$. One then finds that
$\Sigma{^\prime_{FG}}$ goes like $T^2$, while
$\Sigma{^\prime_{\xi,F}}$ and $\Sigma{^\prime_{\xi,B}}$
go like $T$. It may be noted that one cannot obtain the correct high-$T$
behaviour of $\Sigma{^\prime_{\xi,B}}$ without removing the regulator
$\lambda$, and this can be done only after performing the $p_0$-integration,
as in (\ref{eq:23}).

At high $T$, one can neglect the renormalized $\Sigma^{T=0}$
compared to $\Sigma^\prime$ \cite{weldon89}. Actually $\Sigma^{T=0}$
depends on the renormalization scale $\sigma$, but since the dependence is
logarithmic, we can still ignore it vis-a-vis the power law dependences on $T$
in the various parts of $\Sigma^\prime$. We shall, however, see that
it is {\it not necessary} to neglect $\Sigma^{T=0}$ for proving
gauge independence at $K\gg eT$. We shall also use (\ref{eq:azero}) to
incorporate the $\sigma$-dependence in the equation for the effective mass.
\section {Equation governing gauge dependence of dispersion relation at one loop,
          and general considerations} \label{sec:governingeqn}
Inverting Eq.\ (\ref{eq:1}), one obtains 
\bea
a&=&\frac{1}{4K^2}[-\Tr(\rlap/k\Sigma)+\omega\Tr(\rlap/v\Sigma)]\,,
                                                                     \label{eq:24}\\*
b&=&\frac{1}{4K^2}[\omega \Tr(\rlap/k\Sigma)-
        (\omega^2-K^2) \Tr(\rlap/v\Sigma)]\,.                      \label{eq:25}
\eea
The dispersion relation is obtained by putting $a$, $b$ in Eq.\ (\ref{eq:27a}).
Let us now write the function $f$ in Eq.\ (\ref{eq:27a}) as
\bea
f&=&f_{\rm{FG}}+f_{\xi},\\*
f_{\rm{FG}}&=&(\omega-K)(1-a_{\rm{FG}})-b_{\rm{FG}},\\*
f_{\xi}&=&-(\omega-K)a_{\xi}-b_{\xi}.                      \label{eq:29}
\eea
Next using Eqs.\ (\ref{eq:17}), (\ref{eq:21}) and (\ref{eq:23}), one readily sees
 that $k^2$ factors out from the expression for
$\Tr(\rlap/k \Sigma{^\prime_\xi})$.
Therefore in view of Eq.\ (\ref{eq:k^2}), one can write from Eq.\ (\ref{eq:25}) that
\bea
b_\xi=(\omega-K)(\mbox{1-loop function of $\omega$, $K$})
\eea
(note that $b$, being zero at $T=0$, is determined by the $\Sigma^\prime$ part only).
Consequently
\bea
f_\xi= -(\omega-K)[a_\xi + (\mbox{1-loop function of $\omega$, $K$})]
           \label{eq:gov}.
\eea
This is the equation governing the gauge-dependence of the electron dispersion
relation at one-loop. This equation is of the form
\bea
(f_\xi)_{\rm{1-loop}}=f_{\rm{tree}}\times\mbox{[1-loop $\xi$-dependent function]}.
                        \label{eq:gov1}
\eea
As we shall see in the next section, {\it the fact that $f_{\rm tree}$
factors out on the R.H.S. of Eq.\ (\ref{eq:gov1}) is
crucial to the proof of gauge independence.}

Ref.\ \cite{kkr} contains a general, nonperturbative derivation of the identities
determining the gauge dependence of the gluon dispersion relations (see Eqs. (16)
and (17) of \cite{kkr}). To one loop these gauge dependence identities are shown to
reduce to relations (see Eq. (20) of \cite{kkr}) which are analogous to Eq.
(\ref{eq:gov1}) above. {\it It is therefore likely that a general
gauge dependence identity for the electron dispersion relation,
reducing to Eq.\ (\ref{eq:gov1}) at one loop, can be derived and used to
arrive at a general proof of gauge independence of the dispersion relation}
(as in the gluon case).
In this paper,
however, we confine ourselves to one-loop calculations.
\section {Gauge independence of one-loop dispersion relation at momenta much
larger than $\lowercase{e}T$}                 \label{sec:proof-for-high}
Since we shall consider $K\gg eT$ in this section, let us first note that
for {\it sufficiently small e} there always exists a
domain $eT\ll K\ll T$, so that the restriction $K\gg eT$ does not
contradict the
high-$T$ approximation.

The basic premise for the considerations of this section is that when the 
finite-temperature effects are small, the dispersion
relation should involve a deviation from $\omega=K$ primarily by powers of $e^2$.
We have used the phrase ``primarily" because the terms in the deviation may
come multiplied with powers of
${\rm ln}(1/e)$ (originating from powers of ${\rm ln}(\omega-K)$; see, for example,
Eqs.\ (\ref{aandb}) and (\ref{correct})). Now, to first order in $e^2$, Eq.\
(\ref{correct}) becomes $\omega-K=e^2T^2/8K$. This, when used in the remaining term
on the R.H.S. of Eq.\ (\ref{correct}), results in \cite{qed:weldon}
\bea
\omega=K+\frac{{M_0}^2}{K}-\frac{{M_0}^4}{2K^3}\ln\frac{2K^2}
                 {{M_0}^2}+ \cdots   \label{eq:27}
\eea
where ${M_0}^2=e^2T^2/8$. The terms on the R.H.S. of the above expansion clearly indicate 
that the expansion is valid for $K\gg M_0$ i.e. $K\gg eT$. This can be understood in
the following way.
As shown in Ref.\ \cite{qed:weldon} and as we shall see in Sec.\ \ref{sec:proof-for-mass},
$M_0$ is the leading-order effective
 electron mass \cite{symbol} and thus can be considered a measure of
finite-temperature effects. So, $K\gg eT$ ensures that the finite-temperature
{\it effects}
are small (even though $T$ is large), and only then the expansion in $e^2$ given by
Eq.\ (\ref{eq:27}) is valid. (As we will see in Sec.\ \ref{subtle}, such an expansion
does not exist for $K\ll eT$.)

Now we turn to gauge independence.
The expected general form of the one-loop dispersion relation for $K\gg eT$
{\it when the terms subleading in $T$ are kept} is
\bea
\omega=K+e^2f_1(e,K,T)                        \label{eq:26}
\eea
{\it where the $e$-dependence of $f_1$ involves only powers of $\:\ln(1/e)$}.
Equation (\ref{eq:27}) (without the last term on its R.H.S., which is actually of
order $e^4 \ln(1/e)$) is a special case of Eq.\ (\ref{eq:26}),
where $f_1$ turned out
to be totally $e$-independent. Let us now take Eq. (\ref{eq:26}) to be the
relation in the Feynman gauge i.e. assume that it satisfies $f_{\rm FG}=0$.
Then to prove that (\ref{eq:26}) is gauge independent we have to show that
it also satisfies $f_{\rm FG}+f_\xi=0$. It suffices to show that, at
(\ref{eq:26}), $f_\xi$ is of order $e^4$ (with or without
powers of $\ln(1/e)$) \cite{similar}.
This follows readily
from Eq.\ (\ref{eq:gov}), since the portion of the R.H.S. of Eq.\ (\ref{eq:gov})
within the square brackets involves terms of the order of $e^2$ and
$e^2({\rm ln}(1/e))^n$ ($n$ is an integer), and so also does
$(\omega-K)$ at (\ref{eq:26}).

Digressing briefly from {\it high-temperature} QED, we mention that {\it the
above demonstration of gauge independence is actually sufficient also for the
case which involves} not large $T$ but {\it large momenta}. Thus for $K\gg T$,
a one-loop relation of the key form (\ref{eq:26}) should still hold. Note that
the rest of the proof, namely, arriving at Eq.\ (\ref{eq:gov}), did not assume large
$T$ (in particular, we did not neglect the vacuum contribution to $a$). The
proof of gauge independence continues to be valid if, in addition, we have a
chemical potential $\mu$ such that $K\gg \mu$ as well. Then the relevant changes are
that $f_1$ in (\ref{eq:26})
depends on $\mu$ as well, $f_F$ is modified and we have $f_F(-p)$ in place of
$f_F(p)$ in (\ref{eq:21}), but none of these affect the proof outlined above.
\section{Gauge independence of effective mass at one-loop}
\subsection{Subtleties at zero momentum}\label{subtle}
The effective electron mass is the value of $\omega$ at $K=0$. 
(Here we are following the nomenclature used in much of the literature on
high-temperature quantum field theory, including Ref.\ \cite{qed:weldon},
though this quantity has been called the inertial mass earlier in this work.)
The analysis of
the previous section does not prove the gauge independence of the effective mass
because the form (\ref{eq:26}) does not hold near $K=0$. For example, to leading order
in $T$, the one-loop dispersion relation for $K\ll eT$ is \cite{qed:weldon}
\bea
\omega=M_0+\frac{K}{3}+\frac{K^2}{3M_0}+\cdots                  \label{smallK}
\eea
(This is obtained by substituting in Eq.\ (\ref{correct}) a trial solution of
the form (\ref{eq:smallk}) and determining the coefficients.) 
Eq.\ (\ref{smallK}) suggests that 
{\it the leading order values of $a$ and $b$ are not of $O(e^2)$}
(with or without powers of ${\rm ln}(1/e)$) {\it in the $K\ll eT$ limit}; indeed the values
at $K=0$ are $a=-1/3$ and $b=-2M_0/3$ \cite{footnote2}. 
This apparently surprising behaviour can be
understood in a simple way,
without performing a detailed calculation, as follows.

At leading order $a$ and $b$ are $e^2 T^2$ times some function of $\omega$ and $K$
($T^2$ being deduced from ultraviolet power counting in $\Sigma^\prime$).
Therefore, to have the correct dimensions, $a_{K=0}\sim e^2 T^2/{M_0}^2$
and $b_{K=0}\sim e^2 T^2/M_0$ (remembering $M_0=\omega_{K=0}$ at leading order).
Putting these in the dispersion relation (\ref{eq:27a}) at $K=0$, namely,
\bea
M_0-M_0 a_{K=0} -b_{K=0}=0,
\eea
gives us 
 $M_0\sim eT$. Using this, it follows that
$a_{K=0}$ is of $O(1)$ and $b_{K=0}$ is of $O(eT)$.
\subsection{Formula for the effective mass} \label{app:massformula}
The traces $\Tr(\rlap/k \Sigma)$ and
$\Tr(\rlap/v \Sigma)$ are functions of
$\omega$ and $K$. The expressions for them are obtained from
Eqs.\ (\ref{eq:sigmazero}), (\ref{eq:azero}), (\ref{eq:16}),
(\ref{eq:21}) and (\ref{eq:23}). First of all we show that
these traces are both even functions of $K$. For the $\Sigma^{T=0}$ part,
 this is obvious from (\ref{eq:k^2}). For the $\Sigma^\prime$ part, first note that
$k^\mu$ can occur in the expressions for the traces only through $k\cdot p$,
$k\cdot v$ and $k^2$ ($p^\mu$ being the integration variable). Next, let us go to
the rest frame of the medium (in which $K$ is the magnitude of the three-momentum)
and note that
$\vec K$ can now occur in the expressions for the traces only
through $\vec K.\vec P$ and $K^2$. 
Odd power of $K$ can come from
$\vec K.\vec P=KP\cos\theta$ where $\theta$ is the angle between
$\vec K$ and $\vec P$. This is also the only place where $\theta$ occurs
in the integrand. So changing $\theta$
to $\pi-\theta$ together with the change $K\rightarrow -K$, we establish
the even nature as a function of $K$.

For small $K$, therefore, we can write
\bea
\frac{1}{4}\Tr(\rlap/k \Sigma)
&=&h_0+h_1K^2+h_2K^4+\cdots                        \label{smallktrace}\\*
\frac{1}{4}\Tr(\rlap/v \Sigma)
&=&g_0+g_1K^2+g_2K^4+\cdots
\eea
where $h_i$ and $g_i$ are functions of $\omega$. Substitution of the above
expressions into Eqs.\ (\ref{eq:24}) and
 (\ref{eq:25}) gives
\bea
a&=&-\frac{1}{K^2}(h_0 -\omega g_0) -(h_1 -\omega g_1) +O(K^2),
                     \label{eq:f1}\\*
b&=&\frac{\omega}{K^2}(h_0 -\omega g_0) +
  (\omega h_1 -\omega^2 g_1+g_0) +O(K^2).
                      \label{eq:f2}
\eea
For the form factors $a$ and $b$ to remain analytic at $K=0$, the
relation
\bea
h_0=\omega g_0      \label{eq:f3}
\eea
 must hold. We now put (\ref{eq:f1})
and (\ref{eq:f2}), subject to the constraint (\ref{eq:f3}), in
the dispersion relation (\ref{eq:27a}). Then we use the fact that for
any small $K$, the solution
\bea
\omega=M+\omega_1 K+\omega_2 K^2+\cdots    \label{eq:smallk}
\eea
where $M$ and $\omega_i$ are constants, must satisfy the dispersion
relation. We also expand $h_i$ and $g_i$
in $K$ by first doing a Taylor expansion of them around
 $\omega=M$ as functions of $\omega$, and then putting
(\ref{eq:smallk}). Thus
\bea
h_i(\omega)=h_i(M)+h{^\prime_i}(M)\omega_1 K+O(K^2)   \label{eq:47}
\eea
with a similar equation for $g_i$. The equation that now results from
the dispersion relation (\ref{eq:27a}) has a series in powers of $K$
alone on the L.H.S. Equating the coefficient of each power of $K$
separately to zero will determine the constants $M$ and $\omega_i$
of (\ref{eq:smallk}). Thus,
equating the constant term to zero gives
\bea
M=g_0(M).         \label{eq:50}
\eea
Multiplying both sides by $M$, we make use of
\bea
h_0(M)=Mg_0(M)              \label{eq:48}
\eea
(which follows from (\ref{eq:f3})) to finally arrive at
\bea
M^2=h_0(M).                 \label{eq:51}
\eea
One can use either (\ref{eq:51}) or (\ref{eq:50}) as the formula for the effective mass.
We shall use (\ref{eq:51}), which, by using Eq.\ (\ref{smallktrace}),
can be written as \cite{erdas}:
\bea
M^2=\frac{1}{4}\lim_{K\rightarrow 0,\omega\rightarrow M}
           \Tr[\rlap/k\Sigma]\,.                           \label{eq:31}
\eea

We comment that one should not attempt to compare this formula with the formula
for the inertial mass of Sec.\ \ref{qed:dis}, since the latter was deduced by
assuming a straightforward expansion in $e^2$ and neglecting various terms accordingly.
\subsection {Investigation of gauge independence}   \label{sec:proof-for-mass}
Let us now substitute the expression for the self-energy in Eq.\ (\ref{eq:31}).
On using Eqs.\ (\ref{eq:sigmas}) and (\ref{eq:sigmazero}) in Eq.\ (\ref{eq:31}), we obtain
\bea
M^2&=&M^2a^{T=0}(M^2)+{M^\prime}^2,             \label{eq:totalmass}\\*
{M^\prime}^2&\equiv&\frac{1}{4}\lim_{K\rightarrow 0,\omega\rightarrow M}
           \Tr[\rlap/k\Sigma^\prime].      \label{eq:Tmass}
\eea
At $T=0$, $M^\prime=0$ and so Eq.\ (\ref{eq:totalmass}) is correctly satisfied by
$M=0$.

Putting in Eq.\ (\ref{eq:Tmass}) the expressions for $\Sigma{^\prime_{\rm{FG}}}$,
$\Sigma^\prime_{\xi,F}$ and $\Sigma^\prime_{\xi,B}$ given by Eqs.\ (\ref{eq:16}),
(\ref{eq:21}) and (\ref{eq:23}), and noting that
the limit $\omega\rightarrow M$ translates to $k_0\rightarrow M$
in the rest-frame of the medium, we obtain
\bea
{M^\prime}^2&=&M{^\prime_{\rm{FG}}}^2+M{^\prime_{\xi,F}}^2+
             M{^\prime_{\xi,B}}^2,         \label{eq:three}\\*
M{^\prime_{\rm{FG}}}^2&=&2e^2\int\frac{d^4p}{(2\pi)^3}[(p_0+M)f_B(p)+p_0f_F(p)]
               \delta(p^2)\re\frac{1}{2p_0+M+i\frac{\epsilon}{M}},
                                         \label{eq:33}\\*
M{^\prime_{\xi,F}}^2&=&-\xi e^2 M\int\frac{d^4p}{(2\pi)^3}p_0 f_F(p)
               \delta(p^2)\re\frac{1}{(2p_0+M+i\frac{\epsilon}{M})^2},
                                         \label{eq:34}\\*
M{^\prime_{\xi,B}}^2&=& M{^\prime_{\xi,B({\rm I})}}^2+ M{^\prime_{\xi,B({\rm II})}}^2+
                      M{^\prime_{\xi,B({\rm III})}}^2,
                                       \label{eq:mb}\\*
M{^\prime_{\xi,B({\rm I})}}^2&=&\frac{\xi e^2 M}{2}\int\frac{d^3P}{(2\pi)^3}
                \frac{1}{P}\frac{1}{e^{\beta P}-1}\re
                \Bigg(\frac{1}{E_+}+\frac{1}{E_-}\Bigg),   \label{eq:mb1}\\*
M{^\prime_{\xi,B({\rm II})}}^2&=&-\frac{\xi e^2 M}{2}\int\frac{d^3P}{(2\pi)^3}
                \frac{1}{e^{\beta P}-1}\re
                \Bigg[\frac{M}{P}\Bigg(\frac{1}{{E_+}^2}+\frac{1}{{E_-}^2}\Bigg)
                +\Bigg(\frac{1}{{E_+}^2}-\frac{1}{{E_-}^2}\Bigg)\Bigg]\,,\nonumber\\*
                                                         \label{eq:mb2}\\*
M{^\prime_{\xi,B({\rm III})}}^2&=&-\frac{\xi e^2 M^2\beta}{4}\int\frac{d^3P}{(2\pi)^3}
                \frac{1}{P}\frac{e^{\beta P}}{(e^{\beta P}-1)^2}\re
                \Bigg(\frac{1}{E_+}-\frac{1}{E_-}\Bigg).
                                                \label{eq:mb3}
\eea
Here
\bea
E_\pm=M\pm2P+\frac{i\epsilon}{M}.
\eea

We now investigate the high-$T$ behaviour of the three terms on the R.H.S. of
Eq.\ (\ref{eq:three}). This can be arrived at from the ultraviolet behaviour, as explained
towards the end of Sec. \ref{sec:selfenergy}. However there is a
constraint that each term on the R.H.S. of Eq.\ (\ref{eq:three}) is an even function
of $M$. (This can be seen by changing
$p_0$ to $-p_0$ together with the change $M\rightarrow -M$ in (\ref{eq:33}),
in (\ref{eq:34}) and in
\bea
M{^\prime_{\xi,B}}^2=-\xi e^2 M^2\int\frac{d^4p}{(2\pi)^3}
             (p^2+p_0 M)f_B(p)\frac{d\delta(p^2)}{dp^2}
              \re\frac{1}{p^2+2p_0M+M^2+i\epsilon},
\eea
the last equation having been obtained by putting (\ref{eq:simple})
in (\ref{eq:Tmass}).) These considerations,
plus dimensional analysis, tell us that at high $T$, $M{^\prime_{\rm FG}}^2$ goes like
$T^2$, $M{^\prime_{\xi,F}}^2$ like $M^2(\ln|T/M|)^{n_1}$
and $M{^\prime_{\xi,B}}^2$ like $M^2(\ln|T/M|)^{n_2}$
($n_1$, $n_2$ being positive integers).  Since after $T^2$, the next
allowed term in $M{^\prime_{\rm FG}}^2$ is $M^2(\ln|T/M|)^{n_3}$ ($n_3$ a
positive integer), the general expression for ${M^\prime}^2$ at high $T$ can be written as
\bea
{M^\prime}^2=e^2\Bigg[c_0 T^2 +c_3 M^2 \Bigg(\ln\frac{T}{M}\Bigg)^{n_3}
     +\xi M^2 \Bigg(c_1\Bigg(\ln\frac{T}{M}\Bigg)^{n_1}
     +c_2\Bigg(\ln\frac{T}{M}\Bigg)^{n_2}\Bigg)\Bigg]
          \label{eq:39}
\eea
where $c_0$, $c_1$, $c_2$, $c_3$ are constants, and we have dropped the modulus of the
 arguments of the logarithms since $T/M$ is positive. Terms independent
of $T$ have been neglected on the R.H.S. of Eq.\ (\ref{eq:39}).

In principle there could be further constraints ruling out some term(s)
in Eq.\ (\ref{eq:39}), but a detailed calculation, described in Appendix \ref{app:masseqn},
reveals that all these terms are indeed present and that $n_1=1=n_2=n_3$.
Thus we actually have the high temperature equation
\bea
{M^\prime}^2=e^2\Bigg[\frac{ T^2}{8} +\frac{M^2}{8\pi^2}\ln\frac{T}{M}
     +\xi\frac{ M^2}{8\pi^2}\ln\frac{T}{M}\Bigg].              \label{eq:40}
\eea
The detailed calculation may also be viewed as a check on the argument
involving the degree of ultraviolet divergence mentioned before.

Using Eqs.\ (\ref{eq:azero}) and (\ref{eq:40}) in Eq.\ (\ref{eq:totalmass}), we then have
\bea
M^2=e^2\Bigg[\frac{T^2}{8}+(1+\xi)\frac{M^2}{8\pi^2}\ln\frac{T}{\sigma}\Bigg].
                                     \label{eq:newmass}
\eea
On the R.H.S. of Eq.\ (\ref{eq:newmass}) (as also in Eq.\ (\ref{eq:40})),
we have not given the term $\sim e^2 M^2$; all other terms contributing
to the R.H.S. of Eq.\ (\ref{eq:newmass}) (and (\ref{eq:40})) vanish in the limit
of large $T/M$. It is interesting to note that in Eq.\ (\ref{eq:newmass}), the
$\ln(M/\sigma)$ term from $\Sigma^{T=0}$ and the
$\ln(T/M)$ term from $\Sigma^\prime$ have exactly combined to yield
just a $\ln(T/\sigma)$ term, because it has been observed that similar
combination also takes place in the case of the gauge boson self-energy
in the Yang-Mills theory \cite{weldon2}. A discussion of similar behaviour
in the case of three-point function, and, in general, $N$-point function
of gauge boson in the Yang-Mills theory, is to be found in
Ref.\ \cite{brandt}.

The first important observation from Eq.\
(\ref{eq:newmass}) is that $M^2=e^2T^2/8$ to leading order in $T$, which, of course,
is a well-known result \cite{qed:weldon}. Now, the gauge dependent part of
$\Sigma^\prime$ goes like $T$.
Therefore in view of Eq.\ (\ref{eq:31}), one
would expect an $e^2\xi MT$ term in Eq.\ (\ref{eq:newmass}). The absence of such
a term shows that {\it the part of $\Sigma{^\prime_\xi}$ leading in
$T$ does not contribute to $M$}.

To quantify the effect of this absence let us consider the equation
\bea
M^2=e^2(\frac{T^2}{8}+ cMT)                   \label{eq:41}
\eea
where $c$ is a constant. To leading order in $T$, the second term on
 the R.H.S. is negligible, so that $M=O(eT)$.
 This can now be used as an approximation in the second term, to give
\bea
M^2=e^2\frac{T^2}{8}[1+O(e)]
\eea
showing that there is a correction of $O(e^2T)$ to $M$. (This conclusion
 can also be arrived at by solving Eq.\ (\ref{eq:41}) for $M$.) So the absence
of the $e^2 MT$ term means that {\it there is no $O(e^2T)$ correction to the effective mass
in any linear covariant gauge}. It may be noted that the consequences of the
presence or absence of various {\it powers} of $T$ in Eq.\ (\ref{eq:newmass})
are to be taken seriously despite the presence of the scale $\sigma$, since
the $\sigma$ dependence is only logarithmic and is not multiplied with
 any power of $T$.

Finally, (\ref{eq:newmass}) shows that only the subleading $\ln T$ dependence of
$\Sigma{^\prime_\xi}$ contributes to $M$. Using the leading order
result $M=O(eT)$ as before to approximate the remaining terms on the R.H.S. of 
Eq.\ (\ref{eq:newmass}),
we infer that the $\xi$-dependence in Eq.\ (\ref{eq:newmass}) is $O(e^4 T^2)$,
apart from the logarithm.
As one certainly expects $O(e^4T^2)$ contribution to Eq.\ (\ref{eq:newmass}) 
from two loops, {\it it is possible that the $\xi$-dependence which we have
 obtained will be canceled by two-loop contribution}. But it seems that this
issue can be decided only by an actual two-loop calculation. It is however clear
that an $e^2\xi MT$ term, being of $O(e^3T^2)$, was less likely to be canceled
by two-loop contribution. This shows the significance of the absence of such a term.

\chapter{Conclusions}\label{conclu}
In this thesis, we studied some aspects of particle physics in the
presence of a background medium, using the real-time formulation of thermal
field theory. We
studied the $O(e^2)$ corrections to the gravitational
interactions of a charged lepton in the presence of a matter
background. We also investigated gauge independence of one-loop electron
dispersion relation in high-temperature QED. These calculations complement
and extend previous calculations along similar lines, in various useful ways.
The usefulness, as we will explain, is both theoretical and phenomenological.

We first discuss the importance of the various aspects of our 
work from a theoretical point of view. The 
calculation of gravitational couplings in a medium, as presented by us,
employed several thermal field theoretic
techniques that can be useful in other contexts also. For example, we made
explicit use of only the 1PI diagrams
in our calculation of the gravitational mass.
The effect of the one-particle reducible diagrams were
taken into account by choosing each external spinor to be the solution of the
effective Dirac equation in a medium, and multiplying the spinor with the
square-root of the wavefunction renormalization factor. It was then seen that
the wavefunction renormalization factors are instrumental in cancelling an
infrared divergent contribution in the 1PI gravitational vertex function.

Again, we dealt extensively with the limiting case of the finite-temperature 
gravitational vertex function in which the graviton carries zero momentum.
The ambiguity of the finite-temperature Green's functions evaluated at 
zero momentum is a well-known problem in thermal field theory. 
This property is usually due to the fact that the different mathematical
limits correspond to different physical situations, so that
the resolution of the apparent paradox lies in recognizing the
appropriate correspondence with the physical situation at hand.
In our case also, the 1PI gravitational vertex at zero graviton momentum
turned out to contain contributions which superficially seemed to be 
ill-defined, and the problem was resolved by taking the relevant physical situation
into account. Thus, taking the various limits in the
precise order indicated in the operational formula for the gravitational mass,
 given by Eq.\ (\ref{operationalM'}), we obtained unique and well-defined
expressions for all the contributions. We also reproduced these expressions
by using the special regularization technique given in Eq.\ (\ref{eeq:reg}).

On the other hand, our work on high-temperature QED gives credence to the idea that
any gauge dependence present in the self-energy should not show up in the dispersion
relation. This idea motivated us to focus on the subleading part of the self-energy
of the electron at one-loop, which (unlike the leading, $O(T^2)$ part) depends on the 
choice of gauge. 

We obtained an equation which governs the gauge dependence of the one-loop
dispersion relation, stressed its analogy with the corresponding equation in the
gluon case, and consequently pointed out the possibility of a generalization of our
equation to all orders. From this equation obtained by us, the gauge independence
in the $K\gg eT$ limit followed in a straightforward way.
We then showed that the effective mass is not affected by the leading
part (going like $T$) of the
gauge-dependent part of the
self-energy and hence does not receive $O(e^2 T)$
correction in any gauge. While the effective mass was found to be influenced
by the subleading part (going like $\ln T$) of the gauge-dependent part of the 
self-energy, it is
possible that this will be canceled by two-loop contribution. Exploration of this
possibility can be a problem for further investigation.

Our calculations and results are also important from a
phenomenological point of view. We found that the contribution to the 
gravitational mass of the electron from a matter
background with a non-zero chemical potential, such as the Sun,
can be quite appreciable. An interesting problem, combining the
various aspects covered in this thesis, would be the study of the gravitational
interactions of charged leptons in a background at high temperature.
The results can then be estimated for the case of a supernova, which is an
example of such a background. 

Moreover, the
matter-induced corrections to the gravitational mass are different for
the various charged lepton flavors, and are not the same for the
corresponding antiparticles.  There are situations in which mass
differences, intrinsic or induced, have important physical
implications, such as the neutron-proton mass difference in the
context of the nucleosynthesis calculations in the Early Universe.
Although our work has focused in the case of the charged leptons,
similar considerations can be applied to the other fermions as well.
Our calculations have
provided a necessary ingredient for being able to consider them in a
systematic manner, and set the stage for their further study on a firm
basis.

%
%

\appendix
\chapter{Field quantization using discrete momenta}\label{appdisc}
In this Appendix, we relate the quantization with discrete momenta, used
in Sec.\ \ref{th:TE}, to the quantization with continuous momenta, used in Sec.\ 
\ref{th:TP} onward \cite{itz1}.

Beginning with the complex scalar field, we consider the Fourier expansion
given in Eq.\ (\ref{TP1}). The equal-time commutation relations then force
\bea
[a(\vec P),a^\dagger(\vec P\,')]_-=\delta^{(3)}(\vec P-\vec P\,')\,,~
[b(\vec P),b^\dagger(\vec P\,')]_-=\delta^{(3)}(\vec P-\vec P\,')\,,
                                                                  \label{ETCR}
\eea
all other commutators being zero. Also, Eq.\ (\ref{TP1}) leads to 
\bea
H&=&\int d^3P E_P \Big(a^\dagger(\vec P)a(\vec P)+
  b^\dagger(\vec P)b(\vec P)\Big)\,,\nonumber\\*
Q&=&\int d^3P \Big(a^\dagger(\vec P)a(\vec P)-
  b^\dagger(\vec P)b(\vec P)\Big)\,.                              \label{HQab}
\eea

If we instead use quantization in a volume $V$, the plane-wave states can only
possess the discrete momenta
\bea
\vec P=\frac{2\pi}{V^{1/3}}\vec n\,,                          
\eea
where $n_1$, $n_2$, $n_3$ are integers. Integrals are then replaced by sums
\bea
\int\frac{d^3 P}{(2\pi)^3}\rightarrow\frac{1}{V}\sum_{\vec P}       \label{is}
\eea
and delta functions by Kronecker symbols
\bea
(2\pi)^3\delta^{(3)}(\vec P-\vec P\,')\rightarrow V\delta_{\vec P,\vec P\,'}\,.
                                                                     \label{dK}
\eea
Note that the left-hand sides of Eqs.\ (\ref{is}) and (\ref{dK}) taken together
give unity, and so also do the right-hand sides.

Let us now redefine the ladder operators through
\bea
a(\vec P)=\sqrt{\frac{V}{(2\pi)^3}}\,a_{\vec P}\,,~
b(\vec P)=\sqrt{\frac{V}{(2\pi)^3}}\,b_{\vec P}\,.                           \label{aa}
\eea
Eqs.\ (\ref{ETCR}), (\ref{dK}) and (\ref{aa}) give 
\bea
[a_{\vec P},a_{\vec P\,'}]_-=\delta_{\vec P,\vec P\,'}\,,~
[b_{\vec P},b_{\vec P\,'}]_-=\delta_{\vec P,\vec P\,'}\,,           \label{ETCR'}
\eea
while Eqs.\ (\ref{HQab}), (\ref{is}) and (\ref{aa}) give
\bea
H&=&\sum_{\vec P} E_P \Big(a^\dagger_{\vec P}a_{\vec P}+
  b^\dagger_{\vec P}b_{\vec P}\Big)\,,\nonumber\\*
Q&=&\sum_{\vec P} \Big(a^\dagger_{\vec P}a_{\vec P}-
  b^\dagger_{\vec P}b_{\vec P}\Big)\,.                              \label{HQab'}
\eea
(Alternatively, one can use Eqs.\ (\ref{TP1}), (\ref{is}) and (\ref{aa}) to arrive at
\bea
\phi(x)=\sum_{\vec P}\frac{1}{\sqrt {2E_PV}}\Big(a_{\vec P}e^{-ip\cdot x}
        +b{^\dagger_{\vec P}} e^{ip\cdot x}\Big)\,,                         \label{TP1'}
\eea                         
the discretized Fourier expansion for the field. Eqs.\ (\ref{ETCR'}) and (\ref{HQab'})
then follow from this expansion.) Eqs.\ (\ref{ETCR'}) and (\ref{HQab'}) are the key
expressions used in Sec.\ \ref{th:TE}.

Turning next to the fermion field, we consider the Fourier expansion of Eq.\ (\ref{psiFour})
and the spinor normalization given in Eq.\ (\ref{TP12}). The equal-time canonical
anticommutation relations then force
\bea
[c_s(\vec P),c^\dagger_{s'}(\vec P\,')]_+=\delta_{ss'}\delta^{(3)}(\vec P-\vec P\,')\,,~
[d_s(\vec P),d^\dagger_{s'}(\vec P\,')]_+=\delta_{ss'}\delta^{(3)}(\vec P-\vec P\,')\,,
                                                                  \label{ETACR}
\eea
all other anticommutators being zero. (In more general terms, the choice
$N_1=1=N_2$ forces $N_3=1$ in Eqs.\ (\ref{N_1})~--~(\ref{N_3}).) 
Also, Eq.\ (\ref{psiFour}) leads to 
\bea
H&=&\int d^3P E_P \sum_s
\Big(c^\dagger_s(\vec P)c_s(\vec P)+ d^\dagger_s(\vec P)d_s(\vec P)\Big)
                   \,,\nonumber\\*
Q&=&\int d^3P \sum_s
\Big(c^\dagger_s(\vec P)c_s(\vec P)- d^\dagger_s(\vec P)d_s(\vec P)\Big)\,.
                                                                          \label{HQ}
\eea
Use of Eqs.\ (\ref{is}) and (\ref{dK}), and redefinitions similar to those in Eq.\
(\ref{aa}) then give the fermionic counterparts of Eqs.\ (\ref{ETCR'}) and 
(\ref{HQab'}), which were used in Sec.\ \ref{th:TE}.

\chapter{Full propagator and wavefunction renormalization factor
in vacuum}\label{appz}
We start with the definition of the full fermion propagator in vacuum
\bea
iS{^\prime_{f\alpha\beta}}(x)&\equiv&\langle 0|T[\psi_\alpha(x)\bar\psi_\beta(0)] |0\rangle\\
           &=&\theta(x_0)\langle 0|\psi_\alpha(x)\bar\psi_\beta(0)|0\rangle
           -\theta(-x_0)\langle 0|\bar\psi_\beta(0)\psi_\alpha(x)|0\rangle
                                                                         \label{app:S}
\eea
where $\psi(x)$ is the Heisenberg picture field, and $|0\rangle$ the vacuum of the 
full interacting theory. The wavefunction renormalization factor $Z$ is defined through
\bea
\langle 0|\psi_\alpha(0)|f(\vec P,s)\rangle=\sqrt Z\, u_{s\alpha}(\vec P)\,,      
                                                                          \label{app:Z}
\eea
where the 1-particle state $|f(\vec P,s)\rangle$ is again an eigenstate of the full
interacting hamiltonian. The tree-level spinor $u_s(\vec P)$ is actually the analogue
of $U_s(\vec P)$ (see Eq.\ (\ref{U_s})), since the sum of the bare mass and the self-energy 
correction in vacuum is set equal to the physical mass of the fermion.

Eq.\ (\ref{app:Z}) implies that                 
\bea
\langle 0|\psi_\alpha(x)|f(\vec P,s)\rangle=\sqrt Z\, u_{s\alpha}(\vec P) e^{-ip\cdot x}\,,      
                                                                          \label{app:Z1}
\eea
with $p_0=E_P\equiv\sqrt{P^2+m{_f^2}}$. (We used an identity similar to the one in Eq.\ 
(\ref{gr:S3''}).) It then follows that
\bea
\langle f(\vec P,s)|\bar\psi_\alpha(x)|0\rangle=\sqrt Z\, \bar u_{s\alpha}(\vec P) e^{ip\cdot x}\,.      
                                                                          \label{app:Z2}
\eea
Similar relations involving the 1-antiparticle state are $|\bar f(\vec P,s)\rangle$ are
\bea
\langle \bar f(\vec P,s)|\psi_\alpha(x)|0\rangle=\sqrt Z\, v_{s\alpha}(\vec P) e^{ip\cdot x}\,,      
                                                                          \label{app:Z3}\\
\langle 0|\bar\psi_\alpha(x)|\bar f(\vec P,s)\rangle=\sqrt Z\,\bar v_{s\alpha}(\vec P) e^{-ip\cdot x}\,.
                                                                          \label{app:Z4}
\eea
In Eqs.\ (\ref{app:Z3}) and (\ref{app:Z4}) also, $p_0=E_P$. 
It may be noted that in vacuum, $Z$ is independent of momentum and also the same for the
fermion and the antifermion, properties which may not hold in a medium.

We shall take the normalization of 1-particle states to be
\bea
\langle f(\vec P\,',s')|f(\vec P,s)\rangle =
           (2\pi)^3 \delta^{(3)}(\vec P - \vec P\,')\delta_{s,s'}\,,
                                                         \label{statenorm}
\eea
with a similar normalization for $|\bar f(\vec P,s)\rangle$ as well. 
The completeness relation is then given by
\bea
1=|0\rangle\langle 0|+\sum_s\int\frac{d^3P}{(2\pi)^3}|f(\vec P,s)\rangle\langle f(\vec P,s)|      
    +\sum_s\int\frac{d^3P}{(2\pi)^3}|\bar f(\vec P,s)\rangle\langle \bar f(\vec P,s)|      
    +\cdots\,,                                                          \label{complete}
\eea
where the ellipsis stands for multiparticle state contributions, which we shall
not be interested in.

Let us now insert Eq.\ (\ref{complete}) in each of the two terms of Eq.\ (\ref{app:S}).
A look at Eqs.\ (\ref{app:Z1})-(\ref{app:Z4}) then reveals that the 
1-particle intermediate states contribute to the first term in Eq.\ (\ref{app:S})
and the 1-antiparticle intermediate states to the second term (there is no
contribution from $|0\rangle\langle 0|$ since the fermion field cannot have a
vacuum expectation value). Thus,
\bea
iS{^\prime_{f\alpha\beta}}(x)\bigg|_{\rm 1-particle}
           =\theta(x_0)\sum_s\int\frac{d^3P}{(2\pi)^3}\langle 0|\psi_\alpha(x)
            |f(\vec P,s)\rangle\langle f(\vec P,s)|\bar\psi_\beta(0)|0\rangle\,,
                                                                       \label{app:S1}\\
iS{^\prime_{f\alpha\beta}}(x)\bigg|_{\rm 1-antiparticle}
           =-\theta(-x_0)\sum_s\int\frac{d^3P}{(2\pi)^3}\langle 0|\bar\psi_\beta(0)
            |\bar f(\vec P,s)\rangle\langle\bar f(\vec P,s)|\psi_\alpha(x)|0\rangle\,.
                                                                         \label{app:S2}
\eea
In Eq.\ (\ref{app:S1}), use Eqs.\ (\ref{app:Z1}) and (\ref{app:Z2}), and also
\bea
\theta(x_0)=i\int\frac{dp_0}{2\pi}\,\frac{e^{-ip_0x_0}}{p_0+i\epsilon}\,.
                                                                    \label{app:theta}
\eea
Changing $p_0$ to $p_0-E_P$ then leads to 
\bea
S'_f(x)\bigg|_{\rm 1-particle}= \int\frac{d^4p}{(2\pi)^4}e^{-ip\cdot x}
{Z \sum_s u_s(\vec P)\overline u_s(\vec P) \over p_0 - E_p +i\epsilon}\,. \label{app:pole1}
\eea
In Eq.\ (\ref{app:S2}), use Eqs.\ (\ref{app:Z3}) and (\ref{app:Z4}), and also put
$\theta(-x_0)$ from Eq.\ (\ref{app:theta}). Changing $p_0$ to $-p_0-E_P$ 
and $\vec P$ to $-\vec P$ then leads 
to 
\bea
S'_f(x)\bigg|_{\rm 1-antiparticle}= \int\frac{d^4p}{(2\pi)^4}e^{-ip\cdot x}
{Z \sum_s v_s(-\vec P)\overline v_s(-\vec P) 
        \over p_0 + E_p -i\epsilon}\,.                          \label{app:pole2}
\eea

We are now going to check Eqs.\ (\ref{app:pole1}) and (\ref{app:pole2}) at the
tree-level, for which $Z=1$. We first note that Eqs.\ (\ref{app:Z1}) and
(\ref{statenorm}) imply that 
\bea
u{^\dagger_s}(\vec P)u_s(\vec P)=1\,.                                     \label{app:unorm}
\eea
(This can be seen in the following way. Let us introduce the factors $N_1$,
$N_2$, $N_3$, $N_4$ through
\bea
&&u{^\dagger_s}(\vec P)u_{s^\prime}(\vec P)=N_1\delta_{s s^\prime}\,, \label{N_1}\\
&&\psi(x)=\sum_s\int\frac{d^3P}{\sqrt{(2\pi)^3 N_2}}
            \left[c_s(\vec P)u_s(\vec P)e^{-ip\cdot x}+\cdots \right]\,,\\
&&\left[c_s(\vec P), c{_{s^\prime}^\dagger}(\vec P~')\right]_+=N_3\delta_{s s^\prime}\delta^3
                                    (\vec P-\vec P\,^\prime)\,,   \label{N_3}\\
&&|f(\vec P,s)\rangle=\sqrt{N_4}c{_s^\dagger}(\vec P)|0\rangle\,,
\eea
omitting the antiparticle parts for brevity. Then,
$\left[\psi(\vec x,t),\psi(\vec y,t)\right]_+=\delta^3(\vec x-\vec y)$
gives $N_1 N_3=N_2$, Eq.\ (\ref{app:Z1}) (with $Z=1$) gives 
${N_3}^2 N_4=(2\pi)^3 N_2$, and
Eq.\ (\ref{statenorm}) gives $N_3 N_4=(2\pi)^3$. It then follows that $N_1=1$.)

Similarly, $v{^\dagger_s}(p)v_s(\vec P)=1$. For these normalizations, 
the spin-sum relations are given by Eq.\ (\ref{TP18}). Since
\bea
S_f(p_0,\vec P) = \frac{p_0\gamma_0-\vec P\cdot\vec \gamma+m_f}
                 {p{_0^2 }-E{_P^2}+i\epsilon}
             \approx\frac{E_P\gamma_0-\vec P\cdot\vec \gamma+m_f}
                   {2E_P(p_0-E_P+i\epsilon)}
\eea
near the pole $p_0=E_P-i\epsilon$, Eq.\ (\ref{app:pole1}) (with $Z=1$) checks out.
Similarly one can check Eq.\ (\ref{app:pole2}).

In a medium , Eqs.\ (\ref{app:pole1}) and (\ref{app:pole2}) are generalized to Eqs.\
(\ref{qed:Sfull}) and (\ref{SfullV}) respectively, while maintaining the relations
given by Eq.\ (\ref{statenorm}) (see remark after Eq.\ (\ref{gr:S11})) and Eq.\
(\ref{app:unorm}) (see Eq.\ (\ref{Unorm})) which do not depend on the wavefunction
renormalization factor.

\chapter{Charge conjugation and the 1PI gravitational vertex}
\label{appc}
The aim of this Appendix is to derive the charge conjugation property of the
1PI gravitational vertex, given by Eq.\ (\ref{vertexC}), or Eq.\ 
(\ref{vertexC'}), below.

We begin by considering the case of a $C$ invariant Lagrangian and
a $C$ symmetric background. Since $\mu=0$, Eqs.\ 
(\ref{TE1})~--~(\ref{TE3}) reduce to 
\bea
\langle{\cal O}\rangle=\frac{\Tr(e^{-\beta H}{\cal O})}
                        {\Tr e^{-\beta H}}\,.             \label{<O>}
\eea
Using $C^2=1$, this can be rewritten as
\bea
\langle{\cal O}\rangle=\frac{\Tr(Ce^{-\beta H}C\,C{\cal O}C)}
             {\Tr(Ce^{-\beta H}C)}\,.   
\eea
Use of $CHC=H$ (see Eq.\ (\ref{CHC})) and $C^2=1$ then gives
\bea
\langle O\rangle=\langle COC\rangle\,.                           \label{COC}
\eea

It will be instructive to first apply this equation to the definition
of the full fermion propagator $S'_f(p)$, given by
\bea
i(2\pi)^4\delta^{(4)}(p-p')S'_{fnm}(p)=\int d^4 y\, d^4z\,e^{ip\cdot y}
e^{-ip'\cdot z}\langle T[\psi_n(y)\bar\psi_m(z)]\rangle\,,        \label{S'def}
\eea  
where the fermion fields are the Heisenberg picture fields. (This is equivalent 
to the definition
\bea
\langle T[\psi_n(y)\bar\psi_m(z)]\rangle=i\int\frac{d^4k}{(2\pi)^4}
     e^{-ik\cdot(y-z)}S'_{fnm}(k)
\eea
used elsewhere in this work.) Using Eq.\ (\ref{COC}) and $C^2=1$, we can write
\bea
\langle T[\psi_n(y)\bar\psi_m(z)]\rangle
=\langle T[C\psi_n(y)C\,C\bar\psi_m(z)C]\rangle\,.                 \label{<psipsi>}
\eea
Now, $C\psi C$ is given by Eq\ (\ref{qed:psiC}). Also, $C\bar\psi C
=(C\psi C)^\dagger\gamma_0$, since (as explained below Eq.\ (\ref{ccdd})) 
$C$ is hermitian. So, using Eqs.\ (\ref{qed:psiC}),
(\ref{qed:CR}) and (\ref{qed:CU}), we obtain 
\bea
C\bar\psi C=-\psi^{\rm T}{\sf C}^{-1}\,.                          \label{CpsibarC}
\eea
One can now use Eqs.\ (\ref{qed:psiC}) and (\ref{CpsibarC}) to determine the R.H.S.
of Eq.\ (\ref{<psipsi>}). Consequently, the R.H.S. of Eq.\ (\ref{S'def}) equals
\bea
{\sf C}_{nb}({\sf C}^{-1})_{am}\int d^4 z\, d^4y\,e^{-ip'\cdot z}
e^{ip\cdot y}\langle T[\psi_a(z)\bar\psi_b(y)]\rangle\,.
                                                                          \label{di}
\eea
The double integral is just $i(2\pi)^4\delta^{(4)}(p-p')S'_{fab}(-p)$, by the
definition (\ref{S'def}). So the expression given in Eq.\ (\ref{di}) equals
\bea
i(2\pi)^4\delta^{(4)}(p-p')[{\sf C}S'{^{\rm T}_f}(-p){\sf C}^{-1}]_{nm}\,.
                                                                          \label{di1}
\eea       
Since this equals the L.H.S. of Eq.\ (\ref{S'def}), we arrive at
\bea
S'_f(p)={\sf C}S'{^{\rm T}_f}(-p){\sf C}^{-1}\,.                             \label{SC}
\eea
As a check, we note that the use of Eq.\ (\ref{qed:pole}) reduces Eq.\ (\ref{SC}) to
Eq.\ (\ref{qed:Ccon1}), which was deduced by demanding the action to be $C$ invariant.

Now we shall apply Eq.\ (\ref{COC}) to the definition of the 1PI gravitational vertex
function $\Gamma^{\lambda\rho}(p,p')$, given by
\bea
&&(2\pi)^4\delta^{(4)}(p-p'-q)S'_{fnn'}(p)\Gamma{^{\lambda\rho}_{n'm'}}(p,p')
S'_{fm'm}(p')\nonumber\\*
&&=\int d^4x\, d^4 y\, d^4z\,e^{-iq\cdot x}e^{ip\cdot y}
e^{-ip'\cdot z}
\langle T[\widehat T^{\lambda\rho}(x)\psi_n(y)\bar\psi_m(z)]
\rangle\,.                                                           \label{vertexdef}     
\eea  
Using Eq.\ (\ref{COC}), $C^2=1$, and $C\widehat T^{\lambda\rho}C=\widehat T^{\lambda\rho}$,
we can write
\bea
\langle T[\widehat T^{\lambda\rho}(x)\psi_n(y)\bar\psi_m(z)]\rangle=
\langle T[\widehat T^{\lambda\rho}(x)C\psi_n(y)C\,C\bar\psi_m(z)C]\rangle\,.
                                                                    \label{Tpsipsi}
\eea
One can now use Eqs.\ (\ref{qed:psiC}) and (\ref{CpsibarC}) to determine the R.H.S.
of Eq.\ (\ref{Tpsipsi}). Consequently, the R.H.S. of Eq.\ (\ref{vertexdef}) equals
\bea
{\sf C}_{nb}({\sf C}^{-1})_{am}
\int d^4x\,d^4 z\, d^4y\, e^{-iq\cdot x}e^{-ip'\cdot z}
e^{ip\cdot y}\langle T[\widehat T^{\lambda\rho}(x)\psi_a(z)\bar\psi_b(y)]\rangle \,.
                                                                          \label{tri}
\eea
The triple integral is just $(2\pi)^4\delta^{(4)}(p-p'-q)S'_{fan'}(-p')
\Gamma{^{\lambda\rho}_{n'm'}}(-p',-p) S'_{fm'b}(-p)$, by the
definition (\ref{vertexdef}). So the expression given in Eq.\ (\ref{tri}) equals
\bea
(2\pi)^4\delta^{(4)}(p-p'-q)[{\sf C}S'{^{\rm T}_f}(-p)\Gamma^{\lambda\rho{\rm T}}
    (-p',-p)S'{^{\rm T}_f}(-p'){\sf C}^{-1}]_{nm}\,.
                                                                          \label{tri1}
\eea       
Since this equals the L.H.S. of Eq.\ (\ref{vertexdef}), we arrive at
\bea
S'_f(p)\Gamma_{\lambda\rho}(p,p')S'_f(p')={\sf C}S'{^{\rm T}_f}(-p)
\Gamma{_{\lambda\rho}^{\rm T}} (-p',-p)S'{^{\rm T}_f}(-p'){\sf C}^{-1}\,.
\eea
Using Eq.\ (\ref{SC}), this leads to
\bea
\Gamma_{\lambda\rho}(p,p')={\sf C}\Gamma{_{\lambda\rho}^{\rm T}} (-p',-p){\sf C}^{-1}\,.
\eea

Now, consider the case of $C$ symmetric Lagrangian and $C$ asymmetric background. 
Then 
\bea
\langle{\cal O}\rangle=\frac{\Tr(e^{-\beta (H-\mu Q)}{\cal O})}
                        {\Tr e^{-\beta (H-\mu Q)}}\,,          
\eea
instead of Eq.\ (\ref{<O>}).
Since $CQC=-Q$ (see Eq. (\ref{CHC})), we now have
\bea
\langle O \rangle=[\langle COC\rangle]_{\mu\rightarrow -\mu}
\eea
in the place of Eq.\ (\ref{COC}). It is easy to see that this results in
\bea
S'_f(p)&=&[{\sf C}S'{^{\rm T}_f}(-p){\sf C}^{-1}]_{\mu\rightarrow -\mu}\,,\\
\Gamma_{\lambda\rho}(p,p')&=&[{\sf C}\Gamma{_{\lambda\rho}^{\rm T}} 
        (-p',-p){\sf C}^{-1}]_{\mu\rightarrow -\mu}\,.               \label{vertexC}
\eea
Since $\mu$ enters through the thermal propagator of the fermion given by Eqs.\
(\ref{S'}) and (\ref{etaf}), and ${\mu\rightarrow -\mu}$ in the propagator
is equivalent to $v\rightarrow -v$, one can also write Eq.\ (\ref{vertexC})
in the form
\bea
\Gamma_{\lambda\rho}(p,p',v)={\sf C}\Gamma{_{\lambda\rho}^{\rm T}} 
        (-p',-p,-v){\sf C}^{-1}\,,   
                                                                     \label{vertexC'}      
\eea
where we have explicitly indicated the dependence of the vertex function on
the vector $v^\mu$.

\chapter{Transversality of the gravitational vertex}\label{app:transv}
It is useful to verify that the complete vertex function to $O(e^2)$,
obtained in Sec.~\ref{s:indgrav}, satisfies the transversality
conditions
\begin{eqnarray}
q^\lambda \overline U(\vec P') \, \Gamma_{\lambda\rho}(p,p') \, U(\vec P)
=0
= q^\rho \overline U(\vec P') \, \Gamma_{\lambda\rho}(p,p') \, U(\vec P) 
\end{eqnarray}
to this order. The wavefunction renormalization factors associated with the
external lines, being scalars, do not play any role in these relations (in
other words, they just cancel out). 
Also, since the vertex is symmetric in the Lorentz indices
$\lambda,\rho$, either of these relations guarantees the other.
In order to simplify the notation, in this Appendix
we omit the subscript $s$ in the spinors.

In order to verify this relation,
the important point is that we must
include all the terms upto $O(e^2)$. Since the one-loop
terms in the induced vertex are already
$O(e^2)$, for them we can adopt the tree-level definition of
the spinors, i.e.
\begin{eqnarray}
\rlap/p \, u(\vec P) = mu(\vec P)\,, 
\qquad
\overline u(\vec P') \rlap/p' = m \overline u(\vec P')\,, 
\label{treespinors}
\end{eqnarray}
as well as the tree-level on-shell conditions 
\begin{eqnarray}
p^2=p'^2=m^2 \,. 
\label{treedisp}
\end{eqnarray}
In this appendix, as well as in Appendix \ref{app:tough}, we use the 
tree level mass $m$ without any
subscript, implying $m_\ell$, $m_e$ or $m_f$ which should be
understood from the context.  Also note that the photon distribution
function as well as the associated $\delta$-function are even in $k$,
and therefore those terms which are odd in $k$ in the rest 
of the integrand do not contribute.

We first show that the vertex contribution from
Fig.~\ref{f:Ztype}A is transverse by itself. From Eq.\ (\ref{A...}),
\begin{eqnarray}
q^\lambda A_{\lambda\rho\alpha}(k,k-q) = - (q^2-2k\cdot q) (4k_\rho
k_\alpha + k\cdot q \eta_{\rho\alpha}) \,,
\end{eqnarray}
where we omit the terms that are proportional to $q_\alpha$ 
because, in Eq.\ (\ref{Xq}),
they will yield $\rlap/q$ which vanishes between the spinors.
Changing the sign of $q$ in the last equation yields
\begin{eqnarray}
q^\lambda A_{\lambda\rho\alpha}(k+q,k) = (q^2+2k\cdot q) (4k_\rho
k_\alpha - k\cdot q \eta_{\rho\alpha}) \,,
\end{eqnarray}
and as a result $q^\lambda \overline u(\vec P')
\Gamma^{(X)}_{\lambda\rho}(p,p') u(\vec P)$ turns out to be
proportional to $\sum_f Q_f (n_f - n_{\bar f})$, which is zero
to this order.

As for the other diagrams, straightforward algebra gives the
following results:
\begin{eqnarray}
q^\lambda \overline u(\vec P') \Gamma'^{(A1)}_{\lambda\rho} (p,p') u(\vec P) 
&=& -\, {e^2\over 4} \int {d^4k \over (2\pi)^3} \;
\delta(k^2) \eta_\gamma(k) \nonumber\\*
&\times & \overline u(\vec P') 
\Bigg[ {4m k_\rho - (p+5p')_\rho \rlap/k + 2 k\cdot p 
\gamma_\rho \over k\cdot p'} - \Big( p \leftrightarrow p' \Big)
\Bigg] u(\vec P) \nonumber\\
q^\lambda \overline u(\vec P') \Gamma'^{(A2)}_{\lambda\rho} (p,p') u(\vec P) 
&=& {e^2 \over 4} \int {d^4k \over
(2\pi)^3} \; \delta(k^2-m^2) \eta_f(k) \nonumber\\*
&\times & \overline u(\vec P') \Bigg[ 
{4(\rlap/k-2m) k_\rho + \rlap/k (p+p')_\rho - 2 k \cdot p' \gamma_\rho
\over m^2- k\cdot p}\nonumber\\* 
&&- \Big( p \leftrightarrow p' \Big)
\Bigg] u(\vec P) \,, \nonumber\\
q^\lambda \overline u(\vec P') \Gamma'^{(B1)}_{\lambda\rho} (p,p')  u(\vec P) 
&=& e^2 \int {d^4k \over (2\pi)^3} \;
\delta(k^2) \eta_\gamma(k) \nonumber\\*
&\times& \overline u(\vec P') \Bigg[ {m k_\rho - \rlap/k p'_\rho \over
k\cdot p'} - {m k_\rho - \rlap/k p_\rho \over k\cdot p}
\Bigg] u(\vec P) \nonumber\\
q^\lambda \overline u(\vec P') \Gamma'^{(B2)}_{\lambda\rho} (p,p')  u(\vec P) 
&=& -\, e^2 \int {d^4k \over (2\pi)^3} \; \delta(k^2-m^2)
\eta_f(k) \nonumber\\* 
&\times & \overline u(\vec P') 
\left[ {(\rlap/k - 2m) k_\rho + mp_\rho \over m^2 - k\cdot p} -
{(\rlap/k - 2m) k_\rho + mp'_\rho \over m^2 - k\cdot p'} \right]
u(\vec P) \,, \nonumber\\ 
q^\lambda \overline u(\vec P') \Gamma'^{(C1+D1)}_{\lambda\rho} (p,p') u(\vec P)
&=& {e^2 \over 2} \int {d^4k \over (2\pi)^3} \;
\delta(k^2) \eta_\gamma(k) \left( {1 \over k\cdot p'} + {1 \over
k\cdot p} \right) \nonumber\\*
&\times& 
\overline u(\vec P') \Big[  \rlap/k q_\rho + k\cdot q
\gamma_\rho \Big] u(\vec P) \,, \nonumber\\
q^\lambda \overline u(\vec P') \Gamma'^{(C2+D2)}_{\lambda\rho} (p,p') u(\vec P) 
&=& -\, {e^2 \over 2} \int {d^4k \over (2\pi)^3} \; \delta(k^2-m^2)
\eta_f(k)\nonumber\\*
&\times& \left(
{1\over m^2-k\cdot p'} + {1\over m^2-k\cdot p}
\right) \nonumber\\* 
&\times& \overline u(\vec P') \Big[ (\rlap/k-3m) q_\rho + k\cdot q
\gamma_\rho \Big] u(\vec P)  \,. 
\end{eqnarray}
Therefore, 
adding all the one-loop contributions to the vertex, we obtain
\begin{eqnarray}
q^\lambda \overline u(\vec P') \Gamma'^{(1)}_{\lambda\rho} (p,p') u(\vec P) 
&=& {e^2 \over 4} \int {d^4k \over (2\pi)^3} \;
\delta(k^2) \eta_\gamma(k) \nonumber\\*
&\times& 
\overline u(\vec P') \Bigg[ {\rlap/k \over k\cdot p'} \left( 3p_\rho -
p'_\rho \right)  - \Big( p \leftrightarrow p' \Big)
\Bigg] u(\vec P) \,,
\label{qGamma1} \\
q^\lambda \overline u(\vec P') \Gamma'^{(2)}_{\lambda\rho} (p,p') u(\vec P) 
&=& {e^2 \over 4} \int {d^4k \over (2\pi)^3} \;
\delta(k^2-m^2) \eta_f(k) \nonumber\\*
&\times& 
\overline u(\vec P') \Bigg[  
{(\rlap/k -2m)(3p'_\rho - p_\rho) - 2 k\cdot p \gamma_\rho \over m^2 -
k\cdot p } - 
\Big( p \leftrightarrow p' \Big)
\Bigg] u(\vec P) \,.\nonumber\\*
\label{qGamma2}
\end{eqnarray}

We need to add to these the tree-level contribution to the gravitational
vertex that appears in Eq.\ (\ref{lg:Vlamrho}). In this case, we must
include the $O(e^2)$ corrections to the equation for the spinors, which
arise from the self-energy diagrams of Sec.~\ref{qed:calcu}. Thus, for
this part, using Eq.\ (\ref{U_s}) and its hermitian conjugate
\begin{eqnarray}
\overline U(\vec P') \Big( \rlap/p' - m - \Sigma(p') \Big) = 0 \,,
\label{Up'}
\end{eqnarray}
we obtain
\begin{eqnarray}
\overline U(\vec P') \rlap/q  U(\vec P) = \overline
U(\vec P') \Big( \Sigma'(p) - \Sigma'(p') \Big) U(\vec P) \,,
\end{eqnarray}
which in turn yields
\begin{eqnarray}
q^\lambda \overline U(\vec P') \, V_{\lambda\rho}(p,p') \, U(\vec P) =
\frac14 \overline 
U(\vec P') \bigg[ (3p'-p)_\rho \Sigma'(p) 
+ p^2 \gamma_\rho 
- \Big( p \leftrightarrow p' \Big) \bigg]  U(\vec P) \,.
\label{qV}
\end{eqnarray}
This can be cast in a different form by multiplying Eq.\
(\ref{U_s}) from the left by $\overline
U(\vec P')\gamma_\rho(\rlap/p+m)$ and Eq.\ (\ref{Up'}) from the right
by $(\rlap/p'+m)\gamma_\rho U(\vec P)$ and taking the difference of the
resulting equations. This gives
\begin{eqnarray}
(p^2-p'^2) \; \overline U(\vec P') \gamma_\rho U(\vec P) = \overline U(\vec P') \Big[
\gamma_\rho (\rlap/p+m) \Sigma'(p) - \Sigma'(p') (\rlap/p'+m)
\gamma_\rho \Big] U(\vec P) \,,
\label{psq-p'sq}
\end{eqnarray}
and substituting this result into Eq.\ (\ref{qV}), we obtain
\begin{eqnarray}
q^\lambda \overline U(\vec P') \, V_{\lambda\rho}(p,p') \, U(\vec P) &=&
\frac14 \overline 
U(\vec P') \bigg[ (3p'-p)_\rho \Sigma'(p) - (3p-p')_\rho \Sigma'(p')
\nonumber\\* 
&& + \gamma_\rho (\rlap/p+m) \Sigma'(p) - \Sigma'(p') (\rlap/p'+m)
\gamma_\rho   \bigg]  U(\vec P) \,.
\label{qVfinal}
\end{eqnarray}
Since $\Sigma'$ is explicitly of $O(e^2)$
while we are interested in results to $O(e^2)$ only,
we can use the tree-level spinors on the right-hand side.
Using Eq.\ (\ref{treedisp}) in Eq.\ (\ref{Sigma1}),
we can write the self-energy contribution involving the photon
distribution function as
\begin{eqnarray}
\Sigma'_1(p) &=& e^2 \int {d^4k \over (2\pi)^3} \; \delta(k^2)
\eta_\gamma(k) \; {\rlap/ k \over k\cdot p}  \,,
\label{Sigma1short}
\end{eqnarray}
disregarding terms odd in $k$. Similarly, from Eq.\
(\ref{Sigma2}), the part containing the Fermi distribution function
can be written as
\begin{eqnarray}
\Sigma'_2(p) 
&=& -\, e^2 \int {d^4k \over (2\pi)^3} \; 
\delta(k^2-m^2) \eta_f(k) \;
{\rlap/ k - 2m \over m^2 - k\cdot p} \,.
\label{Sigma2short}
\end{eqnarray}
Substituting these forms into Eq.\ (\ref{qVfinal}) and using the 
identities
\begin{eqnarray}
\overline u(\vec P') \gamma_\rho (\rlap/p+m) \rlap/k u(\vec P) 
&=& 2 k\cdot p \; \overline u(\vec P') \gamma_\rho u(\vec P) \nonumber\\* 
\overline u(\vec P') \rlap/k (\rlap/p'+m) \gamma_\rho u(\vec P) 
&=& 2 k\cdot p' \; \overline u(\vec P') \gamma_\rho u(\vec P) \,,
\end{eqnarray}
we see that Eq.\ (\ref{qVfinal}) cancels the contribution from the
loop diagrams given in Eqs.\ (\ref{qGamma1}) and (\ref{qGamma2}) to
this order. This proves the transversality of the
effective vertex.

\chapter{Apparently ill-defined contributions to the gravitational
vertex}\label{app:tough}
\section{The B1 contribution}\label{app:tough:B1}
We start from the formula given in Eq.\ (\ref{GamB1}), from which it
follows that
\begin{eqnarray}
\label{GamB1a}
\Gamma'^{(B1)}_{\lambda\rho} (p,p) &=&  
-\, \frac{e^2}{2} 
\lim_{\vec Q\rightarrow 0}
\int\frac{d^4k}{(2\pi)^3}\delta(k^2) \eta_\gamma(k)
 \Bigg[
\frac{\gamma^\nu (\rlap/p- \rlap/k +m) \gamma^\mu 
C_{\mu\nu\lambda\rho} (k,k - q)}{k\cdot p(2\vec K\cdot\vec Q - Q^2)} \nonumber\\*
&&- \frac{\gamma^\nu(\rlap/{p} - \rlap/{k} - \rlap/{q} + m) 
\gamma^\mu C_{\mu\nu\lambda\rho} (k + q,k)}{(k\cdot p + \vec K\cdot\vec Q)
(2\vec K\cdot\vec Q + Q^2)} \Bigg] \,,
\end{eqnarray}
where we have put
\begin{eqnarray}
q^\mu = (0,\vec Q) \qquad k^\mu = (k_0,\vec K) \,.
\label{q&k}
\end{eqnarray}
In order to take the limit $\vec Q\rightarrow 0$, our strategy is to
expand the coefficients of the factors \hbox{$1/(2\vec K\cdot\vec Q
\pm Q^2)$} in powers of $\vec Q$. Of the resulting terms in the
coefficients, those which are quadratic in $\vec Q$ do not contribute
in the $\vec Q\rightarrow 0$ limit and therefore we need to keep only
the terms that are at most linear in $\vec Q$. 

Using the property $C_{\mu\nu\lambda\rho}(k + q,k) =
C_{\nu\mu\lambda\rho}(k,k + q)$, we can write
\begin{eqnarray}
\label{Cexp}
C_{\mu\nu\lambda\rho} (k,k - q) & = & C_{\mu\nu\lambda\rho} (k,k) - 
C'_{\mu\nu\lambda\rho} (k,q) \nonumber\\
C_{\mu\nu\lambda\rho} (k + q,k) & = & C_{\mu\nu\lambda\rho} (k,k) +
C'_{\nu\mu\lambda\rho} (k,q) \,,
\end{eqnarray}
where
\begin{eqnarray}
C'_{\mu\nu\lambda\rho} (k,q) & = &
\eta_{\lambda\rho}(\eta_{\mu\nu}k\cdot q - q_\mu k_\nu)
- \eta_{\mu\nu}(k_\lambda q_\rho + q_\lambda k_\rho)   +
k_\nu(\eta_{\lambda\mu}q_\rho + \eta_{\rho\mu}q_\lambda)\nonumber\\*
&& + q_\mu(\eta_{\lambda\nu}k_\rho + \eta_{\rho\nu}k_\lambda) -
k\cdot q(\eta_{\lambda\mu}\eta_{\rho\nu} +
\eta_{\lambda\nu}\eta_{\rho\mu}) \,.
\end{eqnarray}
To first order in $Q$, we can also put
\begin{eqnarray}
\frac{1}{k\cdot p + \vec K\cdot\vec Q} = \frac{1}{k\cdot p} - 
\frac{\vec K\cdot\vec Q}{(k\cdot p)^2} \,.
\end{eqnarray}
This enables us to decompose $\Gamma'^{(B1)}_{\lambda\rho} (p,p)$ in
the following four terms:
\begin{eqnarray}
\Gamma'^{(B1a)}_{\lambda\rho}(p) \!\!\!& = & -\, \frac{e^2}{2} 
\lim_{\vec Q\rightarrow 0}
\int {d^4k \over (2\pi)^3} \; \delta(k^2) \eta_\gamma(k) 
{C_{\mu\nu\lambda\rho}(k,k)
\gamma^\nu (\rlap/p- \rlap/k +m) \gamma^\mu \over k\cdot p}
\nonumber\\* 
&& \times \left[\frac{1}{2\vec K\cdot\vec Q - Q^2} -
\frac{1}{2\vec K\cdot\vec Q + Q^2}\right] \nonumber\\[12pt]
\Gamma'^{(B1b)}_{\lambda\rho}(p) & = & -\, \frac{e^2}{2} 
\lim_{\vec Q\rightarrow 0}
\int\frac{d^4k}{(2\pi)^3}\delta(k^2) \eta_\gamma(k)
{C_{\mu\nu\lambda\rho}(k,k) \gamma^\nu\rlap/{q}\gamma^\mu \over 
k\cdot p(2\vec K\cdot Q + Q^2)} \nonumber\\[12pt] 
\Gamma'^{(B1c)}_{\lambda\rho}(p) & = & -\, \frac{e^2}{2} 
\lim_{\vec Q\rightarrow 0}
\int {d^4k \over (2\pi)^3} \; \delta(k^2) \eta_\gamma(k) 
{C_{\mu\nu\lambda\rho}(k,k) \gamma^\nu (\rlap/p- \rlap/k +m) \gamma^\mu 
\over (k\cdot p)^2}
\left[\frac{\vec K\cdot\vec Q}{2\vec K\cdot\vec Q + Q^2}\right]
\nonumber\\[12pt]
\Gamma'^{(B1d)}_{\lambda\rho}(p) & = & -\, \frac{e^2}{2} 
\lim_{\vec Q\rightarrow 0}
\int\frac{d^4k}{(2\pi)^3}\delta(k^2) \eta_\gamma(k)
{ \gamma^\nu(\rlap/p - \rlap/k + m)\gamma^\mu
\over k\cdot p} \nonumber\\*
&&  \times\left[
-\frac{C'_{\mu\nu\lambda\rho}(k,q)}{2\vec K\cdot \vec Q - Q^2}
-\frac{C'_{\nu\mu\lambda\rho}(k,q)}{2\vec K\cdot \vec Q + Q^2}
\right] . 
\end{eqnarray}
We carry out these integrals one by one.

Eliminating the manifestly $k$-odd terms from the integrand and
performing the $k_0$-integration, we obtain
\begin{eqnarray}
\Gamma'^{(B1a)}_{\lambda\rho}(p) 
= e^2 \lim_{\vec Q\rightarrow 0}
\int\frac{d^3K}{(2\pi)^3 2K} f_\gamma(K) 
{4k_\lambda k_\rho \rlap/k \over k\cdot p}
\left[\frac{1}{2\vec K\cdot\vec Q - Q^2} -
\frac{1}{2\vec K\cdot\vec Q + Q^2}\right] \,,
\end{eqnarray}
using $k^2=0$. The expression within the square brackets is finite for
$Q\to0$.  Therefore, in the spinors we can set $p=p'$, and using Eq.\
(\ref{ugu}) we then obtain
\begin{eqnarray}
m'_{(B1a)} = {4e^2 \over m} \lim_{\vec Q\rightarrow 0}
\int\frac{d^3K}{(2\pi)^3} f_\gamma(K) K
\left[\frac{1}{2\vec K\cdot\vec Q - Q^2} -
\frac{1}{2\vec K\cdot\vec Q + Q^2}\right] \,. 
\label{m'B1a}
\end{eqnarray}
We can perform the integration over the angular variables in $\vec K$,
the integral being understood, as usual, in terms of the principal
value part. That gives
\begin{eqnarray}
\int d\Omega 
\frac{1}{2\vec K \cdot \vec Q - Q^2} = 
- \int d\Omega 
\frac{1}{2\vec K \cdot \vec Q + Q^2} = 
-\, \frac{\pi}{K^2} + {\cal O}(Q^2) \,,
\label{angint}
\end{eqnarray}
so that
\begin{eqnarray}
m'_{(B1a)} = - \, {e^2 \over \pi^2 m} 
\int dK \; f_\gamma(K) K 
& = & -\, \frac{e^2T^2}{6m} \,.
\label{B1Aresult}
\end{eqnarray}

As for the next contribution, it is straightforward to verify that
\begin{eqnarray}
(2v^\lambda v^\rho - \eta^{\lambda\rho})\gamma^\nu\rlap/{q}\gamma^\mu
C_{\mu\nu\lambda\rho}(k,k) = 
4\vec K\cdot \vec Q (\rlap/k - 2k\cdot v\rlap/v )\,,
\end{eqnarray}
using $q\cdot v=q_0=0$. So
\begin{eqnarray}
(2v^\lambda v^\rho - \eta^{\lambda\rho})
\Gamma'^{(B1b)}_{\lambda\rho}(p) 
& = & e^2 \int\frac{d^4k}{(2\pi)^3}\delta(k^2) \eta_\gamma(k)
\frac{(-\rlap/k + 2k\cdot v\rlap/v)}
{k\cdot p} \,.
\end{eqnarray}
Now using Eqs.\ (\ref{ugu}) and (\ref{unorm}), carrying out the
integral over $k_0$, and finally putting $P=0$, we get
\begin{eqnarray}
\label{B1Bresult}
m'_{(B1b)} = 
{2e^2 \over m} \int\frac{d^3K}{(2\pi)^3 2K}f_\gamma(K) 
= \frac{e^2 T^2}{12m} \,.
\end{eqnarray}

Similarly,
\begin{eqnarray}
\Gamma'^{(B1c)}_{\lambda\rho}(p) 
& = & -\, \frac{e^2}{2}
\int {d^3K \over (2\pi)^3 2K} \; f_\gamma(K) 
\gamma^\nu (\rlap/p + m) \gamma^\mu C_{\mu\nu\lambda\rho}(k,k)
\frac{1}{(k\cdot p)^2} \,,
\end{eqnarray}
and
\begin{eqnarray}
(2v^\lambda v^\rho -
\eta^{\lambda\rho})\Gamma'^{(B1c)}_{\lambda\rho}(p) &=&  
-{2e^2\over m} \int {d^3K \over (2\pi)^3 2K} \; f_\gamma(K)
\frac{1}{(k\cdot p)^2} 
\nonumber\\* 
&& \times \left[
-(k\cdot p)^2 - 2m^2(k\cdot v)^2 + 4(k\cdot p)(k\cdot v)(p\cdot v)
\right]
\end{eqnarray}
so that
\begin{eqnarray}
\label{B1Cresult}
m'_{(B1c)} = 
-{2e^2\over m} \int {d^3K \over (2\pi)^3 2K} \; f_\gamma(K) 
= -\frac{e^2 T^2}{12m} \,.
\end{eqnarray}

For $\Gamma'^{(B1d)}_{\lambda\rho}$ we first perform the integral over
$k_0$. Remembering that in the remaining integral we can change $\vec
K$ to $-\vec K$ and using the fact that $C'_{\mu\nu\lambda\rho}(-k,q)
= - C'_{\mu\nu\lambda\rho}(k,q)$, we obtain
\begin{eqnarray}
\Gamma'^{(B1d)}_{\lambda\rho}(p) & = &
\frac{e^2}{2} \lim_{\vec Q\rightarrow 0}
\int\frac{d^3K}{(2\pi)^3 2K}f_\gamma(K)
\left(\frac{1}{k\cdot p}\right)\nonumber\\*
& & \times \Bigg\{
\gamma^\nu(\rlap/{p} + m)\gamma^\mu\left[
\frac{1}{2\vec K\cdot \vec Q - Q^2} - 
\frac{1}{2\vec K\cdot \vec Q + Q^2}\right]
[C'_{\mu\nu\lambda\rho}(k,q) -
C'_{\nu\mu\lambda\rho}(k,q)] 
\nonumber\\*
& & +
\gamma^\nu(-\rlap/{k})\gamma^\mu\left[
\frac{1}{2\vec K\cdot \vec Q - Q^2} + 
\frac{1}{2\vec K\cdot \vec Q + Q^2}\right]
[C'_{\mu\nu\lambda\rho}(k,q) + C'_{\nu\mu\lambda\rho}(k,q)]
\Bigg\} \nonumber\\[12pt]
& = & - 
\frac{e^2}{2} \lim_{\vec Q\rightarrow 0}
\int\frac{d^3K}{(2\pi)^3 2K}f_\gamma(K)
\frac{1}{k\cdot p} \nonumber\\*
&&\times\Bigg\{ {\gamma^\nu \rlap/k \gamma^\mu \over 
\vec K\cdot \vec Q}
\left[
C'_{\mu\nu\lambda\rho}(k,q) + C'_{\nu\mu\lambda\rho}(k,q)
\right]
+ O(Q)\Bigg\} \,. 
\end{eqnarray}
We now use
\begin{eqnarray}
(2v^\lambda v^\rho - \eta^{\lambda\rho})\gamma^\nu\rlap/{k}\gamma^\mu
\Big[ C'_{\mu\nu\lambda\rho}(k,q) +
C'_{\nu\mu\lambda\rho}(k,q) \Big] 
= 8(k\cdot v)(\vec K\cdot \vec Q) \rlap/v \,.
\end{eqnarray}
Then, using Eq.\ (\ref{ugu}) and putting $\vec P=0$, we get
\begin{eqnarray}
\label{B1Dresult}
m'_{(B2d)}
= - \frac{4e^2}{m}\int\frac{d^3K}{(2\pi)^3 2K}f_\gamma(K)
= -\frac{e^2 T^2}{6m} \,.
\end{eqnarray}

Adding the results given in Eqs. (\ref{B1Aresult}),
(\ref{B1Bresult}), (\ref{B1Cresult}) and (\ref{B1Dresult}), we get the
total contribution from the B1 term presented in Eq.\
(\ref{B1final}).

\section{The A2 contribution}\label{app:tough:A2}
For this contribution, we start from Eq.\ (\ref{A2}). Using Eq.\
(\ref{q&k}), we can write it as
\begin{eqnarray}
\Gamma'^{(A2)}_{\lambda\rho} (p,p) &=& {e^2 \over 2} 
\lim_{\vec Q \to 0} 
\int {d^4k \over (2\pi)^3} \; \delta(k^2-m^2) \eta_f(k) 
\nonumber\\* && \times 
\Bigg[ {\Lambda_{\lambda\rho} (k,k-q) 
\over (2\vec K\cdot \vec Q - Q^2) (m^2 - k\cdot p)} -
{\Lambda_{\lambda\rho} (k+q,k) \over (2\vec K\cdot \vec Q + Q^2)(m^2 -
k\cdot p')}
\Bigg] \,.
\end{eqnarray}
Following the strategy stated below Eq.\ (\ref{q&k}), let us now write
\begin{eqnarray}
\Lambda_{\lambda\rho} (k,k-q) &=& \Lambda_{\lambda\rho} (k,k) +
\Lambda'_{\lambda\rho} (k,q) \,, \nonumber\\* 
\Lambda_{\lambda\rho} (k+q,k) &=& \Lambda_{\lambda\rho} (k,k) + 
\Lambda''_{\lambda\rho} (k,q) \,,
\end{eqnarray}
and, in the denominator, expand $m^2-k\cdot p'$ in powers of $\vec Q$:
\begin{eqnarray}
{1 \over m^2-k\cdot p'} 
&=& {1 \over m^2-k\cdot p} + {\vec K \cdot \vec Q \over (m^2-k\cdot
p)^2} + O(Q^2) \,.
\end{eqnarray}
Then we can decompose $\Gamma'^{(A2)}_{\lambda\rho} (p,p)$ into the
following terms, omitting higher powers of $Q$ which anyway will not
contribute:
\begin{eqnarray}
\Gamma'^{(A2a)}_{\lambda\rho} (p) &=& {e^2 \over 2} 
\lim_{\vec Q \to 0} 
\int {d^4k \over (2\pi)^3} \; \delta(k^2-m^2) \eta_f(k) 
\nonumber\\* && \times 
{\Lambda_{\lambda\rho} (k,k) \over m^2 - k\cdot p} 
\Bigg(
{1 \over 2\vec K\cdot \vec Q - Q^2} - 
{1 \over 2\vec K\cdot \vec Q + Q^2}
\Bigg) \,, \nonumber\\ 
\Gamma'^{(A2b)}_{\lambda\rho} (p) &=& -\, {e^2 \over 4} 
\int {d^4k \over (2\pi)^3} \; \delta(k^2-m^2) \eta_f(k) 
{\Lambda_{\lambda\rho} (k,k) \over (m^2 - k\cdot p)^2} \,, 
\nonumber\\
\Gamma'^{(A2c)}_{\lambda\rho} (p) &=& {e^2 \over 2} 
\lim_{\vec Q \to 0} 
\int {d^4k \over (2\pi)^3} \; \delta(k^2-m^2) \eta_f(k) 
\nonumber\\* && \times 
{1 \over m^2 - k\cdot p} 
\Bigg(
{\Lambda'_{\lambda\rho} (k,q) \over 2\vec K\cdot \vec Q - Q^2} - 
{\Lambda''_{\lambda\rho} (k,q) \over 2\vec K\cdot \vec Q + Q^2}
\Bigg) \,, 
\end{eqnarray}
We discuss these contributions one by one.

\subsubsection*{The A2a contribution}
Using Eq.\ (\ref{Lambdayy}) and the $\delta$-function appearing in the
integrand, we can write
\begin{eqnarray}
\Gamma'^{(A2a)}_{\lambda\rho} (p) &=& -2e^2 
\lim_{\vec Q \to 0} 
\int {d^4k \over (2\pi)^3} \; \delta(k^2-m^2) \eta_f(k) 
\nonumber\\* && \times 
{k_\lambda k_\rho (\rlap/k - 2m) \over m^2 - k\cdot p} 
\Bigg(
{1 \over 2\vec K\cdot \vec Q - Q^2} - 
{1 \over 2\vec K\cdot \vec Q + Q^2}
\Bigg) \,.
\end{eqnarray}
As argued before Eq.\ (\ref{m'B1a}), we can put $\vec Q=0$ in the
spinors, and use Eq.\ (\ref{ugu}). Performing the $k_0$-integration,
we obtain
\begin{eqnarray}
(2v^\lambda v^\rho - \eta^{\lambda\rho}) 
\Gamma^{(A2a)}_{\lambda\rho}(p) &=& 
\frac{2e^2}{m}\lim_{\vec Q \rightarrow 0}
\int {d^3K \over (2\pi)^3} \; F(\vec K)
\Bigg(
{1 \over 2\vec K\cdot \vec Q - Q^2} - 
{1 \over 2\vec K\cdot \vec Q + Q^2}
\Bigg) \,,\nonumber\\*
\end{eqnarray}
where the expression on the left is understood to equal the one on the
right only between the spinors, and 
\begin{eqnarray}
F(\vec K) = {2E_K^2 - m^2 \over 2E_K}
\left[ \left( 1 + {m^2 \over m^2 - k\cdot p} \right) f_e + 
\left( 1 + {m^2 \over m^2 + k\cdot p} \right) f_{\bar e} \right] \,,
\end{eqnarray}
with $k_0=E_K$. Since the integrand contains $\vec K\cdot \vec P$, and
we must set $\vec P=0$ only after taking the limit $Q\to0$, the
angular integrations cannot be performed using Eq.\ (\ref{angint}). So
we shift the integration variable to $\vec K\pm\frac12\vec Q$ in the
terms having $2\vec K\cdot\vec Q \mp Q^2$ in the denominator. This
gives
\begin{eqnarray}
(2v^\lambda v^\rho - \eta^{\lambda\rho})
\Gamma^{(A2a)}_{\lambda\rho}(p) 
& = & \frac{2e^2}{m}
\lim_{\vec Q\rightarrow 0}\int\frac{d^3K}{(2\pi)^3}
\frac{\vec Q\cdot\vec\nabla_K F}{2\vec K\cdot\vec Q} \,.
\end{eqnarray}
Clearly the magnitude of $\vec Q$ now cancels out.  The derivative
with respect to $\vec K$ can be taken easily, using
\begin{eqnarray}
\vec\nabla_K E_K &=& {\vec K \over E_K} \,, \nonumber\\*
\vec\nabla_K\left(\frac{1}{m^2 \pm k\cdot p}\right) &=& 
\frac{\mp 1}{(m^2 \pm k\cdot p)^2}
\left(E_P\frac{\vec K}{E_K} - \vec P\right) \,.
\end{eqnarray}
The term proportional to $\vec P$ from the last derivative does not
contribute because it multiplies a factor whose integrand is odd in
$\vec K$ at $\vec P=0$.  Putting $\vec P=0$ in the other terms, we
obtain the contribution to the gravitational mass:
\begin{eqnarray}
m'_{(A2a)} = 
\frac{2e^2}{m}\int\frac{d^3K}{(2\pi)^3 2E_K} 
&\times& \Bigg\{ 
{2E_K^2-m^2 \over 2E_K} \bigg( {E_K-2m \over E_K-m} \; {\partial f_e \over
\partial E_K} + {E_K+2m \over E_K+m} \; {\partial f_{\bar e} \over
\partial E_K} \bigg) \nonumber\\* 
&& + {2E_K^2-m^2 \over 2E_K} \bigg( {m \over (E_K-m)^2} f_e - 
{m \over (E_K+m)^2} f_{\bar e} \bigg) \nonumber\\*
&& + {2E_K^2+m^2 \over 2E_K^2} \bigg( {E_K-2m \over E_K-m} f_e
+ {E_K+2m \over E_K+m} f_{\bar e} \bigg)
\Bigg\} \,.
\label{A2afinal}
\end{eqnarray}
%

\subsubsection*{The A2b contribution}
The integral in the (A2b) term is independent of $Q$.  So, in a
straight forward way, we obtain
\begin{eqnarray}
m'_{(A2b)} 
&=& {e^2 \over m^2} \int {d^3K \over (2\pi)^3 2E_K}
\; (2E_K^2 - m^2) \Bigg[ {E_K-2m \over (E_K-m)^2} f_e(E_K)
- {E_K+2m \over (E_K+m)^2} 
f_{\bar e}(E_K)  \Bigg] \,.\nonumber\\*
\label{A2bfinal}
\end{eqnarray}
%

\subsubsection*{The A2c contribution}
For the (A2c) contribution, first we use the expression for 
$\Lambda_{\lambda\rho}$ from Eq.\ (\ref{Lambda}) to find
\begin{eqnarray}
\Lambda'_{\lambda\rho} (k,q) &=& 
\eta_{\lambda\rho} (q^2-2k\cdot q) (\rlap/k-2m) + 
(k_\lambda q_\rho + k_\rho q_\lambda) (\rlap/k - 4m) 
+ k_\lambda \rlap/k \gamma_\rho \rlap/q 
+ k_\rho \rlap/k \gamma_\lambda \rlap/q \,,
\nonumber\\
\Lambda''_{\lambda\rho} (k,q) &=& 
\eta_{\lambda\rho} (q^2+2k\cdot q) (\rlap/k-2m)  
- (k_\lambda q_\rho + k_\rho q_\lambda) (\rlap/k - 4m) 
- k_\lambda \rlap/q \gamma_\rho \rlap/k 
- k_\rho \rlap/q \gamma_\lambda \rlap/k \,,
\label{}\nonumber\\*
\end{eqnarray}
dropping irrelevant $O(q^2)$-terms and using $k^2=m^2$.  In the
$\eta_{\lambda\rho}$ terms, the integrand becomes independent of
$q$. Thus, these terms give a regular contribution. Let us denote it
by (A2r):
\begin{eqnarray}
m'_{(A2r)} &=& {2e^2 \over m} 
\int {d^3K \over (2\pi)^3 2E_K} \;
\Bigg[ {E_K-2m \over E_K-m} f_e(E_K) + 
{E_K+2m \over E_K+m} f_{\bar e}(E_K) \Bigg] \,.
\label{A2rfinal}
\end{eqnarray}

The terms which appear next will be called (A2s). For these, we use
the fact that
\begin{eqnarray}
(2v^\lambda v^\rho - \eta^{\lambda\rho}) (k_\lambda q_\rho + k_\rho
q_\lambda) = -2 k\cdot q = 2 \vec K \cdot \vec Q \,,
\end{eqnarray}
using $q\cdot v =q_0=0$. The $Q\to0$ limit can then be taken easily,
and we obtain
\begin{eqnarray}
m'_{(A2s)} 
&=& - {e^2 \over m} 
\int {d^3K \over (2\pi)^3 2E_K} \;
\Bigg[ {E_K-4m \over E_K-m} f_e(E_K) + 
{E_K+4m \over E_K+m} f_{\bar e}(E_K) \Bigg] \,.
\label{A2sfinal}
\end{eqnarray}

Finally, we come to the terms with three $\gamma$-matrices, which we
denote by (A2t). For these, first we note that
\begin{eqnarray}
(2v^\lambda v^\rho - \eta^{\lambda\rho}) 
(k_\lambda \rlap/k \gamma_\rho \rlap/q +
k_\rho \rlap/k \gamma_\lambda \rlap/q) = 
4 k\cdot v \rlap/k \rlap/v \rlap/q - 2m^2 \rlap/q \,,
\end{eqnarray}
and a similar expression with the other term. Since the $\rlap/q$ term
vanishes between the spinors, we can write
\begin{eqnarray}
m'_{(A2t)} &=& 2e^2 \lim_{P\to 0} \lim_{Q\to 0} \int {d^4k\over (2\pi)^3} 
\; \delta(k^2-m^2) \eta_f(k) \nonumber\\* && \times
{k_0 \over m^2 - k \cdot p} \; {1 \over 2 \vec K \cdot \vec Q} \; 
\overline u(\vec P')
\Big( \rlap/k \rlap/v \rlap/q + \rlap/q \rlap/v \rlap/k \Big) u(\vec P) \,,
\label{A2t}
\end{eqnarray}
omitting the $Q^2$ terms in the denominator since they will not
contribute for $Q\to0$.  Using the identity
\begin{eqnarray}
\gamma_\kappa \gamma_\mu \gamma_\nu 
= \eta_{\kappa\mu} \gamma_\nu + \eta_{\mu\nu}
\gamma_\kappa - \eta_{\kappa\nu} \gamma_\mu - i
\varepsilon_{\kappa\mu\nu\alpha} \gamma^\alpha \gamma_5 \,,
\end{eqnarray}
we obtain
\begin{eqnarray}
\rlap/k \rlap/v \rlap/q + \rlap/q \rlap/v \rlap/k 
= 2 \vec K\cdot \vec Q \rlap/v 
\end{eqnarray}
between the spinors, since $q\cdot v=0$ and $\rlap/q$ terms vanish.
Putting this back into Eq.\ (\ref{A2t}) and using Eqs.\ (\ref{ugu})
and (\ref{unorm}), we obtain
\begin{eqnarray}
m'_{(A2t)} = - {2e^2 \over m} 
\int {d^3K \over (2\pi)^3 2E_K} \;
\Bigg[ {E_K \over E_K-m} f_e(E_K) + 
{E_K \over E_K+m} f_{\bar e}(E_K) \Bigg] \,.
\label{A2tfinal}
\end{eqnarray}
The sum of Eqs.\ (\ref{A2afinal}), (\ref{A2bfinal}), (\ref{A2rfinal}),
(\ref{A2sfinal}) and (\ref{A2tfinal}) gives the total contribution of
the A2 term, given in Eq.\ (\ref{A2final}) in the text.

\section{The X contribution}\label{app:tough:X}
The part of the integral from Eq.\ (\ref{GamX.5}) that we consider
here is given by
\begin{eqnarray}
\label{Aq}
I^{(f)}(Q) 
& = & \int {d^3K\over (2\pi)^3} F(E_K) 
\left( {1\over 2 \vec K\cdot \vec Q - Q^2} -
{1\over 2 \vec K\cdot \vec Q + Q^2} \right) \,,
\end{eqnarray}
where
\begin{eqnarray}
\label{Aq1} 
F(E) \equiv \Big(f_f(E) - f_{\bar f}(E) \Big)(2E^2 - m^2) \,.
\end{eqnarray}
Shifting the variables, the integral can be written as
\begin{eqnarray}
I^{(f)}(Q) & = &
\int {d^3K\over (2\pi)^3} 
{F(E_{\vec K + \frac{1}{2}\vec Q}) - 
F(E_{\vec K - \frac{1}{2}\vec Q}) \over 2\vec K\cdot\vec Q} \,. 
\end{eqnarray}
We have to expand the numerator to $O(Q^3)$ in order to obtain the
integral to $O(Q^2)$. Writing $\partial_i$ to denote a partial
derivative with respect to $K^i$,
\begin{eqnarray}
F(E_{\vec K \pm \frac{1}{2}\vec Q}) = F(E) \pm \frac{1}{2}Q^i\partial_i F
+ \frac{1}{2}\left(\frac{1}{4}Q^i Q^j\right)\partial_i\partial_j F \pm
\frac{1}{3!}\left(\frac{1}{8}Q^i Q^j Q^l\right)
\partial_i\partial_j\partial_l F \,.
\end{eqnarray}
The derivatives we need to use are:
\begin{eqnarray}
\partial_i F & = & K^i \left( {1\over E} {\partial \over \partial E}
\right) F \,, \nonumber\\
\partial_i\partial_j\partial_l F & = & (\delta^{ij}K^l + \delta^{il}K^j + 
\delta^{jl}K^i) \left( {1\over E} {\partial \over \partial E}
\right)^2 F +
K^i K^j K^l \left( {1\over E} {\partial \over \partial E} \right)^3 F \,.
\end{eqnarray}
Using
\begin{eqnarray}
K^i K^j \rightarrow \frac{1}{3}K^2\delta^{ij}
\end{eqnarray}
within the integrand, we have
\begin{eqnarray}
I^{(f)} (Q) = \int {d^3K \over (2\pi)^3}
\Bigg[ \frac12 \left( {1\over E} {\partial \over \partial E} \right) F + 
\frac{1}{3!}\frac{Q^2}{8}\left\{
3 \left( {1\over E} {\partial \over \partial E} \right)^2 F +  
\frac13 K^2 \left( {1\over E} {\partial \over \partial E} \right)^3 F 
\right\} \Bigg] \,.
\end{eqnarray}
Therefore, the quantity that we must substitute in Eq.\ (\ref{XQ-X0})
is
\begin{eqnarray}
I^{(f)}(Q) - I^{(f)}(Q\to 0) & = & 
\frac{1}{3!}\frac{Q^2}{8}\int {d^3K\over (2\pi)^3}
\left\{
3 \left( {1\over E} {\partial \over \partial E} \right)^2 F +  
\frac{K^2}{3} \left( {1\over E} {\partial \over \partial E} \right)^3 F 
\right\} \,.
\label{QX}
\end{eqnarray}
We now use the identity
\begin{eqnarray}
\int_0^\infty dK \; K^n \left( {1\over E} {\partial \over \partial E}
\right)^\nu F = 
- (n-1) \int_0^\infty dK \; K^{n-2} \left( {1\over E} {\partial \over
\partial E} \right)^{\nu-1} F \,,
\end{eqnarray}
which holds for $n\geq2$, so that the surface term vanishes. It is
obtained by using Eq.\ (\ref{EdE=KdK}) and performing a partial
integration.  Using it repeatedly, we can rewrite Eq.\ (\ref{QX}) as
\begin{eqnarray}
\label{Aqlast}
I^{(f)}(Q) - I^{(f)}(Q\to 0) & = & 
- \frac{Q^2}{48\pi^2}\int_0^\infty dK \;
\left( {1\over E} {\partial \over \partial E} \right) F \,.
\end{eqnarray}
Putting this back into Eq.\ (\ref{GamX.5}), we obtain the total X 
contribution given in Eq.\ (\ref{Xfinal}).

\section{Alternative evaluation of the B1 and A2 contributions}
\label{applambda}
If we straightaway put $q=0$ in Eqs.\ (\ref{GamB1}) and (\ref{A2}), they are seen to
involve singular expressions of the form ${\mathscr Pr}(1/x)\delta(x)$ with $x=k^2$
and $x=k^2-m^2$ respectively.  Using Eqs.\ (\ref{Pdelta}), we can obtain the following regularized form 
for such expressions \cite{alternative}:
\bea
{\mathscr Pr}\left(\frac{1}{x}\right)\delta(x)&=&
\lim_{\epsilon\rightarrow 0}\frac{\epsilon x}{\pi(x^2+\epsilon^2)^2}
                          \nonumber\\
     &=&\lim_{\epsilon\rightarrow 0}\lim_{\lambda\rightarrow 0}
\frac{\epsilon (x-\lambda^2)}{\pi[(x-\lambda^2)^2+\epsilon^2]^2}
                                                \nonumber\\
     &=&\lim_{\epsilon\rightarrow 0}\lim_{\lambda\rightarrow 0}
\frac{1}{2}\frac{\epsilon}{\pi}\frac{d}{d\lambda^2}
\frac{1}{ (x-\lambda^2)^2+\epsilon^2}
                                                \nonumber\\
     &=&\frac{1}{2}\lim_{\lambda\rightarrow 0}
\frac{d}{d\lambda^2}\delta(x-\lambda^2).            \label{eeq:reg}
\eea 

We shall find that the use of Eq.\ (\ref{eeq:reg}) leads to expressions
for $m'_{(B1)}$ and $m'_{(A2)}$ which are identical to the ones deduced earlier 
in this Appendix. Let us also compare Eq.\ (\ref{eeq:reg}) with Eq.\ (\ref{Pdel}).
It is clear that Eq.\ (\ref{eeq:reg}) for $x=k^2$ is identical with Eq.\ (\ref{Pdel})
for $m\rightarrow 0$. For the case of $m\neq 0$, Eq.\ (\ref{Pdel}) is quite
convenient to use if the rest of the expression in which it occurs does not involve
$m^2$ (so that $d/dm^2$ can be performed at the end, as in Ref.\ \cite{bedaque}).
But in our problem, the rest of the expression involves not only $m^2$, but $m$
and $m^3$ as well (see Eqs.\ (\ref{eq:l2}) and (\ref{eq:l3}) below). 
Consequently, we have found it convenient to use Eq.\ (\ref{eeq:reg}) with
$x=k^2-m^2$.

It may be noted that it is the form (\ref{eeq:reg}) which was essentially
employed by DHR to regularize ${\mathscr Pr}(1/k^2)\delta(k^2)$ in
$\Gamma^{\prime(B1)}_{\lambda\rho}(p,p)$ \cite{DHRapp}, who, however, made a detour
into the imaginary-time formulation for this purpose.
Here we first give the $m'_{(A2)}$ calculation, and then, for the
sake of completeness, summarize the $m'_{(B1)}$ calculation.


Putting $q=0$ in Eq.\ (\ref{A2}), we get
\bea
\Gamma^{\prime(A2)}_{\lambda\rho}(p,p)&=&2e^2\int\frac{d^4 k}{(2\pi)^3}
     \Lambda_{\lambda\rho}(k,k){\mathscr Pr}(\frac{1}{(k-p)^2}){\mathscr Pr}(\frac{1}{k^2-m^2})  
      \delta(k^2-m^2)\eta_{f}(k)
                                                \nonumber\\
      &=&e^2\int\frac{d^3 K}{(2\pi)^3}\lim_{\lambda\rightarrow 0}
         \frac{d}{d\lambda^2}\int dk_0 \Lambda_{\lambda\rho}(k,k)
          \frac{1}{2m^2+\lambda^2 -2p\cdot k} 
                                                \nonumber\\
         &&\times\delta(k^2-m^2-\lambda^2)\eta_{f}(k),
                                                                   \label{eq:l2}
\eea
using Eq.\ (\ref{eeq:reg}) and also setting $k^2=m^2+\lambda^2$ as
dictated by the delta function.

Making use of the delta function and Eqs.\ (\ref{ugu}) and (\ref{unorm}), we obtain,
from Eq.\ (\ref{Lambdayy}),
\bea
\!\!\!(2v^\lambda v^\rho -\eta^{\lambda\rho})\bar u(\vec P)\Lambda_{\lambda\rho}(k,k)u(\vec P)
=-8k_0^3+20mk_0^2-8m^3+2k_0(2m^2+\lambda^2-2mk_0)                           \label{eq:l3}
\eea
at $v^\mu=(1,\vec 0)$ and $p^\mu=(m,\vec 0)$.                

Use of Eq.\ (\ref{eq:l3}), and of $p^\mu=(m,\vec 0)$, in Eq.\ (\ref{eq:l2}) then
leads to 
\bea
m^\prime_{(A2)}=e^2(-8 I_1 +20m I_2 -8m^3 I_3 +2 I_4),   \label{eq:l4}
\eea
where
\bea
I_1&=&\int[d^4 k]_\lambda\frac{k_0^3}{2m^2+\lambda^2-2mk_0},\label{eq:l5} \\
I_2&=&\int[d^4 k]_\lambda\frac{k_0^2}{2m^2+\lambda^2-2mk_0}, \\ 
I_3&=&\int[d^4 k]_\lambda\frac{1}{2m^2+\lambda^2-2mk_0},     \\
I_4&=&\int[d^4 k]_\lambda k_0.  \label{eq:l8}
\eea
We have used the notation
\bea 
\int[d^4k]_\lambda \mbox{function}(k_0,\lambda)&\equiv&
    \int\frac{d^3 K}{(2\pi)^3}\lim_{\lambda\rightarrow 0}
         \frac{d}{d\lambda^2}\int dk_0 \delta(k_0^2-E_{K\lambda}^2)
     \eta_f(k) \nonumber\\
      &&\times\mbox{function}(k_0,\lambda),           \label{eq:l9}
\eea
with
\bea
E_{K\lambda}\equiv (K^2+m^2+\lambda^2)^{1/2}.       \label{eq:l10}
\eea
Performing the $k_0$ integration first, and then removing the regulator
$\lambda$ in Eqs.\ (\ref{eq:l5}) -- (\ref{eq:l8}) (it is useful to make note of
\bea
\lim_{\lambda\rightarrow 0}\frac{d}{d\lambda^2} f_{e,\bar e}
       (E_{K\lambda})=\frac{1}{2E_K}\frac{\partial}{\partial E_K}
        f_{e,\bar e}(E_K),
\eea
for this purpose), we arrive at
\bea
I_1&=&-\frac{1}{8m^2}\int\frac{d^3 K}{(2\pi)^3}
    \Bigg[\frac{E_K+2m}{E_K-m}
    f_e(E_K)-\frac{E_K-2m}{E_K+m}f_{\bar e}(E_K)\nonumber\\
    &&+mE_K\Bigg(\frac{1}{E_K-m}\frac{\partial f_e}{\partial E_K}
    +\frac{1}{E_K+m}\frac{\partial f_{\bar e}}{\partial E_K}
    \Bigg)\Bigg],
\eea
\bea
I_2&=&-\frac{1}{8m^2}\int\frac{d^3 K}{(2\pi)^3}
    \Bigg[\frac{1}{E_K}\Bigg(\frac{E_k+m}{E_k-m}
    f_e(E_K)+\frac{E_K-m}{E_K+m}f_{\bar e}(E_K)\Bigg)\nonumber\\
    &&+m\Bigg(\frac{1}{E_K-m}\frac{\partial f_e}{\partial E_K}
    -\frac{1}{E_K+m}\frac{\partial f_{\bar e}}{\partial E_K}
    \Bigg)\Bigg],
\eea 
\bea    
I_3&=&-\frac{1}{8m^2}\int\frac{d^3 K}{(2\pi)^3}
    \Bigg[\frac{1}{E_K^3}( f_e(E_K)+f_{\bar e}(E_K))\nonumber\\
    &&+\frac{m}{E_K^2}\Bigg(\frac{1}{E_K-m}\frac{\partial f_e}
    {\partial E_K}
    -\frac{1}{E_K+m}\frac{\partial f_{\bar e}}{\partial E_K}
    \Bigg)\Bigg],
\eea
\bea
I_4=\frac{1}{4}\int\frac{d^3 K}{(2\pi)^3}
    \frac{1}{E_K}\frac{\partial}{\partial E_K}
    ( f_e-f_{\bar e}).
\eea
Substitution in Eq.\ (\ref{eq:l4}) finally gives $m^\prime_{(A2)}$,
which is found to be the same as the result given by Eq.\ (\ref{A2final}). 
 
For $\Gamma^{\prime(B1)}_{\lambda\rho}(p,p)$, given by Eq.\ (\ref{GamB1}),
similar steps are to be
carried out. Thus the analogues of Eqs.\ (\ref{eq:l2}), (\ref{eq:l3})
and (\ref{eq:l4}) are 
\bea
\Gamma^{\prime(B1)}_{\lambda\rho}(p,p)
      &=&e^2\int\frac{d^3 K}{(2\pi)^3}\lim_{\lambda\rightarrow 0}
         \frac{d}{d\lambda^2}\int dk_0 \gamma^\nu(\rlap /p-\rlap /k
         +m)\gamma^\mu C_{\mu\nu\lambda\rho}(k,k)\nonumber\\
         &&\times \frac{1}{\lambda^2 -2p\cdot k} 
         \delta(k^2-\lambda^2)\eta_{\gamma}(k),
                                           \label{eq:l16}
\eea
\bea
\!\!\!(2v_\lambda v_\rho-\eta_{\lambda\rho}) \bar u(\vec P)\gamma^\nu(\rlap /p-\rlap /k
         +m)\gamma^\mu u(\vec P)C_{\mu\nu\lambda\rho}(k,k)
         =2\lambda^2(k_0-2m)+4mk_0^2-8k_0^3,  
\eea
\bea
m^\prime_{(B1)}=e^2(2J_1 +4mJ_2 -8J_3).
\eea
Here
\bea
J_1&=&\int[d^4 k]_\lambda\frac{\lambda^2(k_0-2m)}{\lambda^2-2mk_0}, \\
J_2&=&\int[d^4 k]_\lambda\frac{k_0^2}{\lambda^2-2mk_0}, \\ 
J_3&=&\int[d^4 k]_\lambda\frac{k_0^3}{\lambda^2-2mk_0}.     
\eea
The notation $[d^4 k]_\lambda$ is now defined by Eq.\ (\ref{eq:l9}) with
$\eta_f\rightarrow\eta_\gamma$, and the notation $E_{K\lambda}$ by Eq.\ (\ref{eq:l10})
with $m=0$.

The evaluation of the above three integrals is straightforward. Each one 
reduces to the standard integral
\bea
\int_0^\infty dK\:K f_\gamma(K)=\frac{\pi^2 T^2}{6},
\eea
while $J_3$ also involves
\bea
\int_0^\infty dK\:K^2\frac{e^{\beta K}}{(e^{\beta K}-1)^2}=-\frac{\partial}
                    {\partial\beta} \int_0^\infty dK\:K f_\gamma(K)
\eea
We then have $J_1=-T^2/24m$, $J_2=-T^2/48m^2$ and $J_3=T^2/48m$, so that
the value of $m^\prime_{(B1)}$ is the same as the one stated in Eq.\ (\ref{B1final}).

Finally, we comment that the regularization presented above is not useful 
for evaluating the $X$-contribution, as we need the $O(Q^2)$ contribution
from the fermion loop in that case, and not the the value at $q=0$.

\chapter{Thermal contribution to the equation for the effective mass
         at high temperature} \label{app:masseqn}
In this appendix we arrive at Eq.\ (\ref{eq:40}), which expresses the
high $T$
contribution to the equation for the effective
mass, by explicit calculation of the expressions given in Eqs.\ (\ref{eq:33}),
(\ref{eq:34}), (\ref{eq:mb1}), (\ref{eq:mb2}) and (\ref{eq:mb3}).

We begin with $M{^\prime_{\rm FG}}^2$, given by Eq.\ (\ref{eq:33}). After the $p_0$
integration is carried out, the integrand is a function of $P$ (the
magnitude of the three-momentum) alone.
Then integrating over the angles trivially, we are left with a one-dimensional
 integral. On doing some rearrangement, and making use of the identity
\bea
f_B(P)-f_F(P)=2f_B(2P)              \label{eq:identity}
\eea
(here and elsewhere in this appendix, $f_F(P)$ and $f_B(P)$ denote (\ref{eq:9})
and (\ref{eq:13}) with $|p_0|$ replaced by $P$), we obtain
\bea
M{^\prime_{\rm{FG}}}^2=\frac{e^2}{2\pi^2}\int_0^\infty dP\:P[f_B(P)+f_F(P)]
   -\frac{e^2 M^2}{4\pi^2}\int_0^\infty dP\:Pf_B(P)\re
          \frac{1}{P^2-M^2-i\epsilon}\,.     \label{eq:52}
\eea
Now the first part of the R.H.S. of Eq.\
(\ref{eq:52}) can be evaluated exactly by using
\bea
\int_0^\infty dP\:P f_B(P)&=&\frac{\pi^2 T^2}{6},\\*
\int_0^\infty dP\:P f_F(P)&=&\frac{\pi^2 T^2}{12}.
\eea
The second part of the R.H.S. of Eq.\ (\ref{eq:52}) involves the integral
\bea
I&\equiv&\int_0^\infty dP\:P\frac{1}{e^{P/T}-1} {\mathscr Pr}\frac{1}{P^2-M^2 }
                                          \label{eq:55} \\
 &=&\int_0^\infty dx\:x\frac{1}{e^x-1} {\mathscr Pr}\frac{1}{x^2-(M/T)^2 }.
                                           \label{eq:Integral}
\eea
Here $\mathscr Pr$ denotes the principal value. The functional
dependence of $I$ on $T/M$  was determined on a computer
 for large $T/M$, yielding
\bea
I\approx -\frac{1}{2}\ln\frac{T}{M}.      \label{eq:55a}
\eea
Note that this logarithmic behaviour is in accord with the ultraviolet behaviour
 of (\ref{eq:55}) without the distribution function. Thus at high T
\bea
M{^\prime_{\rm FG}}^2=e^2\Big(\frac{T^2}{8}+\frac{M^2}{8\pi^2}\ln\frac{T}{M}\Big)
                                   \label{eq:56}
\eea
neglecting terms independent of $T$.

Next we turn to $M{^\prime_{\xi,F}}^2$, given by Eq.\ (\ref{eq:34}). Integrating
over $p_0$ and the angles, we are led to
\bea
M{^\prime_{\xi,F}}^2&=&\frac{\xi e^2 M^2}{8\pi^2}\int_0^\infty dP\:P
      f_F(P)\re\frac{1}{P^2-\frac{M^2}{4}-\frac{i\epsilon}{2}}\nonumber\\*
       &&+\frac{\xi e^2 M^4}{32\pi^2} \int_0^\infty dP\:P
      f_F(P)\re\frac{1}{(P^2-\frac{M^2}{4}-\frac{i\epsilon}{2})^2}.
                                  \label{eq:57}
\eea
Moving on to $M{^\prime_{\xi,B}}^2$, integration over the angles in Eqs.\ (\ref{eq:mb1}),
(\ref{eq:mb2}) and (\ref{eq:mb3}) lead to
\bea
M{^\prime_{\xi,B({\rm I})}}^2&=&-\frac{\xi e^2 M^2}{8\pi^2}\int_0^\infty dP\:P
      f_B(P)\re\frac{1}{P^2-\frac{M^2}{4}-\frac{i\epsilon}{2}},\\*
M{^\prime_{\xi,B({\rm II})}}^2&=&-\frac{\xi e^2 M^4}{32\pi^2} \int_0^\infty dP\:P
      f_B(P)\re\frac{1}{(P^2-\frac{M^2}{4}-\frac{i\epsilon}{2})^2},\\*
M{^\prime_{\xi,B({\rm III})}}^2&=&-\frac{\xi e^2 M^2\beta}{8\pi^2}\int_0^\infty dP\:
          P^2 \frac{e^{\beta P}}{(e^{\beta P}-1)^2}
          \re\frac{1}{P^2-\frac{M^2}{4}-\frac{i\epsilon}{2}}.
\eea
Using the identity (\ref{eq:identity}) we then arrive at
\bea
M{^\prime_{\xi,F}}^2+M{^\prime_{\xi,B({\rm I})}}^2+
                       M{^\prime_{\xi,B({\rm II})}}^2 
              &=&-\frac{\xi e^2 M^2}{4\pi^2}\int_0^\infty dP\:P
               f_B(P)\re\frac{1}{P^2-M^2-2i\epsilon} \nonumber  \\*
               &&   -\frac{\xi e^2 M^4}{4\pi^2} \int_0^\infty dP\:P
               f_B(P)\re\frac{1}{(P^2-M^2-2i\epsilon)^2}.\nonumber\\*
                    \label{eq:1111}
\eea
While the first part again involves (\ref{eq:55}), the second part, being
ill-defined, is regularized in a way similar to (\ref{eq:reg}):
\bea
\int_0^\infty dP\:P f_B(P)\re\frac{1}{(P^2-M^2-2i\epsilon)^2}
 =\lim_{\lambda\rightarrow M}\frac{\partial}{\partial\lambda^2}
\int_0^\infty dP\:P f_B(P)\re\frac{1}{P^2-\lambda^2-2i\epsilon}\,.\nonumber\\*
                                \label{eq:58}
\eea
The integral on the R.H.S. of Eq.\ (\ref{eq:58}) is similar to $I$ in Eq.\ (\ref{eq:55})
and equals $-\frac{1}{2}\ln (T/\lambda)$ for high $T$. So the second part of
(\ref{eq:1111}) equals $-\xi e^2 M^2/16\pi^2$ which can be neglected as it
is independent of $T$. Thus we are left with the first part of
 (\ref{eq:1111}), so that
\bea
M{^\prime_{\xi,F}}^2+M{^\prime_{\xi,B({\rm I})}}^2+
                       M{^\prime_{\xi,B({\rm II})}}^2
=\frac{\xi e^2 M^2}{8\pi^2}\ln\frac{T}{M}.  \label{eq:59}
\eea
In $M{^\prime_{\xi,B({\rm III})}}^2$, we encounter the integral
\bea
J\equiv\frac{1}{T}\int_0^\infty dP\:P^2
\frac{e^{P/T}}{(e^{P/T}-1)^2}
 {\mathscr Pr}\frac{1}{P^2-{M_1}^2}   \label{eq:62}
\eea
where $M_1=M/2$. Converting $J$ into a function of $T/M_1$ alone
(as we did in the case of $I$ in Eq.\ (\ref{eq:Integral})),
the functional dependence was found out for large $T/M_1$ using a
 computer, which gave $J\approx -\frac{1}{2}$.
  This constancy again agrees
with the expectation from the argument involving the ultraviolet behaviour.
Alternatively one can note that
\bea
J=T\frac{\partial}{\partial T}\int_0^\infty dP\:P\frac{1}{e^{P/T}-1}
       {\mathscr Pr}\frac{1}{P^2-{M_1}^2}.
\eea
Then using our knowledge of $I$ (see Eqs.\ (\ref{eq:55}) and {\ref{eq:55a})),
we obtain $J\approx -\frac{1}{2}$ at large $T$.
Thus $M{^\prime_{\xi,B({\rm III})}}^2$ equals $\xi e^2 M^2/16\pi^2$
and can be neglected.

Therefore in the high-$T$ limit, ${M^\prime}^2$ receives contribution from (\ref{eq:56})
and (\ref{eq:59}), thereby leading us to Eq.\ (\ref{eq:40}).


%
%
%
\begin{plist}

\begin{enumerate}
\item 
     I. Mitra, J. F. Nieves and P. B. Pal, ``Gravitational couplings of
     charged leptons in a medium,"
     Phys. Rev {\bf D64}, 085004 (2001) [hep-ph/0104248].
\item
     I. Mitra, ``Gauge independence of limiting cases of one-loop electron
     dispersion relation in high-temperature QED," Phys. Rev. {\bf D62},
     045023 (2000) [hep-ph/9905362].
\item
     I. Mitra, A. DasGupta, and B. Dutta-Roy,
     ``Regularization and renormalization in scattering from Dirac delta
     potentials," Am. J. Phys. {\bf 66}, 1101 (1998): Not included in this
     thesis.
\end{enumerate}
\end{plist}


\begin{thebibliography}{99}
\bibitem{notation}
      It may be noted that Refs.\ \cite{qed:weldon} and \cite{qed:indra} use upper-case
      letter for the electron four-momentum, and define $a$ and $b$ with an extra minus 
      sign.
\bibitem{footnote3}
     One way of making $\partial f/ \partial\lambda^2$ at $\lambda=0$ finite
     in the present case is to use in it
     a regularization similar to the one given in (\ref{eq:reg}), and keep the
     regulator (small but) non-zero.
\bibitem{symbol}
     We reserve the symbol $M$ for the effective electron mass including the
     subleading $T$-dependence.
\bibitem{similar}
     Similar idea has been used for neutrino dispersion relation in
     Ref.\ \cite{josenieves}.
\bibitem{alternative}
     Alternatively, take the real part of the relation (3.104) of M.\ Le Bellac (Ref.\
     \cite{lebellac}), and use Eq.\ (\ref{eq:delta}) of the present work in it.
\bibitem{DHRapp}
      The finite--temperature part in Eq.\ (A13) of the Appendix of Ref.\ \cite
      {DHR84} is the same as our Eq. (\ref{eq:l16}).
\end{thebibliography}
\end{document}